\documentclass[12pt]{article}
\pdfoutput=1
\usepackage{epsf,amsfonts,amssymb,epsfig,amsmath,graphics,cite}
\usepackage[dvipsnames]{xcolor}
\usepackage{hyperref,graphicx,subfig}
\usepackage{booktabs}
\usepackage{float}
\usepackage{verbatim}
\usepackage{mathrsfs}
\usepackage{empheq}
\usepackage{enumerate}
\usepackage{bbm}
\usepackage{soul}
 
\makeatletter

\newcommand{\Rmnum}[1]{\expandafter\@slowromancap\romannumeral #1@}
\makeatother

\newcommand{\nn}{\notag }
\def\be{\begin{equation}}
\def\ee{\end{equation}}

\newcommand{\ii}{\mathrm{i}}
\newcommand{\ex}{\mathrm{e}}
\newcommand{\diff}{\mathrm{d}}
\newcommand{\dd}{\mathrm{d}}
\newcommand{\R}{\mathbb{R}}
\newcommand{\Z}{\mathbb{Z}}
\newcommand{\vol}{\mathrm{vol}}
\newcommand{\Vol}{\mathrm{Vol}}
\newcommand{\C}{\mathbb{C}}
\newcommand{\cG}{\mathcal{G}}
\newcommand{\cL}{\mathcal{L}}
\newcommand{\identity}{\mathbbm{1}}

\newcommand{\mc}[1]{\mathcal{#1}}

\newcommand{\rcharge}{q}
\newcommand{\hook}{\mathbin{\rule[.2ex]{.4em}{.03em}\rule[.2ex]{.03em}{.9ex}}}
\newcommand{\fleft}{\left.}
\newcommand{\fright}{\, \right|_F\ }

\newcommand{\Qmat}{Q}
\newcommand{\Imat}{I}

\newcommand{\MM}{\mathcal{M}}
\newcommand{\varphinew}{\phi}

\newcommand{\INN}{\langle N, N \rangle}
\newcommand{\InN}{\langle N, n \rangle}
\newcommand{\Inn}{\langle n, n \rangle}

\numberwithin{equation}{section}       

\begin{document}

\pagenumbering{Alph}
\begin{titlepage}


\vskip 1cm

\begin{center}


{\Large \bf Equivariant localization for AdS/CFT}

\vskip 1cm
{Pietro Benetti Genolini,$^{\mathrm{a}}$ Jerome P. Gauntlett,$^{\mathrm{b}}$ and James Sparks$^{\mathrm{c}}$}

\vskip 1cm

${}^{\mathrm{a}}$\textit{Department of Mathematics,
King's College London,\\
Strand, London, WC2R 2LS, U.K.\\}

\vskip 0.2cm

${}^{\mathrm{b}}$\textit{Blackett Laboratory, Imperial College, \\
Prince Consort Rd., London, SW7 2AZ, U.K.\\}

\vskip 0.2cm

 ${}^{\mathrm{c}}$\textit{Mathematical Institute, University of Oxford,\\
Andrew Wiles Building, Radcliffe Observatory Quarter,\\
Woodstock Road, Oxford, OX2 6GG, U.K.\\}

\vskip 0.2 cm

\end{center}

\vskip 0.5 cm

\begin{abstract}
\noindent We explain how equivariant localization may be 
applied to AdS/CFT  to compute various BPS observables in gravity, 
such as central charges and conformal dimensions of 
chiral primary operators, without solving the supergravity equations. 
The key ingredient is that 
supersymmetric AdS solutions with an R-symmetry 
are equipped with a set of equivariantly closed forms. 
These may in turn be used to impose flux quantization and 
compute observables for supergravity solutions, using only topological information 
and the Berline--Vergne--Atiyah--Bott fixed point formula. 
We illustrate the formalism by considering 
$AdS_5\times M_6$ and $AdS_3\times M_8$ solutions
of $D=11$ supergravity. 
As well as recovering results for many classes of well-known supergravity  
solutions, without using any knowledge of their explicit form, 
we also compute central charges for which explicit supergravity solutions have not 
been constructed.

\end{abstract}

\end{titlepage}

\pagenumbering{arabic}

\pagestyle{plain}
\setcounter{page}{1}
\newcounter{bean}
\baselineskip18pt

\tableofcontents


\section{Introduction}\label{sec:intro}

The systematic analysis of the geometry that
underlies the AdS/CFT correspondence in the supersymmetric context
is an important ongoing investigation in holography. In order to geometrically characterize general classes of SCFTs that have a holographic dual, 
one considers spacetimes of the form $AdS\times M$ in either $D=10$ or $D=11$ supergravity, where $M$ is a compact manifold, together with a
warped product metric and general fluxes, both of which are preserved by the symmetries of the $AdS$ spacetime.
Demanding that some supersymmetry is preserved then leads to the existence of
certain Killing spinors on $M$ which, possibly combined with the equations of motion and/or the Bianchi identities
for the fluxes, defines the supersymmetric geometry on $M$. The Killing spinors define a $G$-structure on $M$ \cite{Gauntlett:2002sc}, and the geometry can be effectively studied by analyzing Killing spinor bilinears.

The first such analysis was carried out for the general class of $AdS_5\times M_6$ solutions of $D=11$ supergravity in \cite{Gauntlett:2004zh}. The geometry on $M_6$ was precisely characterized and, moreover, was then used to construct infinite classes of explicit $AdS_5\times M_6$ solutions, dual to $\mathcal{N}=1$ SCFTs in $d=4$. 
Subsequently, similar analyses have been carried out for many different cases in various dimensions (some early examples include \cite{Lin:2004nb,Gauntlett:2005ww,Kim:2005ez,Gauntlett:2006qw,Gauntlett:2006ux,Kim:2006qu,DHoker:2007zhm,Gauntlett:2007ph})
and various important insights have been made. 

In this paper, which complements our recent papers 
\cite{BenettiGenolini:2023kxp,BenettiGenolini:2023yfe}, we reveal an important general structure that these geometries possess 
which has been overlooked for nearly twenty years. Specifically, provided that the preserved supersymmetry includes an R-symmetry,
{\it i.e.} the dual SCFT admits an R-symmetry, 
we show there is a natural equivariant calculus involving a subset of the Killing spinor bilinears. Furthermore, this calculus can be used to compute physical quantities of the dual SCFTs without knowing the full supergravity solution, and thus defines a canonical way to
do various computations off-shell. 

When the dual SCFT has an R-symmetry the
geometry on $M$ has a canonical Killing vector $\xi$ that can be constructed as a bilinear from the Killing spinor. It is then natural to introduce the equivariant exterior derivative
$\dd_\xi\equiv \dd-\xi\hook$\, . This derivative acts on polyforms, {\it i.e.} sums of differential forms of different degrees, and satisfies $\dd_\xi^2=-\mathcal{L}_\xi$, where 
$\mathcal{L}_\xi$ is the Lie derivative with respect to $\xi$. In the case that the compact manifold $M$ has even
dimension, by explicit computation for several different set-ups,
we show that various equivariantly closed polyforms $\Phi$, satisfying $\dd_\xi\Phi=0$, can be constructed from 
the subset of the Killing spinor bilinears which are invariant under the action of $\mathcal{L}_\xi$. 
Furthermore, we show that the integrals of the polyforms compute physical quantities of the dual SCFT, such as
the central charge and flux integrals. The Berline--Vergne--Atiyah--Bott (BVAB) fixed point formula \cite{BV:1982,Atiyah:1984px}
can then be used to compute these physical quantities as a sum of contributions arising from the fixed point set of the action of $\xi$. 

Before discussing the examples that we consider in this paper, we further clarify in what sense 
our procedure is off-shell.
For a given class of $AdS\times M$ geometries we start with an action for some fields on $M$ (metric, warp factor function and forms for the fluxes), obtained by dimensional reduction from $D=10,11$ supergravity, and hence, by definition, $AdS\times M$ solutions extremize this action. We are interested in supersymmetric solutions, and for such $AdS\times M$ geometries
it is known that supersymmetry, sometimes supplemented by the equations of motion for the fluxes, actually implies the equations of motion. Thus, if one imposes \emph{all} of the supersymmetry equations, plus the equations of motion for the fluxes, one will also have an on-shell solution. Instead, what we show is that the supersymmetry equations imply that a certain \emph{subset} of differential forms that we construct as bilinears of the Killing spinor are equivariantly closed, and this data thus imposes a subset of the equations of motion. As we show, and remarkably,  these differential forms are then sufficient to compute various physical quantities of interest, such as fluxes and central charges using the BVAB fixed point formula, including the action on $M$ itself. In fact in the examples
in this paper the central charge is proportional to the partially on-shell action. Thus, in our computations we do not impose all of the supersymmetry equations, and our final answers for the central charge (say) is an off-shell result that requires, generically, extremization over a finite number of constant parameters in order to obtain the on-shell result. Thus, our results provide a precise holographic analogue of field theory extremization principles, {\it e.g.} \cite{Intriligator:2003jj,Benini:2012cz,Benini:2013cda}.

There have been previous holographic incarnations of extremization principles that arise in supersymmetric conformal field theories. For example, substantial progress has been made for $AdS_5\times SE_5$ solutions with just five-form flux and $AdS_4\times SE_7$ solutions with just electric four-form flux, where $SE_{2n+1}$ is a Sasaki--Einstein space \cite{Acharya:1998db}. In particular, it has been shown that various physical quantities of the dual SCFTs can be computed by suitably going
off-shell and solving a novel variational problem over Sasaki metrics \cite{Martelli:2005tp,Martelli:2006yb}. More recently, starting with \cite{Couzens:2018wnk},
there has been similar progress for $AdS_3\times GK_7$ solutions of type IIB with general five-form flux \cite{Kim:2005ez} and 
$AdS_2\times GK_9$ solutions of $D=11$ with electric four-form flux \cite{Kim:2006qu}, where $GK_{2n+1}$ is a GK geometry \cite{Gauntlett:2007ts}. For both the SE and the GK geometry the precise class of geometries that one is extremizing over has been clearly identified and it would be desirable to have an analogous understanding of this for the more general $AdS\times M$ geometries that we study using equivariant localization, but we leave that for future work.

In this paper we will illustrate our new calculus for three classes of solutions in $D=11$ supergravity. 
We first consider the class of $AdS_5\times M_6$ solutions of \cite{Gauntlett:2004zh}, expanding and extending the
discussion in \cite{BenettiGenolini:2023kxp}. We show in detail how the new formalism can be used to obtain
the central charge and scaling dimensions of certain chiral primaries for various different classes of
examples of $M_6$. In several classes, the explicit solutions are known and we can compare our results with the known answer, finding exact agreement. This detailed exercise reveals how the new equivariant calculus can be used and, in particular, we will see that the details depend on which class of examples of $M_6$ is being considered. We also carry out similar computations for classes of solutions which are not known in explicit form, thus providing results for new supergravity solutions, just assuming that they exist. 
We also note that we consider examples where the dual $d=4$ SCFT has $\mathcal{N}=2$ supersymmetry and hence a non-Abelian R-symmetry; however in deriving the central charge we only utilise an Abelian R-symmetry.

Our next class of examples consist of $AdS_3\times M_8$ solutions. A sub-class of geometries
were first analysed in \cite{Gauntlett:2006qw} and this was later extended to a general classification
in \cite{Ashmore:2022ydf}. Here we will display the equivariant polyforms for the sub-class considered in \cite{Gauntlett:2006qw}, for which it
is known that there is a rich class of solutions.  We certainly expect that our results can be generalized
to the general class of \cite{Ashmore:2022ydf}, but we shall leave that to future work. We will again deploy the new calculus
to recover some known results, as well as obtain some new results, focusing on cases
where $M_8$ is an $S^4$ fibration over a four-dimensional base $B_4$, associated with M5-branes wrapped
on $B_4$. In \cite{BenettiGenolini:2023yfe} we consider examples where the R-symmetry has isolated fixed points and show that
one can obtain off-shell expressions for the central charge expressed in terms of `gravitational blocks' \cite{Hosseini:2019iad}, while here we consider a complementary class of examples. 

Finally, we also consider the $AdS_7\times S^4$ solution, dual to the $(2,0)$ SCFT in $d=6$.
Of course, this solution is explicitly known and one can immediately obtain the central charge of the dual field theory by a straightforward computation.
However, it is illuminating to analyse it from the new point of view as it underscores the universality of the approach.

In this paper we focus on $AdS\times M$ solutions, but our new formalism based on equivariant localization is more general, with wider applications. Indeed in 
\cite{BenettiGenolini:2023kxp} we also discussed supersymmetric solutions of various gauged supergravity theories
which admit an R-symmetry where the equivariant calculus can also be deployed to compute quantities of 
physical interest. In particular, these results 
provided a systematic derivation 
of the result for the on-shell action in minimal, gauged, Euclidean supergravity in $D=4$ that was derived 
in \cite{BenettiGenolini:2019jdz} as well as generalizing this to $D=6$.

Before concluding this introduction we briefly compare our work on $AdS\times M$ solutions
with that of \cite{Martelli:2023oqk}, which
also discussed equivariant localization in holography. A detailed discussion
of the so-called \emph{equivariant volume} of symplectic toric orbifolds was made in \cite{Martelli:2023oqk} and,
in particular, it was shown that various known
 \emph{off-shell}  holographic results for Sasaki--Einstein geometry and GK geometry could be recast in an elegant way in this language. Furthermore, by rephrasing certain off-shell field theory results it was suggested that, more generally, the equivariant volume should play a role in analysing certain classes of supergravity solutions associated with wrapped M5-branes and wrapped D4-branes. However,  the  crucial step of how to go off-shell in a supergravity context was not given. 
 By contrast, we emphasize that our approach does provide a canonical procedure for going off-shell, utilizing
 the structure of spinor bilinears, and furthermore this does not rely on the specific setting of symplectic toric orbifolds considered in \cite{Martelli:2023oqk}. 
Thus, our approach should be applicable more universally to all holographic geometries with an R-symmetry, as illustrated here and in various other examples in \cite{BenettiGenolini:2023kxp, BenettiGenolini:2023yfe}.

The plan of the rest of the paper is as follows. In section \ref{sec:EquivariantLocalization} we briefly review the localization formula of BVAB, as well
as discuss some simple examples.
In sections \ref{sec:M6} and \ref{sec:M8} we consider $AdS_5\times M_6$ and  $AdS_3\times M_8$ solutions, respectively,
and we conclude with some discussion in section \ref{sec:disc}. We also have three appendices.
In appendix \ref{app:SpinorRegularity} we examine how spinors with a definite charge under a Killing vector behave near a fixed point,
and how this is correlated with the chirality of the spinor. 
In appendix \ref{app:AdS7} we discuss the $AdS_7\times S^4$ solution using equivariant localization.
Finally, in appendix \ref{app:homology} we prove some homology relations for manifolds which are the total space of even-dimensional sphere 
bundles, focusing on $S^2$ and $S^4$ bundles, that we use in the main text.

\section{The equivariant localization theorem}
\label{sec:EquivariantLocalization}

The key mathematical result we use in this paper is the equivariant localization formula of Berline--Vergne--Atiyah--Bott (BVAB)\cite{BV:1982,Atiyah:1984px}. We begin by introducing some notation, reviewing the formula\footnote{See \cite{BV} for a pedagogical introduction.} and illustrating how it can be used with some simple examples.

We consider compact Riemannian manifolds $(M,g)$ of dimension $2n$, with a $U(1)$ isometric action generated by a Killing vector $\xi$. 
On the space of polyforms, which are formal sums of differential forms of various degrees, one introduces the equivariant differential
\begin{equation}
	\dd_\xi \equiv \dd - \xi \hook \, . 
\end{equation}
This differential squares to minus the Lie derivative along $\xi$: $\dd_\xi^2 = - \cL_\xi$. Thus, on the space of invariant polyforms $\{\Phi\}$, with $\cL_\xi {\Phi} = 0$, the differential above is nilpotent and we can construct an equivariant cohomology.
In the special case that $\xi$ generates a (locally) free $U(1)$ action on $M$, where all the orbits of $\xi$ are circles, this cohomology is simply the de Rham cohomology of the quotient $M/U(1)$. However, we are interested in the case that $\xi$ does not act freely, and there is a non-trivial fixed point set $F\equiv \{\xi=0\}\subset M$. 

The integral of a polyform is defined to be the integral of its top component. We also define the inverse of a polyform via a formal geometric series
\begin{equation}
\label{eq:GeometricSeries}
	(1-\Phi)^{-1} = \sum_{a = 0}^\infty \Phi^a \, ,
\end{equation}
where $\Phi^a$ is understood to be a wedge product of $a$ copies of $\Phi$, 
and note this series necessarily truncates.

The BVAB theorem expresses the integral of an equivariantly closed form $\Phi$ on $M$ in terms of contributions from the fixed point set $F$ of the group action. The connected components of $F$ have even codimension, so the normal bundle $\mc{N}$ to $F$ in $M$ is an even-dimensional orientable bundle. 
Then the following relation holds
\begin{equation}
\label{eq:BV_v1}
	\int_M \Phi = \int_{F} \frac{1}{d_F} \frac{f^* \Phi}{e_\xi(\mc{N})} \, .
\end{equation}
Here $f:F \hookrightarrow M$ is the embedding of the fixed point locus, $e_\xi(\mc{N})$ is the equivariant Euler form of the normal bundle, and $d_F\in \mathbb{N}$ is the order of the orbifold structure group of $F$ (in cases where $F$ and $M$ have orbifold 
singularities). {Thus, if a $2n$-form is the top form of an equivariantly closed polyform, then the integral over $M_{2n}$ of the 
$2n$-form is given by a sum of integrals of lower rank over the components of the fixed point set.}

To be more concrete, consider a connected component of $F$ with codimension $2k$. Then, $\xi$ generates a linear isometric action on the rank-$2k$ normal bundle, so there are local coordinates in which it has the form
\begin{equation}
	\xi = \sum_{i=1}^k \epsilon_i \, ( - x_{2i} \partial_{x_{2i-1}} + x_{2i-1} \partial_{x_{2i}} ) = \sum_{i=1}^k \epsilon_i \, \partial_{\varphi_i} \, .
\end{equation} 
Here $(x_{2i},x_{2i-1})$ are Cartesian coordinates on the $i$th two-plane $\R^2\subset \R^{2k}$ in a normal space, with
 $\varphi_i$ corresponding polar angles with period $\Delta\varphi_i=2\pi$. For generic\footnote{While this assumption will be sufficient for the examples studied in this paper, there are certainly cases where $\mathcal{N}$ does not 
split into a sum of complex line bundles. But in that case one should simply revert to the general 
formula \eqref{eq:BV_v1}, rather than apply \eqref{eq:BV_v3}. } $\epsilon_i$, $\mc{N}$ splits into a sum of complex line bundles $\mc{N} \cong L_1 \oplus \cdots \oplus L_k$, with $L_i$  rotated by $\partial_{\varphi_i}$, and we can simplify the expressions in \eqref{eq:BV_v1}. Specifically, since
the equivariant Euler class satisfies the Whitney sum formula, we can reduce it to the product of the equivariant Euler classes of each summand $L_i$, given by the sum of the Chern class of the bundle (representing the ordinary curvature), and the infinitesimal action generated by $\xi$ to obtain
\begin{equation}
\label{eq:EquivariantEuler}
	e_\xi(\mc{N}) = \prod_{i=1}^k e_\xi(L_i) = \prod_{i=1}^k {\rm Pf} \left( \frac{ R(L_i) + \cL_\xi |_{L_i} }{2\pi} \right) = \prod_{i=1}^k \left[ c_1(L_i) + \frac{\epsilon_i}{2\pi} \right] \, ,
\end{equation}
where $R$ denotes  a curvature two-form. 
In the last step we used the restriction of the Lie derivative to the normal bundle
\begin{equation}
\label{eq:LieDerivativeAction}
	\cL_\xi |_{L_i} = \begin{pmatrix} 0 & \epsilon_1 & \cdots & 0 & 0 \\ -\epsilon_1 & 0 & \cdots & 0 & 0 \\ & & \ddots & & \\ 0 & 0 & \cdots & 0 & \epsilon_k \\ 0 & 0 & \cdots & -\epsilon_k & 0 \end{pmatrix}\,,
\end{equation}
and the definition of the Pfaffian of a skew-symmetric matrix in the form above is
\begin{equation}
	{\rm Pf}(\cL_\xi |_{L_i}) = \epsilon_1 \cdots \epsilon_k \, .
\end{equation}
The inverse of the equivariant Euler form \eqref{eq:EquivariantEuler} can then be found using the geometric series \eqref{eq:GeometricSeries} and we have
\begin{equation}
	\frac{1}{e_\xi(\mc{N})} = \frac{(2\pi)^k}{\prod_{i=1}^k \epsilon_i} \left[ 1 - \sum_{1 \leq i \leq k} \frac{2 \pi}{\epsilon_i} c_1(L_i) + \sum_{1 \leq i \leq j \leq k} \frac{(2\pi)^2}{\epsilon_i \epsilon_j} c_1(L_i) \wedge c_1(L_j) + \cdots \right] \, .
\end{equation}

Thus, the integral of an equivariantly closed form on a $2n$-dimensional manifold, can receive contributions from
connected components of the fixed point set of various dimensions. 
Putting these points together, we can therefore write the BVAB formula in the following form 
\begin{align}
\label{eq:BV_v3}
	\int_M \Phi &= \sum_{\text{dim $0$}} \frac{1}{d_{F_0}} \frac{(2\pi)^n}{\epsilon_1 \cdots \epsilon_n} \Phi_0 + \sum_{\text{dim $2$}} \frac{1}{d_{F_2}}\frac{(2\pi)^{n-1}}{\epsilon_1 \cdots \epsilon_{n-1}} \int \Big[ \Phi_2 - \Phi_0 \sum_{1 \leq i \leq n-1} \frac{2\pi}{\epsilon_i} c_1(L_i) \Big] \nn \\
	&  + \sum_{\text{dim $4$}} \frac{1}{d_{F_4}} \frac{(2\pi)^{n-2}}{\epsilon_1\cdots \epsilon_{n-2}} \int \Big[ \Phi_4 - \Phi_2 \wedge \sum_{1 \leq i \leq n-2} \frac{2\pi}{\epsilon_i} c_1(L_i) \nn \\
	& \qquad \qquad \qquad \qquad \qquad \quad + \Phi_0 \sum_{1 \leq i \leq j \leq n-2} \frac{(2\pi)^2}{\epsilon_i \epsilon_j}  c_1(L_i) \wedge c_1(L_j) \Big] + \dots  \, , 
\end{align}
where the normal space to a subspace $F_{2n-2k}$ is $\R^{2k}/\Gamma$, where $\Gamma \subset SO(2k)$ is the finite group of order $d_{F_{2n-2k}}$ defining the orbifold structure (which 
is the trivial group in the case of smooth manifolds).
Notice that in writing \eqref{eq:BV_v3} we have suppressed the dependence 
of both the weights $\epsilon_i$ and the line bundles $L_i$ on the particular connected 
component of $F$. If we fix a particular connected component, 
and (continuing to abuse the notation) also refer to that as $f:F\hookrightarrow M$, then a more 
precise notation would be $\epsilon_i(F\hookrightarrow M)$ and 
$L_i(F\hookrightarrow M)$, where both are associated to the normal bundle 
$\mathcal{N}=\mathcal{N}(F\hookrightarrow M)$. We shall occasionally use this notation for clarity. 

We remark that entirely analogous formulae to  \eqref{eq:BV_v1}, \eqref{eq:BV_v3} hold for \emph{invariant submanifolds} of $M$. That is, 
if $M'\subset M$ is an even-dimensional submanifold of $M$ that is invariant under the action generated by $\xi$ (so $\xi$ is everywhere tangent to $M'$), then $\Phi$ pulls back 
to an equivariantly closed form $\Phi'$ on $M'$. The BVAB theorem then immediately applies to the integral of $\Phi'$ over the manifold $M'$, with 
a corresponding fixed point set $F'=\{\xi=0\}\subset M'$. 
{Thus, the integral of a $2n'$-form over a $2n'$-dimensional submanfiold $M'$ is given by a sum of lower-dimensional integrals over the fixed point
set, provided that the $2n'$-form is the top form of an equivariantly closed polyform.}  

We now illustrate how to use the BVAB formula by reviewing three simple examples: $S^2$, $S^4$ and $\mathbb{CP}^2$. Aspects of the $S^2$ and $S^4$ examples 
will appear in the supergravity solutions considered later.

\subsubsection*{\texorpdfstring{$S^2$}{S2} example}\label{sec:s2example}
The metric on the round two-sphere with unit radius can be written
\begin{equation}
	\dd s^2 = \dd \vartheta^2 + \sin^2 \vartheta \, \dd \varphi^2 \, ,
\end{equation}
where $\vartheta \in [0,\pi]$ and $\varphi$ is a periodic coordinate with $\Delta\varphi=2\pi$. We consider the Killing vector $\xi = b \, \partial_\varphi$, where $b$ is a non-zero constant, which has
isolated fixed points at the north and south poles $\vartheta=0, \pi$, respectively. Here $|b|$ is also the weight of the linearization of the action on the $\R^2$ normal to the poles, but the sign at the two poles must be opposite in order to have a consistent orientation on the entire two-sphere. In particular, $\xi$ rotates the plane tangent to $S^2$ at the north pole counter-clockwise, so the weight of the isometric linear action \eqref{eq:LieDerivativeAction} is 
$\epsilon^N\equiv\epsilon(N\hookrightarrow S^2)=+b$ at the north pole $N$ and $\epsilon^S\equiv \epsilon(S\hookrightarrow S^2)=-b$ at the south pole $S$. The integral of 
$\frac{1}{b}\, \vol$ can be computed by first completing the volume two-form to a equivariantly closed polyform:
\begin{equation}
	\Phi = \frac{1}{b} \sin\vartheta \, \dd\vartheta \wedge \dd\varphi + \cos\vartheta \equiv \Phi_2 + \Phi_0 \, ,
\end{equation}
with $\dd_\xi\Phi=0$.
We then use the BVAB formula \eqref{eq:BV_v3}, which in this case receives contributions only from the north and south poles, to get 
\begin{equation}
	\int_{S^2} \frac{1}{b}\, \vol = \int_{S^2} \Phi = 2\pi \left( \frac{\Phi_0 \rvert_N}{\epsilon^N} + \frac{\Phi_0 \rvert_S}{\epsilon^S} \right) = \frac{4\pi}{b}\,.
\end{equation}

Notice that we can replace $\Phi_0\to \Phi_0+c$, for an arbitrary constant $c$, without spoiling the condition $\dd_\xi\Phi=0$.
However, this constant ``gauge" freedom drops out of the final integral.

\subsubsection*{\texorpdfstring{$S^4$}{S4} example}\label{sec:s4example}
The metric on the round four-sphere with unit radius can be written
\begin{align}
\diff s^2_{S^4} = \diff\alpha^2 + \sin^2\alpha \left( \diff\vartheta^2 + \sin^2 \vartheta \, \diff \varphi_1^2 + \cos^2\vartheta \, \diff\varphi_2^2\right) \, ,
\end{align}
where $\alpha\in[0,\pi]$, $\vartheta\in[0,\pi/2]$ and $\varphi_1$, $\varphi_2$ have period $2\pi$. 
The poles of the $S^4$ are at $\alpha=0, \pi$, while there 
are two embedded copies of $S^2$ at $\vartheta=0, \pi/2$. There is a $U(1)$ isometry generated by the Killing vector
\begin{align}
\label{eq:xiS4}
\xi = b_1 \partial_{\varphi_1} + b_2 \partial_{\varphi_2} \, ,
\end{align}
which, when both $b_i\ne 0$ (which we henceforth assume), precisely fixes the poles. It is then straightforward to verify that the polyform $\Phi = \Phi_4 + \Phi_2 + \Phi_0$, with
\begin{align}
	\Phi_4 & = \frac{1}{b_1b_2}\vol = \frac{1}{(b_1b_2)}\sin^3\alpha \sin\vartheta \cos\vartheta \, \dd\alpha \wedge \dd \vartheta \wedge \dd \varphi_1 \wedge \dd \varphi_2 \, ,\nn \\
	\Phi_2 &= -\tfrac{1}{2}\sin^3\alpha \left( \tfrac{1}{b_1} \sin^2\vartheta
\, \diff\varphi_1 + \tfrac{1}{b_2} \cos^2\vartheta \, \diff\varphi_2 \right) \wedge \diff \alpha\, , \nn\\
\qquad \Phi_0 & = \tfrac{1}{24}(9\cos\alpha - \cos 3\alpha)\, ,
\end{align}
is equivariantly closed, $\dd_\xi \Phi = 0$. Importantly, 
while $\varphi_1$ and $\varphi_2$ are ill-defined 
at 
$\vartheta=0$ and $\vartheta=\pi/2$, respectively, 
notice that $\Phi$, and in particular $\Phi_2$, is globally defined.

One can thus compute the volume of the four-sphere by applying the BVAB formula \eqref{eq:BV_v3}. 
Since the fixed point set is the two poles, there are two contributions and we obtain 
\begin{equation}\label{boris}
	\int_{S^4} \frac{1}{b_1b_2} \vol = \int_{S^4}\Phi = (2\pi)^2 \left( \frac{\Phi_0\rvert_N}{\epsilon_1^N \epsilon_2^N} + \frac{\Phi_0\rvert_S}{\epsilon_1^S \epsilon_2^S} \right) \, , 
\end{equation}
where $\epsilon_i^N$ and $\epsilon_i^S$ are the weights of the linear isometric action on the normal space to the north pole $N$ and south pole $S$, respectively. From \eqref{eq:xiS4}, we find that at the north pole 
$\epsilon_1^N\epsilon_2^N = b_1b_2$, but at the south pole, because of the change of orientation on the normal $\R^4$ necessary for the consistency of the orientation of  $S^4$, $\epsilon_1^S \epsilon_2^S = - b_1b_2$.
It is worth emphasizing that it is only the orientation of the entire $\R^4$ that is fixed, not that of the individual summands in $\R^4 = \R^2 \oplus \R^2$.
Equation \eqref{boris} then gives the expected result
\begin{equation}
	\int_{S^4}\Phi = \frac{8\pi^2}{3 b_1 b_2} \, .
\end{equation}

Again, notice if we replace $\Phi_0\to \Phi_0+c$ with an arbitrary constant $c$, we maintain $\dd_\xi\Phi=0$
and $c$ simply drops out of the final integral.

\subsubsection*{\texorpdfstring{$\mathbb{CP}^2$}{CP2} example}
Our final example, which displays more structure, is $\mathbb{CP}^2$. The standard Fubini--Study metric is 
\begin{align}
\label{eq:CP2Metric}
\dd s^2= \dd \vartheta^2+\frac{1}{4}\sin^2\vartheta (\sigma_1^2+\sigma_2^2)+\frac{1}{4}\sin^2\vartheta\cos^2\vartheta \, \sigma_3^2\,,
\end{align}
where $\sigma_i$ are left-invariant one-forms on $SU(2)=S^3$, with 
$\dd\sigma_i= -\tfrac{1}{2}\epsilon_{ijk}\sigma_j\wedge \sigma_k$, and $\vartheta \in [0,\pi/2]$. There is an isometry generated by the Reeb vector of the Hopf fibration on $S^3$ and using Euler angle coordinates on $S^3$, we can write $\xi = \partial_\psi$, where $\psi \in [0,4\pi]$. The square norm of $\xi$ is $\lVert \xi \rVert^2 = \frac{1}{4} \sin^2 \vartheta \cos^2\vartheta$, which vanishes at $\vartheta = 0, \pi/2$. At $\vartheta = 0$ there is an isolated fixed point, where the entire $S^3$ collapses and regularity of the metric is guaranteed by the periodicity of $\psi$ being $4\pi$ (as on $S^3$). On the other hand, at $\vartheta = \frac{\pi}{2}$ the two-sphere fixed by $\xi$, and with 
volume form proportional to $\sigma_1\wedge \sigma_2$, has finite size and radius $\frac{1}{2}$.

One then proves that the polyform $\Phi = \Phi_4 + \Phi_2 + \Phi_0$ is equivariantly closed, where 
\begin{align}
	\Phi_4 &= \vol = \frac{1}{8}\sin^3\vartheta \cos\vartheta \, \dd\vartheta \wedge \sigma_1\wedge\sigma_2\wedge \sigma_3 \, , \nn \\
	\Phi_2 &= \frac{1}{8}\sin^3\vartheta\cos\vartheta \, \dd\vartheta\wedge \sigma_3\,, \nn \\
\Phi_0 &= -\frac{1}{32}(\sin^4\vartheta+c) \, .
\end{align}
Here we have explicitly included a real constant $c$ in $\Phi_0$; we will see that the integral of $\Phi$ is independent of $c$, but in a slightly less trivial way than the previous two examples.

In this case, the BVAB formula \eqref{eq:BV_v3} receives contributions from the isolated fixed point and the fixed two-sphere, which are a nut and a bolt, in the terminology of \cite{Gibbons:1979xm}. The contribution from the former is
\begin{equation}
	\left[ \int_{\mathbb{CP}^2}\Phi \right]_{\rm nut} = \frac{(2\pi)^2}{\epsilon_1 \epsilon_2}\left. \Phi_0 \right|_{\vartheta\to 0} = - \frac{\pi^2 c }{8 \epsilon_1 \epsilon_2} \, ,
\end{equation}
where $\epsilon_1 \epsilon_2 = \frac{1}{4}$. On the other hand, the contribution from the bolt is
\begin{align}
	\left[ \int_{\mathbb{CP}^2} \Phi \right]_{\rm bolt} &= \frac{2\pi}{\epsilon} \int_{\rm bolt} \left[\left. \Phi_2 \right|_{\vartheta\to \frac{\pi}{2}} - \left.\Phi_0 \right|_{\vartheta \to \frac{\pi}{2}} \frac{2\pi}{\epsilon} c_1(L) \right] \nn \\
	&= \frac{ \pi^2(1+c)}{8\epsilon^2} \int_{\rm bolt}  c_1(L) \, ,
\end{align}
where $\epsilon= \frac{1}{2}$ is the weight of the linear action on the normal bundle $L$ to the $\mathbb{CP}^1=S^2$ bolt. Just as in the previous cases, we have fixed a natural orientation, and in this case it is such that the first Chern number of the normal bundle to the $\mathbb{CP}^1$ bolt is $+1$. Therefore, overall we have
\begin{equation}
\int_{\mathbb{CP}^2}\Phi = \frac{\pi^2}{2} \, ,
\end{equation}
which matches the volume computed from the metric \eqref{eq:CP2Metric}, and is independent of $c$, as it should be.

\section{\texorpdfstring{$AdS_5\times M_6$}{AdS5xM6} solutions}\label{sec:M6}

In this section we consider supersymmetric 
$AdS_5\times M_6$ solutions of $D=11$ supergravity, generically dual to $\mathcal{N}=1$ SCFTs in $d=4$, as first analysed in 
\cite{Gauntlett:2004zh}. We begin 
by constructing a set of equivariantly closed forms, and examine some general properties 
of these forms when restricted to fixed point sets under the 
R-symmetry Killing vector $\xi$. 
In the subsequent subsections we then apply this to a wide 
variety of examples, including the holographic duals
to M5-branes wrapped on Riemann surfaces 
$\Sigma_g$ \cite{Bah:2012dg} (which includes the Maldacena--N\'u\~nez solutions \cite{Maldacena:2000mw}
as special cases) and spindles \cite{Ferrero:2021wvk}, 
together with all the explicit families of solutions found 
in \cite{Gauntlett:2004zh}. As well as recovering 
results for known supergravity solutions using this new  
technology, crucially, without using the explicit forms of those solutions, we also compute (off-shell) central charges and other observables
in cases for which solutions have not yet 
been constructed.

\subsection{Equivariantly closed forms}\label{sec:M6forms}

 The $D=11$ metric takes the warped product form
\begin{align}
\label{eq:M6Ansatz}
\diff s^2_{11} = \ex^{2\lambda}[\diff s^2({AdS_5})+\diff s^2({M_6})]\, ,
\end{align}
where we take $AdS_5$ to have unit radius, and will assume that 
$M_6$ is compact without boundary. In order to preserve 
the symmetries of $AdS_5$ the four-form flux $G$ 
and  warp factor function $\lambda$ are defined on $M_6$. 
Note that in this paper we find it convenient to absorb the overall length scale of the $D=11$
metric into the warp factor $\ex^{2\lambda}$. 
The flux quantization condition is
\begin{align}\label{fluxnormn}
\frac{1}{(2\pi\ell_p)^3}\int_{C_4} G \in \mathbb{Z}\, ,
\end{align}
for any four-cycle, $C_4$, and $\ell_p$ is the $D=11$ Planck length.

Supersymmetry requires the existence of a Dirac Killing spinor, $\epsilon$, on 
$M_6$.
From this one can construct the following real bilinear forms\footnote{In the notation of 
\cite{Gauntlett:2004zh}, we have $\epsilon=(\epsilon^+)^{\mathrm{there}}$, $K=(\tilde K^{1})^\text{there}$, $\xi^\flat=\frac{1}{3} (\tilde K^{(2)})^\text{there}$ and we have also set $m^\text{there}=1$.
We also note that
in this section we use $\Gamma_{012\dots10}=+1$ as in \cite{Gauntlett:2004zh}.
}
\begin{align}\label{bildef}
 & 1 =  \bar\epsilon\epsilon\, , \quad \sin\zeta\equiv -\ii\bar\epsilon\gamma_7\epsilon\, , \quad 
K \equiv \bar\epsilon \gamma_{(1)}\epsilon\, , \quad  
\xi^\flat \equiv \tfrac{1}{3}\bar\epsilon \gamma_{(1)}\gamma_7\epsilon\, , 
\nonumber \\
& Y\equiv -\ii \bar\epsilon \gamma_{(2)}\epsilon\, , \quad  Y' \equiv \bar\epsilon\gamma_{(2)}\gamma_7\epsilon\, ,
\end{align}
where $\gamma_7\equiv \gamma_{123456}$. 
In particular from the Killing spinor equation one finds that $\bar\epsilon\epsilon$ is a constant, 
which we have normalized to 1, and the vector field 
$\xi$ dual to the one-form bilinear $\xi^\flat$ is Killing. 
The factor of $\tfrac{1}{3}$ has been introduced into the definition in \eqref{bildef} so that 
the R-charge of the Killing spinor under $\xi$ is $\rcharge=\tfrac{1}{2}$: 
\begin{align}
\mathcal{L}_\xi\epsilon=\frac{\ii}{2}\epsilon\,.
\end{align}
The Killing spinor is globally defined and so too are all of the bilinear forms in \eqref{bildef}.
We can introduce a local coordinate  $\psi$ so that $\xi=\partial_\psi$, and define the function
\begin{align}\label{ydef}
y \equiv \tfrac{1}{2}\ex^{3\lambda}\sin\zeta\, , \qquad \mbox{where}\ \diff y = \ex^{3\lambda} K\, ,
\end{align}
which was also used as a canonical coordinate in \cite{Gauntlett:2004zh}. 

{We emphasize at this point that the differential forms in \eqref{bildef} are all invariant under the action of $\xi$, {\it i.e.} the Lie derivative
acting on the bilinears vanishes, and they comprise a subset of all the bilinears that can be constructed. As we now show, 
the differential conditions satisfied by this subset of bilinears, which follow from the Killing spinor equation, are sufficient to construct
several equivariantly closed polyforms. 

Proceeding, from \cite{Gauntlett:2004zh} we have the following contractions}
\begin{align}
\qquad \xi \hook \diff\lambda = 0\, , \quad   \xi \hook G = - \tfrac{1}{3} \diff ( \ex^{3\lambda} Y') \, ,  \quad  \xi \hook Y' = - \tfrac{1}{3}K \, .
\end{align}
The first two equations imply that $\mathcal{L}_\xi \lambda=0=\mathcal{L}_\xi G$, 
showing that $\xi$ generates a symmetry of the full solution. Moreover, 
these equations can be used to establish that the polyform
\begin{equation}
\label{eq:M6_PhiG}
    \Phi^G \equiv G - \tfrac{1}{3} \ex^{3\lambda} Y' + \tfrac{1}{9} y \,,
\end{equation}
is equivariantly closed under $\diff_\xi$. Using the further 
equations
\begin{align}\label{dY}
\diff(\ex^{6\lambda}Y) = 0\, , \quad \xi \hook Y = \tfrac{2}{3}y\, \ex^{-3\lambda}K\, ,
\end{align}
from \cite{Gauntlett:2004zh}, likewise we have the equivariantly closed form
\begin{align}
\label{eq:M6_PhiY}
	\Phi^Y \equiv \ex^{6\lambda}Y + \tfrac{1}{3} y^2\, .
\end{align}
Both of these will be used later when quantizing the flux $G$. 
We also note from \cite{Gauntlett:2004zh} that we have
\begin{align}\label{bilinstargads5}
\ex^{3\lambda}* G = 3\diff (\ex^{6\lambda}\xi^\flat) - 4\ex^{6\lambda}Y\, ,
\end{align}
where the Hodge star is with respect to the metric $\diff s^2(M_6)$,
and there is also an equivariantly closed from involving $ * G$:
\begin{align}
\Phi^{* G} & = \ex^{3\lambda} * G -\tfrac{1}{3}\ex^{6\lambda}\, .
\end{align}
Note that $\Phi^{* G}$ and $\Phi^Y$ are related via:
\begin{align}\label{eq:M6_Phi*G}
\Phi^{* G} &=-4\Phi^Y+3 \diff_\xi(\ex^{6\lambda}\xi^\flat)\,.
\end{align}

It is next useful to introduce the local $SU(2)$ structure defined by a Dirac spinor in six dimensions. This consists of a real two-form $J$, a complex two-form $\Omega$ (which will play no role in what follows), and two orthonormal one-forms $\ex^5$ and $\ex^6$. In terms of the spinor bilinears already introduced, we have \cite{Gauntlett:2004zh}
\begin{align}
\label{eq:su2structure_M6}
	 K &= \cos\zeta \, \ex^5 \, ,\qquad \qquad\quad \xi^\flat = \tfrac{1}{3} \cos\zeta \, \ex^6 \, , \nn\\
	Y &= J - \sin\zeta \, \ex^5 \wedge \ex^6 \, , \quad Y' = - \sin\zeta \, J + \ex^5 \wedge \ex^6 \, .
\end{align}
The volume form on $M_6$ is $\vol = \tfrac{1}{2} J\wedge J \wedge \ex^5 \wedge \ex^6$, and taking the Hodge dual 
one finds that
\begin{equation}
\label{eq:su2structure_M6_2}
	*\, Y = J \wedge \ex^5 \wedge \ex^6 - \tfrac{1}{2} \sin\zeta \, J \wedge J \, .
\end{equation}

The further bilinear equation 
\begin{equation}
	\diff \left( *\,  \ex^{9\lambda} Y \right) = - 12 \ex^{9\lambda} * \xi^\flat\,,
\end{equation}
then implies that the following polyform is equivariantly closed:
\begin{equation}
\label{eq:M6_Phia}
	\Phi \equiv \ex^{9\lambda} \vol + \tfrac{1}{12} \ex^{9\lambda} * Y - \tfrac{1}{36} y \ex^{6\lambda} Y - \tfrac{1}{162 } y^3 \, . 
\end{equation}
This is particularly significant, since the $a$ central charge for such a solution is \cite{Gauntlett:2006ai}\footnote{The $a$ central charge is given by $a=\pi L^3/(8G_5)$ where $L$ is the radius of the $AdS_5$ vacuum and $G_5$
is the five-dimensional Newton constant \cite{Henningson:1998gx}. To compute $G_5$ and thus find \eqref{aloc}, one simply reduces the $D=11$ supergravity action using the ansatz \eqref{eq:M6Ansatz}.}
\begin{align}\label{aloc}
a = \frac{1}{2(2\pi)^6\ell_p^9} \int_{M_6}\Phi\, ,
\end{align}
and this then localizes, {{\it i.e.} the integral over $M_6$ can be written as a sum of lower-dimensional integrals along the fixed point set of $\xi $ using the BVAB formula (see \eqref{aformula} below).} 
The final bilinear equation we shall need from \cite{Gauntlett:2006ai} is
\begin{equation}
\label{eq:M6_BilinearEqn_Useful}
	\diff \left( \ex^{6\lambda} Y \wedge \xi^\flat \right) = \tfrac{2}{3} y \, G + \tfrac{1}{3} \ex^{6\lambda} \, Y \wedge Y \, ,
\end{equation}
which will again be helpful for imposing flux quantization on $G$.
For later use, notice that since the bilinears are globally defined forms on $M_6$, the left hand side is globally exact.

{The existence of the equivariantly closed polyforms, $\Phi^G$, $\Phi^{*G}$, $\Phi^Y$ and
$\Phi$, has arisen just by imposing a subset of the supersymmetry conditions. It was shown in \cite{Gauntlett:2006ai}
that if one imposes \emph{all} of the supersymmetry conditions, one automatically imposes the equations of motion and
one is necessarily on-shell. Since we have not done that, the existence of these equivariantly closed
forms is an off-shell result, a point we return to in section \ref{sec:actioncalc}.
}

\subsection{Localization and fixed point sets}\label{sec:localize}

Integrals of  the equivariantly closed forms $\Phi^G$, $\Phi^Y$, $\Phi$ defined in 
\eqref{eq:M6_PhiG}, \eqref{eq:M6_PhiY}, \eqref{eq:M6_Phia} will localize 
to the fixed point set $F$ of $\xi$. On the other hand, from \eqref{eq:su2structure_M6} 
we have
\begin{align}
\|\xi\| = \tfrac{1}{3}|\cos\zeta\, |\, ,
\end{align}
implying that 
\begin{align}\label{sinzeta}
\fleft\sin\zeta \fright = \pm 1 \, .
\end{align}
Here we have introduced the notation $\fleft \cdot \fright$ to denote 
 evaluation on (a component of) the fixed point set $F$. 
The sign in \eqref{sinzeta} is  correlated with the chirality of the spinor at that fixed point, 
since correspondingly $\fleft-\ii \gamma_7\epsilon\fright =\pm \epsilon$. 
Since from \eqref{eq:su2structure_M6}  also $\fleft K \fright = 0$, we deduce 
from \eqref{ydef} that both $y$ and hence $\ex^{3\lambda}$ are locally constant 
on $F$, with $\fleft \ex^{3\lambda}\fright =\pm 2y$, and hence constant on each connected component. Introducing the rescaled two-form\footnote{This was denoted 
$\omega=\hat{J}^{\mathrm{there}}$ in \cite{Gauntlett:2004zh}.}
\begin{align}
\omega \equiv \ex^{6\lambda} J\, ,
\end{align} 
we conclude from \eqref{eq:su2structure_M6}, \eqref{eq:su2structure_M6_2} that 
\begin{align}
\fleft\ex^{6\lambda} Y \fright = \omega\, , \quad 
\fleft\ex^{3\lambda}Y' \fright = -\frac{1}{2y}\omega\, , \quad \fleft
 \ex^{9\lambda}* Y \fright =-\frac{1}{4y}\omega\wedge\omega\, .
\end{align}
From \eqref{dY} it follows that $\fleft\diff\omega\fright =0$. 
Thus $\omega$ defines a cohomology class on each connected component of $F$, which 
will again be useful in what follows. Finally, notice that the left-hand side of \eqref{eq:M6_BilinearEqn_Useful} is globally exact, so on a closed four-dimensional fixed point set $F_4$ we find
\begin{equation}
\label{eq:M6_GBolt}
	\int_{F_4} G = - \int_{F_4} \frac{1}{8 y^3} \omega \wedge \omega \, ,
\end{equation}
where as noted above $y$ is necessarily constant on $F_4$, and may thus 
be taken out of the integral. The flux integral in \eqref{eq:M6_GBolt} 
is then a topological invariant, depending only on the cohomology class 
$[\omega]\in H^2(F_4,\R)$ and the constant $\left. y \, \right|_{F_4}$.

Using the above we can now write down a general localization 
formula for the central charge. Combining \eqref{eq:M6_Phia},  \eqref{aloc} and 
applying the BVAB fixed point formula \eqref{eq:BV_v3}, we have
\begin{align}\label{aformula}
 a =  \frac{1}{2(2\pi)^6\ell_p^9}\Bigg\{& \sum_{F_4}\frac{1}{d_{F_4}}
 \int_{F_4} \Big[-\frac{2\pi}{\epsilon_1}\frac{1}{48y}\omega\wedge\omega 
+ \left(\frac{2\pi}{\epsilon_1}\right)^2\frac{y}{36}\omega \wedge c_1(L_1)\nonumber\\
 &   -
\left(\frac{2\pi}{\epsilon_1}\right)^3\frac{y^3}{162}c_1(L_1)\wedge c_1(L_1)\Big]
+\sum_{F_2}\frac{1}{d_{F_2}}\int_{F_2}\Big[-\frac{(2\pi)^2}{\epsilon_1\epsilon_2}\frac{y}{36}\omega \nonumber\\ 
 &  +\frac{(2\pi)^3}{\epsilon_1\epsilon_2}\frac{y^3}{162}\left(\frac{c_1(L_1)}{\epsilon_1}+ \frac{c_1(L_2)}{\epsilon_2}\right)\Big] - \sum_{F_0} \frac{1}{d_{F_0}}\frac{(2\pi)^3}{\epsilon_1\epsilon_2\epsilon_3}\frac{y^3}{162}\Bigg\}\, .
\end{align} 
Notice this depends on the weights $\epsilon_i$ and Chern numbers $c_1(L_i)$, which are topological invariants, but also the constant values of $y$ 
on each connected component of $F$, and  the cohomology class 
$[\omega]\in H^2(F,\R)$. 
As remarked in section \ref{sec:EquivariantLocalization}, 
in writing \eqref{aformula} we have suppressed the dependence of 
$\epsilon_i$, $L_i$ (and also $y$, $[\omega]$) on the particular 
connected component of the fixed point set, and  for clarity we shall 
sometimes restore that dependence explicitly in the examples that follow. 
We shall see, remarkably, that flux quantization may be used 
to effectively eliminate all the data entering \eqref{aformula} in terms of flux quantum numbers, which 
are again global invariants that specify the solution. 

Using the formula for $\Phi^G$ in  \eqref{eq:M6_PhiG} we may write a similar 
formula for the integral of $G$ over a $\xi$-invariant submanifold 
$C_4$ and hence impose Dirac quantization.
It is helpful to consider two separate cases. The first is when the entire $C_4$ is fixed by the action of the
Killing vector and then from \eqref{eq:M6_GBolt} we have
\begin{align}\label{Glocalize1}
\frac{1}{(2\pi\ell_p)^3}\int_{C_4} G = - \frac{1}{(2\pi\ell_p)^3}
\int_{C_4}\frac{1}{8y^3}\omega\wedge\omega  \equiv N_{C_4} \in \mathbb{Z}\, .
\end{align}
The second case is when the fixed point set consists of either or both of  $C_2, C_0$ and then
\begin{align}\label{Glocalize2}
\frac{1}{(2\pi\ell_p)^3}\int_{C_4} G = \frac{1}{(2\pi\ell_p)^3}\Bigg\{&  
\sum_{C_2}\frac{1}{d_{C_2}}\int_{C_2}\Big[ \frac{2\pi}{\epsilon_1}\frac{1}{6y}\omega
-\left(\frac{2\pi}{\epsilon_1}\right)^2\frac{y}{9}c_1(L_1)\Big]\nonumber\\
 +&\sum_{C_0}\frac{1}{d_{C_0}}\frac{(2\pi)^2}{\epsilon_1\epsilon_2}\frac{y}{9}\Bigg\}
 \equiv N_{C_4} \in \mathbb{Z}\, .
\end{align}
Here $C_2, C_0\subset C_4$ are the fixed submanifolds inside $C_4$, with 
$L_1$ being the normal bundle of $C_2$ in $C_4$. 
Note in both cases that the integer $N_{C_4}$ only depends on the homology class
 $[C_4]\in H_4(M_6,\Z)$, since $G$ is closed. 

We can also compute other observables using the same techniques. 
One example is the conformal dimension $\Delta$ of chiral primary operators in the dual~$\mathcal{N}=1$ SCFT, corresponding to M2-branes wrapping supersymmetric two-cycles $\Sigma_2 \subset M_6$. Such M2-branes are  calibrated by $Y'$, 
\emph{i.e.} $\vol|_{\Sigma_2}=Y'|_{\Sigma_2}$, and their conformal dimension may be computed in the gravity dual via \cite{Gauntlett:2006ai}
\begin{equation}
\label{eq:M6_ConformalDimension}
	\Delta(\Sigma_2) = \frac{1}{(2\pi)^2\ell_p^3} \int_{\Sigma_2} \ex^{3\lambda} Y' \ .
\end{equation}
Since $\ex^{3\lambda}Y'$ is the top-form of the restriction to $\Sigma_2$ of the four-form flux polyform $\Phi^G$ in \eqref{eq:M6_PhiG}, this integral can also be evaluated by localization 
when $\Sigma_2$ is $\xi$-invariant. We compute 
\begin{align}\label{Deltaformula}
\Delta(\Sigma_2) = \frac{1}{(2\pi)^2\ell_p^3}\Big(-\int_{\Sigma_2}\frac{1}{2y}\omega
-\sum_{\Sigma_0}\frac{1}{d_{\Sigma_0}}\frac{2\pi}{\epsilon_1}\frac{y}{3}\Big)\, .
\end{align}
Here $\Sigma_0\subset \Sigma_2$ are the  fixed points inside 
$\Sigma_2$ and, as for the flux integral, the first term is only present (and by itself) 
when the entire $\Sigma_2$ is fixed, while the second term is present (and by itself)
when the fixed point set is $\Sigma_0\subset \Sigma_2$.

\subsection{M5-branes wrapped on \texorpdfstring{$\Sigma_g$}{Sigmag}}\label{sec:BBBW}

As a first example  of this new technology we consider the supergravity solutions constructed in 
\cite{Bah:2012dg}. These describe the near-horizon limits of 
$N$ M5-branes wrapped over a Riemann surface $\Sigma_g$ 
inside a local Calabi--Yau three-fold. The latter is the total space 
of the bundle\footnote{Comparing to the notation in 
\cite{Bah:2012dg} we have $p_1=-p_{\mathrm{there}}$, $p_2=-q_{\mathrm{there}}$.} $\mathcal{O}(-p_1)\oplus \mathcal{O}(-p_2)\rightarrow \Sigma_g$, 
where this is Calabi--Yau provided 
\begin{align}\label{p1p2}
p_1+p_2 = 2(1-g)\, .
\end{align}
The near-horizon limit of the wrapped M5-branes is then an $S^4$ bundle over 
$\Sigma_g$:
\begin{align}
S^4\, \hookrightarrow\, M_6\, \rightarrow\, \Sigma_g\, .
\end{align}
More precisely, $S^4\subset \C_1\oplus\C_2\oplus\R=\R^5$, where the 
two copies of $\C_i$ are twisted using the line bundles $\mathcal{O}(-p_i)$, $i=1,2$, respectively. 
There are two special cases: when $p_1=p_2=-(g-1)$ and when $p_1=0$, $p_2=-2(g-1)$ corresponding
to the supergravity solutions of \cite{Maldacena:2000mw} when $g>1$ and dual to $\mathcal{N}=1$ and
$\mathcal{N}=2$ SCFTs in $d=4$, respectively.

We now show how to compute the central charge of such solutions, 
making just one more assumption. 
We first introduce 
vector fields $\partial_{\varphi_i}$ rotating the $\C_i$ above, and hence 
acting on the $S^4$ fibre over the Riemann surface. 
Our extra assumption is
that $\partial_{\varphi_i}$ are Killing vectors in the full solution. 
We then write the 
R-symmetry vector field as
\begin{align}\label{bbbwkv}
\xi = \sum_{i=1}^2 b_i \partial_{\varphi_i}\, .
\end{align}
As explained in 
appendices \ref{app:SpinorRegularity} and \ref{app:AdS7}, 
with an appropriate choice of sign conventions, regularity 
of the spinor at the north pole of the $S^4$ requires 
the sum of the weights there to satisfy $b_1+b_2 = 1$, and then by continuity this fixes the sum to be 1 everywhere.\footnote{The dual picture is that imposing the R-charge of the spinor under the isometry 
to be $\frac{1}{2}$ fixes the R-charge of the holomorphic volume form on the original Calabi--Yau three-fold to be 1, and then by continuity the weights of the isometry in the near horizon geometry should also sum to 1.} We can thus write
\begin{align}\label{bepsilon}
b_1 = \tfrac{1}{2}(1+\varepsilon)\, , \quad b_2 = \tfrac{1}{2}(1-\varepsilon)\, .
\end{align}
The parameter $\varepsilon$ may be found for the (on-shell) solutions in 
\cite{Bah:2012dg}, but for now we leave it arbitrary. 

Following a similar argument in \cite{BenettiGenolini:2023kxp}, we begin 
by choosing an arbitrary point on $\Sigma_g$, 
and consider a linearly embedded $S^2_i\subset \C_i\oplus \R \subset \R^5$ in the  fibre over it.
We choose the submanifold $S^2_i$ to be invariant under the action of $\xi$.
The homology class of this $S^2_i$ is trivial, 
so it follows using localization for $\Phi^Y$ in \eqref{eq:M6_PhiY} that
\begin{align}\label{yS2}
0 = \int_{S^2_i} \Phi^Y  = \frac{2\pi}{b_i}\left(\frac{y_N^2}{3}- \frac{y_S^2}{3}\right)\, ,
\end{align}
where the $N$ and $S$ subscripts refer to the north and south poles in the fibre sphere $S^4$, respectively. 
To obtain \eqref{yS2}, we used the fact that the weights at these north and south poles are 
$\epsilon_i^N=b_i$, 
$\epsilon_i^S=-b_i$ respectively, 
where here notice that $i=1,2$ labels 
different submanifolds $S^2_i$, with the normal spaces at each pole of those 
two-spheres being $\R^2$. 
Note also the change of relative orientation of those spaces, as discussed in the $S^2$ example in section \ref{sec:s2example}.
Equation \eqref{yS2} immediately implies that $|y_N| = |y_S|$. 

Next consider flux quantization of $G$ through a copy of the fibre $S^4$, 
again at an arbitrary point on $\Sigma_g$. In the notation 
of section \ref{sec:localize} we have $C_4=S^4$, and 
the fixed points $C_0$ under $\xi$ are precisely the north 
and south poles $N$, $S$ of the $S^4$. Denoting the flux as $N=N_{S^4}$ from 
\eqref{Glocalize2} 
we have
\begin{align}\label{NBBBW}
 N = \frac{1}{(2\pi \ell_p)^3}\left[\frac{(2\pi)^2}{b_1b_2}\frac{y_N}{9} -
\frac{(2\pi)^2}{b_1b_2}\frac{y_S}{9}\right]\, ,
\end{align}
where 
we have used $\epsilon_1^N\epsilon_2^N=b_1b_2$, 
$\epsilon_1^S\epsilon_2^S=-b_1b_2$.
This is the same as in the $S^4$ example considered in section \ref{sec:s4example}.
For $N> 0$ we must then have $y_N=-y_S>0$, which allows us to solve for $y_N$:
\begin{align}\label{ynyssigmads5}
y_N =-y_S= 9\pi \ell_p^3 \, b_1b_2 N\, .
\end{align}
Notice that {\it a priori} the value of $y_N$ could have depended 
on the point chosen on the Riemann surface. The fact it does not reflects 
the fact that copies of $S^4$ over different points on $\Sigma_g$ are 
homologous, and the flux quantization condition \eqref{NBBBW} 
depends only on the homology class.

There are two other four-cycles of interest, namely the total spaces 
of the $S^2_i$ bundles over $\Sigma_g$, with the fibres 
$S^2_i$  as defined in \eqref{yS2}. Denoting these four-manifolds 
by $C^{(i)}_4$, as explained in appendix \ref{app:homology} we have the following homology relations
\begin{align}\label{hom4}
C^{(1)}_4 = p_2 [S^4]\, , \quad C^{(2)}_4 = p_1[S^4] \ \in \ H_4(M_6,\Z)\, ,
\end{align}
where the factors of $p_2$, $p_1$ arise from the first Chern classes of the normal bundles 
of $C^{(1)}_4$ and $C^{(2)}_4$ inside $M_6$, respectively. 
From \eqref{Glocalize2} we have 
\begin{align}\label{gary}
N_{C_4^{(i)}} = \frac{1}{(2\pi \ell_p)^3}\Bigg\{& \int_{\Sigma_g^N} \Big[
\frac{2\pi}{b_i}\frac{1}{6y_N}\omega - \left(\frac{2\pi}{b_i}\right)^2
\frac{y_N}{9}p_i\Big]\nonumber\\
& +\int_{\Sigma_g^S} \Big[
-\frac{2\pi}{b_i}\frac{1}{6y_S}\omega - \left(\frac{2\pi}{-b_i}\right)^2
\frac{y_S}{9}(-p_i)\Big]\Bigg\}\, .
\end{align}
Here $\Sigma_g^N$, $\Sigma_g^S$ are the fixed 
copies of the base Riemann surface, at the north and south poles 
of the fibre sphere $S^2_i$. In our conventions for the assignment of north and south pole labels, these have normal bundles
$L_i(\Sigma_g^N\hookrightarrow C_4^{(i)})=\mathcal{O}(p_i)$ 
and  $L_i(\Sigma_g^S\hookrightarrow C_4^{(i)})=\mathcal{O}(-p_i)$ inside 
the four-manifolds $C_4^{(i)}$, $i=1,2$,  
which give the respective factors of $p_i$ and $-p_i$ in 
\eqref{gary}. The corresponding weights are
$\epsilon_i^N=b_i$ and $\epsilon_i^S=-b_i$, respectively 
({\it c.f.} the discussion for the $S^2$ example in section \ref{sec:s2example}).
From \eqref{hom4} we also have
\begin{align}\label{N12}
N_{C_4^{(1)}} = p_2 N\, , \quad N_{C_4^{(2)}} = p_1 N\, .
\end{align}
Finally, notice that $[\Sigma_g^N]=[\Sigma_g^S]\in H_2(M_6,\Z)$, 
and integrating the closed form $\ex^{6\lambda}Y$ over both cycles (which 
are also fixed point sets)
immediately implies that $\int_{\Sigma_g^N}\omega = 
\int_{\Sigma_g^S}\omega$. The above equations are then easily solved to find
\begin{align}
\label{eq:omegaSigmaN}
\int_{\Sigma_g^N} \omega = 108\pi^3 \ell_p^6\, b_1b_2(b_1 p_2 + b_2 p_1)N^2\, .
\end{align}

With all of this in hand, we may simply write down the central charge 
using \eqref{aformula}. This reads 
\begin{align}\label{aBBBWstep}
a = & \frac{1}{2(2\pi)^6\ell_p^9}\Bigg\{ \int_{\Sigma_g^N}\Bigg[
-\frac{(2\pi)^2}{b_1b_2}\frac{y_N}{36}\omega + 
\frac{(2\pi)^3}{b_1b_2}\frac{y_N^3}{162}\left(\frac{c_1(\mathcal{O}(p_1))}{b_1}+\frac{c_1(\mathcal{O}(p_2))}{b_2}\right)\Bigg]\nonumber\\
& + \int_{\Sigma_g^S}\Bigg[
\frac{(2\pi)^2}{b_1b_2}\frac{y_S}{36}\omega -
\frac{(2\pi)^3}{b_1b_2}\frac{y_S^3}{162}\left(\frac{c_1(\mathcal{O}(p_1))}{b_1}+\frac{c_1(\mathcal{O}(p_1))}{b_2}\right)\Bigg]\Bigg\}\, .
\end{align}
Here we have used $\epsilon_1^N\epsilon_2^N=b_1b_2$, 
$\epsilon_1^S\epsilon_2^S=-b_1b_2$ for the product 
of weights on the normal spaces $\R^4=\R^2\oplus\R^2$ 
for each of $\Sigma_g^N$, $\Sigma_g^S$, respectively. 
As discussed for the simple $S^4$ example 
in section~\ref{sec:EquivariantLocalization}, there is no invariant 
way to assign orientations to each $\R^2$ factor. Correspondingly, 
the normal bundle to $\Sigma_g^N$ can be taken to be either of the upper or lower signs in 
$\mathcal{N}_N\equiv\mathcal{N}(\Sigma_g^N\hookrightarrow M) =
\mathcal{O}(\pm p_1)\oplus\mathcal{O}(\pm p_2)$,  while the normal bundle to 
$\Sigma_g^S$ may be taken to be either of $\mathcal{N}_S\equiv\mathcal{N}(\Sigma_g^S\hookrightarrow M) =
\mathcal{O}(\pm p_1)\oplus\mathcal{O}(\mp p_2)$. Moreover, these 
signs are then correlated with $(\epsilon_1^N=\pm b_1,\epsilon_2^N=\pm b_2)$, and $(\epsilon_1^S=\pm b_1,\epsilon_2^S=\mp b_2)$, 
in such a way that the ratios $c_1(L_i)/b_i$ in \eqref{aBBBWstep} 
are (necessarily) independent of the choice.\footnote{This non-uniqueness 
is an artefact of splitting the normal bundles into a sum of complex line bundles, 
but computationally this is still convenient.}

Using \eqref{ynyssigmads5}, \eqref{eq:omegaSigmaN} and $\int_{\Sigma_g} c_1(\mathcal{O}(p)) = p$, the expression in \eqref{aBBBWstep} immediately simplifies to
\begin{align}\label{aBBBW}
a & = -\frac{9}{8}b_1 b_2 (b_1p_2  + b_2p_1)N^3\nonumber\\
& = \frac{9}{32}(1-\varepsilon^2)[g-1+(1-p_2-g)\varepsilon]N^3\, .
\end{align}
Here in the second equality we have used 
\eqref{p1p2}, \eqref{bepsilon}. 

This expression precisely agrees with the \emph{off-shell} trial $a$-function in the dual field theory, 
as a function of the parameter $\varepsilon$. Specifically, compare \eqref{aBBBW} 
to equation (2.16) of \cite{Bah:2012dg}, where $z_{\mathrm{there}}=1+\frac{p_2}{g-1}$.
The on-shell central charge is obtained by extremizing over the undetermined parameter $\varepsilon$
and the result agrees with that obtained from the explicit supergravity solutions
in \cite{Bah:2012dg}. 

For example, for
the special case when $p_1=p_2$, on-shell we find $\varepsilon=0$, so $b_1=b_2=\frac{1}{2}$, and a central charge
\begin{equation}
	a = \frac{9}{32}( g -1 ) N^3\,,
\end{equation}
which agrees with the explicit supergravity solutions dual to
$\mathcal{N}=1$ SCFTs of \cite{Maldacena:2000mw}, which only exist when $g>1$.
Similarly, for the special case when $p_1=0$, on-shell we find $\varepsilon=1/3$, so $b_1=\frac{2}{3}$, $b_2=\frac{1}{3}$,
and a central charge
\begin{equation}
	a = \frac{1}{3}  (g-1)N^3\,,
\end{equation}
in agreement with the supergravity solutions dual to
$\mathcal{N}=2$ SCFTs of \cite{Maldacena:2000mw}, which again only exist when $g>1$.
Note that for this case the dual $\mathcal{N}=2$ SCFT has a non-Abelian R-symmetry, but we have only
utilised an Abelian R-symmetry in order to derive the result.

The fact that we obtain the \emph{off-shell} central charge via localization
arises from the fact that the ingredients entering the localization calculations are only
a subset of the full supersymmetry conditions. If all of the supersymmetry conditions are imposed
we automatically impose the equations of motion \cite{Gauntlett:2004zh}. However, if only a subset are imposed then there are some 
remaining variations to consider. Remarkably, as we show in section \ref{sec:actioncalc}, these remaining
variations are associated with varying the off-shell central charge.

We now consider M2-branes wrapped on various calibrated cycles and compute 
the dimension of the dual chiral primaries using \eqref{Deltaformula}. We first consider 
the cycles $\Sigma_g^N$, $\Sigma_g^S$.
From \eqref{eq:su2structure_M6} we can infer that these will be calibrated cycles provided
that $Y'|_{\Sigma_g^{N,S}}=\mp J|_{\Sigma_g^{N,S}}$, which can be argued to be the case for
the explicitly known solutions of \cite{Bah:2012dg}.
Choosing an appropriate orientation for $\Sigma_g^{N}$, using \eqref{ynyssigmads5} and \eqref{eq:omegaSigmaN}, \eqref{Deltaformula} then gives
\begin{align}\label{DeltaBBBW}
\Delta(\Sigma_g^N) 
=  -\frac{1}{(2\pi)^2\ell_p^3}\int_{\Sigma_g^N} 
\frac{1}{2y_N}\omega & = -\frac{3}{2}(b_1 p_2 + b_2p_1)N\nonumber\\
& = \frac{3}{2}[g-1+(1-p_2-g)\varepsilon]N\, ,
\end{align}
Similarly, with an appropriate orientation 
 $\Delta(\Sigma_g^N) = \Delta(\Sigma_g^S)$.
These results
agree with the \emph{off-shell} expressions
obtained in field theory in equation (5.18) of \cite{Bah:2012dg}. Furthermore, after substituting
the value of $\varepsilon$ that extremizes $a$ in \eqref{aBBBW} we find exact agreement
with the result (4.10) computed in \cite{Bah:2012dg} using 
the explicit supergravity solution. 

Another possibility, which has not been previously considered, is to wrap 
M2-branes on the invariant, homologically trivial, submanifolds $S^2_i$, defined just above \eqref{yS2}. 
Now since $\xi$ is tangent to $S^2_i$, the $S^2_i$ will be calibrated by $Y'$
provided that  the volume form on $S^2_i$ is, up to sign, given by $\ex^5\wedge \ex^6$. To see this
we recall \eqref{eq:su2structure_M6} and that $\ex^6\hook  J=0$ so that
$Y'|_{S^2_i}=\ex^5\wedge \ex^6|_{S^2_i}$.
Choosing a suitable orientation to get a positive result,
we can then compute $\Delta(S^2_i)$ using localization via \eqref{Deltaformula}. We have fixed points at the north and south poles
of $S^2_i$, and find
\begin{align}\label{Deltaformula2}
\Delta(S^2_i) &=+ \frac{1}{(2\pi)^2\ell_p^3}\Big(\frac{2\pi}{\epsilon^N_i}\frac{y_N}{3}+\frac{2\pi}{\epsilon^S_i}\frac{y_S}{3}\Big)
=\frac{3b_1 b_2}{b_i}N\, ,
\end{align}
where, as in \eqref{yS2},  we used
$\epsilon_i^N=b_i$, $\epsilon_i^S=-b_i$ and also
\eqref{ynyssigmads5}. 
It would be interesting to verify the calibration condition using the results of appendix D of 
\cite{Bah:2012dg} for the explicit supergravity solutions,
and also to identify the operators in the dual field theory proposed in \cite{Bah:2012dg}.

\subsection{M5-branes wrapped on a spindle}\label{sec:spindle}

In this section we consider
$N$ M5-branes wrapped over a spindle. 
The full supergravity solutions were constructed in 
\cite{Ferrero:2021wvk}, with $M_6$ being the total 
space of an $S^4$ orbibundle fibred over a spindle $\Sigma=\mathbb{WCP}^1_{[n_+,n_-]}$.  
The latter is topologically a two-sphere, but with conical deficit angles 
$2\pi(1-1/n_\pm)$ at the poles. The key difference, compared to the 
previous subsection, is that here the R-symmetry vector $\xi$ 
generically also mixes with the spindle direction $\Sigma$. 
As a result the fixed point sets are all isolated and consequently
the localization formulae 
are used rather differently than in the previous subsection. On the other 
hand, the final results of the two subsections should coincide after setting $g=0$ and 
$n_+=n_-=1$, respectively, so that $\Sigma=S^2$, and we will see that
this is indeed the case.

The physical set-up in this section is very similar to that in the last 
section: we are considering the near-horizon limit of $N$ M5-branes 
wrapped on a spindle surface $\Sigma=\mathbb{WCP}^1_{[n_+,n_-]}$ inside 
a local Calabi--Yau three-fold. The latter is the total space of the bundle 
$\mathcal{O}(-p_1)\oplus \mathcal{O}(-p_2)\rightarrow \Sigma$, 
where in order for the total space to be Calabi--Yau (giving a topological twist as in the known supergravity solution) 
we have 
\begin{align}\label{p1p2sum}
p_1 + p_2 = n_+ + n_-\, . 
\end{align}
The near-horizon limit is then an $S^4$ orbibundle over $\Sigma$, 
where again $S^4\subset \C_1\oplus\C_2 \oplus \R=\R^5$, and the two 
copies of $\C_i$ are twisted using the complex line orbibundles $\mathcal{O}(-p_i)$, $i=1,2$, 
respectively. One technical difference in this case is that the fibres 
over the poles of the spindle base $\Sigma$ are in general
orbifolds $S^4/\Z_{n_\pm}$, where the action of $\Z_{n_\pm}$ is 
determined by the twisting parameters $p_i\in\Z$. This is discussed 
in detail in \cite{Ferrero:2021etw} for circle orbibundles (or their 
associated complex line orbibundles), and that discussion carries 
over straightforwardly for each of $\mathcal{O}(-p_i)$, with an 
induced action on the $S^4$ fibres. However, we will not 
need any of these details in what follows. 
We  write the R-symmetry vector as 
\begin{align}\label{kvspin}
\xi = \sum_{i=1}^2 b_i \partial_{\varphi_i}+\varepsilon\, \partial_{\varphi_3} \, ,
\end{align}
where $\partial_{\varphi_i}$ rotate the two copies of $\C_i$, as before,
 while 
$\partial_{\varphi_3}$ is a lift of the vector field that rotates the spindle, where we use the construction of such a basis in~\cite{Boido:2022mbe}. 
We assume that $\partial_{\varphi_i}$ and $ \partial_{\varphi_3}$ are Killing vectors 
of the full solution. 

Consider first fixing one of the poles on $\Sigma$, say the plus pole with orbifold group $\Z_{n_+}$, 
and consider a linearly embedded $S^2_i\subset \C_i\oplus \R \subset \R^5$ in the covering space of the fibre over it. 
The same argument as in \eqref{yS2} then
gives
\begin{align}
0 = \int_{S^2_i} \Phi^Y  = \frac{2\pi}{b_i^+}\left
[\frac{(y^+_N)^2}{3} - \frac{(y^+_S)^2}{3}\right]\, ,
\end{align}
where the $N$ and $S$ subscripts again refer to the poles in the fibre sphere $S^4$, 
and we 
have used 
$\epsilon_i^N=b_i^+$, $\epsilon_i^S=-b_i^+$, where we shall determine the
weights $b_i^+$
below. 
This immediately implies that $|y^+_N| = |y^+_S|$. A similar argument using the minus pole of the spindle 
allows us to conclude $|y^\pm_N| = |y^\pm_S|$.

We next consider flux quantization through the fibres $S^4/\Z_{n_\pm}$ 
over the poles of $\Sigma$. Similar to \eqref{NBBBW}, we can use 
equation \eqref{Glocalize2} to obtain
\begin{align}\label{Npm}
N_\pm = & \ \frac{1}{(2\pi \ell_p)^3} \int_{S^4/\Z_{n_\pm}} G =
\frac{1}{(2\pi \ell_p)^3}\frac{1}{n_\pm}\left[ \frac{(2\pi)^2}{b_1^\pm b_2^\pm }\frac{y^\pm_N}{9} -  \frac{(2\pi)^2}{b_1^\pm b_2^\pm }\frac{y^\pm _S}{9}\right]\, ,
\end{align}
and note that the orbifold factors of $d=n_\pm$ follow since 
the north and south poles of $S^4/\Z_{n_\pm}$ are both 
orbifold loci.
With $N_\pm>0$, this fixes the signs to be $y^\pm_N=-y^\pm_S>0$. Moreover, 
from the homology relation between these cycles we deduce
\begin{align}
N \equiv n_+ N_+ = n_- N_-\, ,
\end{align}
which can be obtained from \cite{Boido:2022mbe} and we are assuming that we are preserving supersymmetry via the
twist 
{\it viz.} \eqref{p1p2sum}, as in the known supergravity solutions \cite{Ferrero:2021wvk}.
These equations imply
\begin{align}\label{yNpm}
y_N ^\pm =  9\pi \ell_p^3\, b_1^\pm b_2^\pm N\, .
\end{align}

So far this is very similar to the analysis in section \ref{sec:BBBW}, dressed with some 
orbifold factors. However, due to the fixed point set on $M_6$ being 
isolated for $\varepsilon\neq 0$, the rest of the computation now proceeds 
differently. In particular we may already move directly to 
the central charge \eqref{aformula}:
\begin{align}
a=  -\frac{(2\pi)^3}{2(2\pi)^6\ell_p^9}\bigg[& \frac{1}{n_+}\frac{1}{(-\varepsilon/n_+) b_1^+ b_2^+}\frac{(y^+_N)^3-(y^+_S)^3}{162}  \nonumber\\ & \quad  + \frac{1}{n_-}\frac{1}{(\varepsilon/n_-) b_1^- b_2^-}\frac{(y^-_N)^3-(y^-_S)^3}{162}\bigg]\, . 
\end{align}
Here there are $4=2\times 2$ isolated fixed points: the two north and south poles $N$, $S$ 
of the fibre spheres, over the two poles of the spindle. Notice that 
the weights of $\partial_{\varphi_3}$ on the tangent spaces to the spindle poles are 
precisely $\mp 1/n_\pm$, since these are orbifold loci, and we have again 
included the factors of $d=n_\pm$.
Using \eqref{yNpm} this simplifies to
\begin{align}\label{ablock}
a = \frac{9[(b_1^+ b_2^+)^2 - (b_1^- b_2^-)^2]}{16\varepsilon} N^3\, .
\end{align}

Remarkably this expression takes a ``gravitational block'' form (see \cite{Hosseini:2019iad}), involving a difference of M5-brane anomaly polynomials\footnote{{\it c.f.} the expression for the central charge for the $(2,0)$ SCFT in $d=6$ given in \eqref{accs4ex}.} in the numerator, one associated to each $\pm$ pole of $\Sigma$.
The appearance of gravitational blocks for this case is related\footnote{In \cite{BenettiGenolini:2023yfe} we show that the central charge for a class of $AdS_3\times M_8$ solutions, of the type discussed in section \ref{sec:M8}, also leads to an expression
for the central charge in terms of gravitational blocks when $M_8$ is an $S^4$ bundle over
a toric base space, again with only isolated fixed points.} to the fact that $\xi$ has isolated fixed points, 
arising from the mixing with the $U(1)$ symmetry of the spindle as in \eqref{kvspin}. Indeed, by contrast,
the expression \eqref{aBBBW} for the Riemann surface case, $\Sigma_g$, displays a different structure. For the special case when $\Sigma_g \cong S^2$ one can consider deforming the expression for the Killing vector \eqref{bbbwkv} by allowing a mixing with the azimuthal symmetry of the sphere. However, this Abelian isometry sits inside the non-Abelian $SO(3)$ isometry,
and for the superconformal 
R-symmetry one expects, and in fact one finds, trivial mixing.

In order to evaluate \eqref{ablock} further we need to describe the fibration 
structure in more detail. The normal bundle to the M5-brane wrapped on 
$\Sigma$ is 
$\mathcal{N}=\mathcal{N}(\Sigma\hookrightarrow M_6) = \mathcal{O}(-p_1)\oplus \mathcal{O}(-p_2)$, 
with $p_1,p_2$ constrained via \eqref{p1p2sum}.
Compare to \eqref{p1p2} in the case that $g=0$ and $n_+=n_-=1$, 
so that $\Sigma=S^2$. 
The weights $b_i^\pm$ may then be computed using the results 
in  \cite{Boido:2022mbe}. We have 
\begin{align}\label{b1b2sum}
b_1^\pm + b_2^\pm = 1 \mp \frac{\varepsilon}{n_\pm}\, ,\quad 
b_i^+ - b_i^- = -\frac{p_i}{n_+n_-}\varepsilon\, ,
\end{align}
the first equation coming from the charge of the holomorphic $(3,0)$-form 
on the Calabi--Yau, and the second equation being (3.24) of \cite{Boido:2022mbe} 
(with $p_i=-p_i^{\mathrm{there}}$ 
and $\varepsilon=b_0^\text{there}$)). We may then solve these constraints 
by introducing new variables $\varphinew_i$ via 
\begin{align}\label{bchangephi}
b_i^\pm = \tfrac{1}{2}\left(\varphinew_i \mp \frac{p_i}{n_+n_-}\varepsilon\right)\, ,
\end{align}
with the constraint 
\begin{align}\label{phicon}
\varphinew_1 + \varphinew_2 = 2 + \frac{n_+-n_-}{n_+n_-}\varepsilon\, .
\end{align}
The central charge is then
\begin{align}\label{a6dfinal}
a = -\frac{9[(\varphinew_1p_2 + \varphinew_2p_1)(p_1p_2\varepsilon^2+n_+^2n_-^2\varphinew_1\varphinew_2)]}{64n_+^3n_-^3}N^3\, .
\end{align}
This derives the conjectured off-shell gravitational block formula in \cite{Faedo:2021nub}, 
where we have corrected the overall sign. 
In that reference it was shown extremizing  $a$ over the variables 
$\varphinew_i$ (subject to \eqref{phicon}) gives the central charge 
of the explicit supergravity solutions constructed in \cite{Ferrero:2021wvk}. 
Moreover, \eqref{a6dfinal} agrees \emph{off-shell} with the 
trial $a$-function in field theory, obtained by integrating 
the M5-brane anomaly polynomial over the spindle. 
As in the previous subsection, the reason that we have obtained the off-shell central charge via
localization is because we have only imposed a subset of the supersymmetry conditions, and moreover,
any remaining variations to go on-shell are associated with varying the central charge, as we explain in
section \ref{sec:actioncalc}.

It is instructive to see how \eqref{a6dfinal} for the spindle case reduces 
to \eqref{aBBBW} for the $S^2$ case of the previous subsection in the appropriate limit. In this section 
we should first set $n_+=n_-=1$, so that $\Sigma=S^2$, 
and then in order to relate to the variables to those in section \ref{sec:BBBW}, we should 
identify $\varphinew_1=1+\varepsilon$, $\varphinew_2=1-\varepsilon$, satisfying
\eqref{phicon}; if one takes $\varepsilon\rightarrow 0$ in \eqref{bchangephi} we obtain \eqref{bepsilon}. 
Taking the equivariant parameter of the spindle $\varepsilon\rightarrow 0$, 
one easily verifies that \eqref{a6dfinal} agrees with 
\eqref{aBBBW} with $g=0$. 

Let us next consider the conformal dimensions  of chiral primary operators in the dual SCFT that are
associated with M2-branes 
wrapped over the copies of the spindle $\Sigma_{N}$, $\Sigma_S$ 
at the poles of the $S^4$. 
Assuming these cycles are calibrated\footnote{Given that $\xi$ is tangent to the spindle, 
from \eqref{eq:su2structure_M6} we need to check that the volume form on these submanifolds are given by $\pm \ex^5\wedge \ex^6$.
This could be checked for the explicit supergravity solutions using the results of \cite{Ferrero:2021wvk,Ferrero:2021etw}.} by $Y'$,
applying \eqref{Deltaformula} gives for $\Sigma_N$,
\begin{align}\label{Deltaspindle}
\Delta(\Sigma_N) &
= 
\frac{-1}{(2\pi)^2\ell_p^3}\left[\frac{1}{n_+}\frac{2\pi}{(-\varepsilon/n_+)}\frac{y^+_N}{3}+\frac{1}{n_-}\frac{2\pi}{(\varepsilon/n_-)}\frac{y^-_N}{3}\right]\, , \nonumber\\
  &= \frac{3(b_1^+b_2^+ - b_1^-b_2^-) }{2\varepsilon}N = -\frac{3(\varphinew_1p_2 + \varphinew_2p_1)}{4n_+n_-}N\, ,
\end{align}
with the same result for $\Sigma_S$ (up to orientation), $\Delta(\Sigma_N) =   \Delta(\Sigma_S)$.
Evaluating this on the extremal values $\varphinew_i^*$, $\varepsilon^*$, 
one can verify 
the result agrees with that computed using the explicit 
supergravity solutions in \cite{Ferrero:2021wvk}, which provides strong evidence that the cycles are indeed calibrated. Notice 
also that \eqref{Deltaspindle} agrees with \eqref{DeltaBBBW} 
in the appropriate limit, even though it was calculated using 
localization quite differently. 

Finally, notice that in section \ref{sec:BBBW} we needed to impose 
flux quantization of $G$ through the four-cycles 
$C_4^{(i)}$ in order to compute the central charge, 
while here we did not. On the other hand, we may compute 
the fluxes through the analogous cycles in the spindle solutions
also using localization. Thus, let $C_4^{(i)}$ be the 
total space of the $S^2_i$ bundles over the spindle, $i=1,2$. 
Using \eqref{Glocalize2} we can write down 
\begin{align}
N_{C_4^{(i)}} & = \frac{1}{(2\pi\ell_p)^3}\Bigg\{\frac{1}{n_+}\frac{(2\pi)^2}{(-\varepsilon/n_+)b^+_i}\frac{y_N^+-y_S^+}{9}+\frac{1}{n_-}\frac{(2\pi)^2}{(\varepsilon/n_-)b^-_i}\frac{y_N^--y_S^-}{9}\Bigg\}\,,
\end{align}
which using \eqref{yNpm} gives
\begin{align}\label{ads5spinhom}
N_{C_4^{(1)}} = \frac{b_2^--b_2^+}{\varepsilon} N  = \frac{p_2}{n_+n_-}N\, , \quad N_{C_4^{(2)}} = \frac{b_1^--b_1^+}{\varepsilon} N = \frac{p_1}{n_+n_-}N\, ,
\end{align}
where we have used \eqref{b1b2sum} in the second equalities. 
One can check that this reduces to \eqref{N12} in the case $n_+=n_-=1$, 
exactly as it should, and \eqref{ads5spinhom} is a manifestation of the spindle generalization
of the homology relations \eqref{hom4}. 

\subsection{\texorpdfstring{$S^2$ bundle over $B_4$}{S2 bundle over B4}}\label{sec:B4}

Our final set of examples use the same technology, but are also rather 
different in detail. The ansatz we make  for the topology covers all of the 
supergravity solutions found in the original
reference \cite{Gauntlett:2004zh},  but also  many 
more cases for which explicit solutions have not been constructed.

We take $M_6$ to be the total space of an $S^2$ bundle over a base 
four-manifold $B_4$:
\begin{align}\label{M6B4}
S^2\, \hookrightarrow \, M_6 \, \rightarrow\, B_4\, .
\end{align}
Here $B_4$ is {\it a priori} an arbitrary closed four-manifold, where 
the R-symmetry Killing vector $\xi$ is assumed to rotate just the $S^2$ fibre and not act on $B_4$. 
In \cite{Gauntlett:2004zh} all solutions for which a certain natural almost 
complex structure on $M_6$ is integrable were found in closed form, 
and all have the structure \eqref{M6B4}, 
with, moreover, $B_4$ either K\"ahler--Einstein or the product of two Riemann surfaces with constant curvature metrics.  
In particular, 
for those cases $B_4$ is complex, and 
the $S^2$ bundle is that associated to the anti-canonical line bundle $\mathcal{L}$
over $B_4$. More precisely, here we view $S^2\subset\C\oplus \R=\R^3$, and 
use $\mathcal{L}$ to fibre the copy of $\C$ over $B_4$ to construct the bundle in 
\eqref{M6B4}. 
We shall impose these conditions on the 
topology of $M_6$ from the outset in this section, but without assuming any metric on $B_4$,  
and will see that 
not only do we reproduce all of the results for the explicit solutions 
found in  \cite{Gauntlett:2004zh}, but we also compute
closed-form expressions for various BPS quantities for considerable generalizations of those solutions, 
assuming the latter exist.

For simplicity, we continue by assuming that $\xi$ just acts on the $S^2$ fibre.
The fixed point set of $\xi$ then consists of two copies of $B_4$ at the north and south 
poles of the $S^2$ fibre. We denote these by $F_4=B_4^\pm$
(in contrast to the $N$ and $S$ notation used in the previous subsections). The normal bundles 
are then respectively
\begin{align}\label{Lpm}
L^+ = \mathcal{L}^{-1}\, , \qquad L^- = \mathcal{L}\, .
\end{align}
Noting the discussion around equation \eqref{eq:M6_GBolt}, we observe that  $\omega$ defines a cohomology class 
on each of $B_4^\pm$. First pick representative two-manifolds $\Gamma_\alpha\subset B_4$ for a basis 
for the free part of $H_2(B_4,\Z)$, and denote the copies of  these 
in $B_4^\pm$ by $\Gamma_\alpha^\pm$, respectively.
We may then write
\begin{align}
c_\alpha^\pm \equiv \int_{\Gamma_\alpha^\pm} \omega \, ,
\end{align}
with the real constants $c_\alpha^\pm$ then parametrizing the cohomology classes. 
The $[\Gamma_\alpha^\pm]\in H_2(M_6,\Z)$ in turn satisfy the homology relation, proved in appendix \ref{app:homology} (see \eqref{m6casenpmnmhom}),
\begin{align}\label{twocycles}
[\Gamma_\alpha^-] -[\Gamma_\alpha^+] = n_\alpha [S^2_{\mathrm{fibre}}]  \in H_2(M_6,\Z)\, ,
\end{align}
where we have defined the Chern numbers
\begin{align}\label{nalpha}
n_\alpha \equiv \int_{\Gamma_\alpha}c_1(\mathcal{L})\in \Z\, .
\end{align}
These arise since 
$\mathcal{L}^{-1}$ and $\mathcal{L}$ are the normal bundles $L^\pm$ 
in \eqref{Lpm}, and are part of the topological data that specifies the solution. Integrating the equivariantly closed form
 $\Phi^Y$ in \eqref{eq:M6_PhiY} over the two-cycle in \eqref{twocycles} and using localization on the right-hand side immediately 
gives 
\begin{align}\label{cdiff}
c_\alpha^- - c_\alpha^+  = \frac{2\pi}{\epsilon}\left(\frac{y_+^2}{3}-\frac{y_-^2}{3}\right)n_\alpha\, ,
\end{align}
where the weight of the R-symmetry vector on the $\pm$ poles 
of the fibre $S^2$ is $\pm\epsilon$, where in turn 
the R-charge of Killing spinor fixes $\epsilon=1$.\footnote{The R-charge 
of the Killing spinor is $\rcharge=\tfrac{1}{2}$ under $\xi=\partial_\psi$, but this 
is also precisely the charge of a spinor on the tangent space $\C$ to a pole of $S^2$ that is regular at the origin, where 
$\xi=\partial_\psi$ rotates $\C$ with weight one. This fixes $\psi$ to have period 
$2\pi$ hence  $\epsilon=\pm1$, and in our conventions $\epsilon=+1$. 
}
Again, we recall that $y$ is necessarily constant on each copy of $B_4^\pm$, 
and we have denoted those constants by $y_\pm$ in \eqref{cdiff}.

Next we turn to flux quantization for $G$. Equation \eqref{eq:M6_GBolt}
immediately gives
\begin{align}\label{Npmflux}
N_\pm \equiv \frac{1}{(2\pi\ell_p)^3}\int_{B_4^\pm} G = -\frac{1}{(2\pi\ell_p)^3}
\frac{1}{8y_\pm^3}\Imat_{\alpha\beta} c_\alpha^\pm c_\beta^\pm {\equiv - \frac{1}{(2\pi\ell_p)^3}\frac{1}{8y_\pm^3} \langle c^\pm, c^\pm \rangle } \, ,
\end{align}
where 
$\Imat=(I_{\alpha\beta})$ is the inverse of the intersection form 
for the four-manifold $B_4$, and we introduced the notation $\langle \cdot, \cdot \rangle$ for the bilinear form defined by $\Imat$. The intersection form is a unimodular 
integer-valued symmetric matrix, and hence its inverse has the same property. 
Further background and discussion of this may be found in appendix~\ref{app:homology}. 
We also have four-cycles $C_4^{(\alpha)}$ that 
are the total spaces of the $S^2$ bundle over $\Gamma_\alpha\subset B_4$ 
in the base. Using \eqref{Glocalize2} gives
\begin{align}\label{Nalpha}
N_\alpha \equiv \frac{1}{(2\pi\ell_p)^3}\int_{C_4^{(\alpha)}} G
& = \frac{1}{(2\pi\ell_p)^3}\Big[ \frac{2\pi}{\epsilon}\frac{1}{6y_+}c_\alpha^+ +
\left(\frac{2\pi}{\epsilon}\right)^2\frac{y_+}{9}n_\alpha\nonumber\\
&\qquad \qquad \qquad  -\frac{2\pi}{\epsilon}\frac{1}{6y_-}c_\alpha^- -\left(\frac{2\pi}{\epsilon}\right)^2\frac{y_-}{9}n_\alpha
\Big]\, ,
\end{align}
where we have used \eqref{Lpm} and \eqref{nalpha}.  Using
also \eqref{cdiff} this equation may then  be solved 
for $c_\alpha^+$, giving
\begin{align}\label{calphap}
c_\alpha^+ = \frac{2\pi y_+[(y_--y_+)^2 n_\alpha + 36\pi \ell_p^3\epsilon^2 y_- N_\alpha]}{3(y_--y_+)\epsilon}\, .
\end{align}
Substituting this into $N_\pm$ in \eqref{Npmflux} gives
\begin{align}\label{Npmmess}
N_\pm = -\frac{(y_--y_+)^2}{144\pi\ell_p^3\epsilon^2 y_\pm}
{\Inn \mp \frac{y_\mp}{2y_\pm}}
{\InN} - \frac{9\pi\ell_p^3\epsilon^2y_\mp^2}{y_\pm(y_--y_+)^2}
{\INN} \, .
\end{align}
These should be read as simultaneous equations to solve 
for $y_\pm$, given flux data $N_\pm$ and $N_\alpha$. On the other hand, 
the four-cycles 
$[B_4^\pm]$, $C_4^{(\alpha)}$ are not independent in $H_4(M_6,\Z)$. The corresponding 
homology relation, proved in appendix \ref{app:homology}, implies the following relation between the fluxes must hold
\begin{align}\label{homrel}
N_- - N_+ ={\InN} \, .
\end{align}
This should be  regarded as a topological constraint, fixing $N_-$ given a choice of $N_+$ and $N_\alpha$.  

Having solved \eqref{Npmmess}, one can then compute the central charge 
using \eqref{aformula}:
\begin{align}
a = &\  \frac{1}{2(2\pi)^6\ell_p^9}\Bigg\{-\frac{2\pi}{\epsilon}\frac{1}{48y_+}
{\langle c^+, c^+ \rangle } -\left(\frac{2\pi}{\epsilon}\right)^2\frac{y_+}{36}
{\langle c^+, n \rangle} - \left(\frac{2\pi}{\epsilon}\right)^3 \frac{y_+^3}{162}
{\Inn} \nonumber\\
& \qquad +\frac{2\pi}{\epsilon}\frac{1}{48y_-}
{\langle c^-, c^- \rangle} + \left(\frac{2\pi}{\epsilon}\right)^2\frac{y_-}{36}
{\langle c^-, n \rangle} +\left(\frac{2\pi}{\epsilon}\right)^3\frac{y_-^3}{162}
{\Inn} \Bigg\}\, .
\end{align}
Substituting the equations \eqref{cdiff}, \eqref{calphap} for $c_\alpha^\pm$, this simplifies to
\begin{align}\label{aB4final}
a = -\frac{(y_--y_+)^4 
{\Inn} + 3888\pi^2\ell_p^6\epsilon^4
y_-y_+ 
{\INN}}
{20736\pi^3\ell_p^9\epsilon^3(y_--y_+)}\, .
\end{align}

Turning to conformal dimensions of wrapped M2-branes, 
first applying \eqref{Deltaformula} to $\Sigma_2=S^2_{\mathrm{fibre}}$ gives
\begin{align}\label{Deltafibre}
\Delta(S^2_{\mathrm{fibre}}) & = -\frac{1}{(2\pi)^2\ell_p^3}\left(\frac{2\pi}{\epsilon}
\frac{y_+}{3} -\frac{2\pi}{\epsilon}
\frac{y_-}{3} \right)\, .
\end{align}
On the other hand, for $\Sigma_2=\Gamma_\alpha^\pm$ we instead have,
with appropriate orientations\footnote{As we will see, these overall signs give rise to $\Delta>0$ in known explicit solutions.},
\begin{align}\label{DeltaGamma}
\Delta(\Gamma_\alpha^\pm) =  \mp\frac{1}{(2\pi)^2\ell_p^3}\frac{1}{2y_\pm}c_\alpha^\pm\, .
\end{align}

We now solve \eqref{Npmmess}, subject to the 
constraint \eqref{homrel}, and  then substitute into
\eqref{aB4final}, \eqref{Deltafibre}, \eqref{DeltaGamma} 
to obtain completely general results. To do so it is convenient to solve the constraint by writing 
\begin{align}
N_-&=\frac{1}{2}\InN+M\,,\nn\\
\label{eq:filippo}
N_+&=-\frac{1}{2}\InN+M\,,
\end{align}
which defines the flux number $M$.
Assuming $\INN>0$, we then find that there are two solutions for $y_\pm$:
\begin{align}
y_\pm=&\pm\frac{9\pi\ell_p^3 \INN}{[\Inn\INN+12M^2]^2}
\Big\{ 2[\Inn\InN\INN\pm8\InN^2 M\nn\\
&\qquad\qquad\qquad\qquad\qquad\qquad
\mp\Inn\INN M +12\InN M^2\mp12M^3]\\
&+(\Inn\INN\pm8M\InN+12M^2)\sqrt{4\InN^2-\Inn\INN-12M^2}\Big\}\nn\,,
\end{align}
and 
\begin{align}
y_\pm=&\pm\frac{9\pi\ell_p^3 \INN}{[\Inn\INN+12M^2]^2}
\Big\{ 2[\Inn\InN\INN\pm 8\InN^2 M\nn\\
&\qquad \qquad\qquad\qquad\qquad\qquad\mp\Inn\INN M +12\InN M^2\mp12M^3]\\
&-(\Inn\INN\pm 8M\InN+12M^2)\sqrt{4\InN^2-\Inn\INN-12M^2}\Big\}\nn\,.
\end{align}
These then give rise to the following two expressions for the central charge, respectively:
\begin{align}\label{centchgegenk}
a=\frac{9\INN^2}{8(\Inn\INN+12M^2)^2}\Big[\InN(8\InN^2-3\Inn\INN-36M^2)\nn\\
\pm[4\InN^2-\Inn\INN-12M^2]^{3/2}\Big]\,.
\end{align}
As we will see, the first solution is associated with a positive central charge for known examples, so we now continue with
this branch.

Turning to conformal dimensions of wrapped M2-branes,  \eqref{Deltafibre} gives
\begin{align}\label{s2delta}
\Delta(S^2_{\mathrm{fibre}}) =-\frac{3\INN\Big(2\InN+\sqrt{4\InN^2-\Inn\INN-12M^2}\Big)}{\Inn\INN+12M^2}\, .
\end{align}
Instead, wrapping $\Gamma_\alpha^\pm$,  we compute 
\begin{align}\label{othergammadelta}
&\Delta(\Gamma_\alpha^{\pm})  = 
\frac{3}{2\Inn\INN+24M^2}\Big[\INN\Big(\sqrt{4\InN^2-\Inn\INN-12M^2} \nn\\
& \qquad \qquad +2\InN\Big)n_\alpha+\Big(-\Inn\INN\pm4\InN M-12M^2
\nn \\
&\qquad \qquad  \pm2M\sqrt{4\InN^2-\Inn\INN-12M^2}\Big)N_\alpha\Big]\,.
\end{align}

The expressions for the central charge and conformal dimensions given in \eqref{centchgegenk} and
\eqref{s2delta}, \eqref{othergammadelta} are new general results that go well beyond
known explicit supergravity solutions. We conclude this subsection by briefly 
making some checks for some specific cases where explicit supergravity solutions are known.

\subsubsection{\texorpdfstring{$B_4=KE_4^+$: simple class}{B4 KE4}}

An interesting family of explicit solutions found in \cite{Gauntlett:2004zh} involves
taking $B_4$ to be a positively curved K\"ahler--Einstein four-manifold $KE_4$. 
The central charge of these solutions was computed in \cite{Gauntlett:2006ai}.
The fluxes $N_\alpha$ for these solutions are not arbitrary, but, by assumption, are constrained
to be proportional to the Chern numbers:
\begin{align}
N_\alpha = k n_\alpha\, ,
\end{align}
for some constant $k$. Following the notation of \cite{Gauntlett:2006ai}, we can then write
\begin{align}
I_{\alpha\beta}n_\alpha n_\beta&\equiv \bar M \in \Z\, ,
\end{align}
where $\bar M$ is a topological invariant of $KE_4$
(and note that $\bar M=M^{\mathrm{there}}$ in \cite{Gauntlett:2006ai}),
and hence we also have $\InN=k\bar M$ as well as $\INN=k^2\bar M$.
To compare with \cite{Gauntlett:2006ai} we further consider the class with $N_-=-N_+$, which corresponds to setting the flux number $M=0$.
In order to match the notation in \cite{Gauntlett:2006ai} we define
\begin{align}
k\equiv-\frac{2N}{h}\, ,
\end{align}
where $h\equiv \mathrm{gcd}[\bar M,2I(B_4)]$, where $I(B_4)\in\mathbb{N}$ is 
the Fano index of $B_4$. 
Substituting these fluxes into the general formulae above using (as below) the first branch, we obtain 
the central charge
\begin{align}
a = \frac{9(3\sqrt{3}-5)}{h^3}\bar M N^3\, ,
\end{align}
which agrees with the explicit result in \cite{Gauntlett:2006ai}. 
Furthermore, from \eqref{s2delta}, \eqref{othergammadelta} we compute
\begin{align}
\Delta(S^2_{\mathrm{fibre}}) = \frac{6(2-\sqrt{3})}{h}N\, , \qquad
\Delta(\Gamma_\alpha^\pm) = \frac{3(\sqrt{3}-1)}{h} Nn_\alpha\, ,
\end{align}
which again agree with \cite{Gauntlett:2006ai}. 

\subsubsection{\texorpdfstring{$B_4=S^2\times \Sigma_g$}{B4 S2 Sigmag}}
We also consider the case $B_4=S^2\times \Sigma_g$ when the base is a product of a two-sphere with a Riemann surface of genus $g$,
for which explicit solutions were given in \cite{Gauntlett:2004zh}.
The 
(inverse) intersection matrix and Chern numbers are then given by
\begin{align}
I= \begin{pmatrix} 0 & 1 \\ 1 & 0\end{pmatrix}\, ,\quad n_1 = 2\,,\qquad n_2=2(1-g_2)\equiv \chi\, .
\end{align}
We can now write 
\begin{align}
\Inn=4\chi\, , \quad\INN=2N_1N_2\,  ,\quad \InN=2N_2+\chi N_1\, .
\end{align}
The central charge then reads 
\begin{align}
a=&
\frac{9N_1^2N_2^2}{4(2\chi N_1N_2+3M^2)^2}\left(4N_2^2+2\chi N_1N_2+\chi^2N_1^2-3M^2\right)^{3/2}\nn\\
&+\frac{9N_1^2N_2^2(2N_2+\chi N_1)}{8(2\chi N_1N_2+3M^2)^2}(8N_2^2+2\chi N_1 N_2+2\chi^2 N_1^2-9M^2)\,.
\end{align}
This result precisely coincides\footnote{We should identify 
$N_1=-N_{S^2}^{\rm there}$, $N_2=-N_{\Sigma}^{\rm there}$.} with the large $N$ limit of the central charge computed in \cite{Bah:2019rgq}, who obtained their result
by consideration of anomaly inflow. 
Expressions for $\Delta(S^2_{\mathrm{fibre}})$ and $\Delta(\Gamma_\alpha^\pm)$ can similarly be obtained from \eqref{s2delta}, \eqref{othergammadelta}, and these comprise new results
for this class. Below we will write explicit expressions for these for a further sub-class associated with explicit supergravity solutions
considered in~\cite{Gauntlett:2006ai}.

Specifically, we next further restrict to the $B_4=S^2\times S^2$ case, correspondingly setting 
$g_2=0$ and hence $\chi=2$. We also restrict to the sub-class with $M=0$.
We write 
\begin{equation}
		N_1 \equiv -q N \, , \qquad N_2 \equiv -p N \, ,
\end{equation}
and find that the central charge can be written as
\begin{equation}
	a = -\frac{9}{16} {N}^3 (p+q) \left(2 p^2+p q+2 q^2\right) + \frac{9}{8} N^3 \left(p^2+p q+q^2\right)^{3/2} \, .
\end{equation}
This precisely coincides with the result computed from the explicit solution in \cite{Gauntlett:2006ai}.
The conformal dimensions of the chiral primaries associated to M2-branes wrapping supersymmetric cycles take the form
\begin{align}
	\Delta(S^2_{\rm fibre}) &= \frac{3}{2} \left( p+q - \sqrt{p^2+pq+q^2} \right) N \, , \nn \\
	\Delta(\Sigma_1^\pm) &= \frac{3}{2} \left( - p + \sqrt{p^2+pq+q^2} \right) N \, , \nn \\
	\Delta(\Sigma_2^\pm) &= \frac{3}{2} \left( - q + \sqrt{p^2+pq+q^2} \right) N \, ,
\end{align}
which again match the results of \cite{Gauntlett:2006ai} for $p,q,N>0$.

Finally, we connect with the $Y^{p,q}$ class of explicit solutions associated with $B_4=T^2\times S^2$.
We now set  
$g_2=1$ and so $\chi=0$ and write 
\begin{align}
N_1=-N,\quad N_2=-p,\quad M=q\,.
\end{align}
The central charge is then given by
\begin{align}
a=\frac{p^2}{4q^4}{\left[-8p^3+9pq^2+(4p^2-3q^2)^{3/2} \right]} N^2\, ,
\end{align}
which agrees with the central charge of the explicit supergravity solutions
\cite{Gauntlett:2006ai}. 
One can check that the expressions for the dimensions of the chiral primaries associated to M2-branes wrapping supersymmetric cycles also agree with those
in \cite{Gauntlett:2006ai}.
This class of explicit solutions are the M-theory duals of the well-known 
$AdS_5\times Y^{p,q}$ Sasaki--Einstein solutions of type IIB string theory \cite{Gauntlett:2004yd}. 
It is amusing to note that in this case the central charge 
may similarly be computed in type IIB without knowledge of the explicit supergravity solution, 
instead employing volume minimization \cite{Martelli:2005tp}. This is 
 in the same spirit as the present paper, but the details
in M-theory and type IIB are very different.

\subsection{Action calculation}
\label{sec:actioncalc}

In this section we show that the expression for the off-shell central charge \eqref{aloc} is in fact proportional to an effective action obtained by reducing the eleven-dimensional action using the ansatz \eqref{eq:M6Ansatz}. This establishes, independently of the analysis of the supersymmetry equations, that the result obtained by computing the integral \eqref{aloc} using equivariant localization is indeed off-shell, in the sense that it only corresponds to imposing a subset of the equations of motion. Furthermore, because of the relation to the effective action, extremizing the central charge integral over any undetermined parameters
provides a further necessary condition for imposing the $D=11$ equations of motion, and gives the on-shell result for
the central charge.

For the general class of $AdS_5\times M_6$ solutions of $D=11$ supergravity considered in
\cite{Gauntlett:2004zh}, one can show that the $D=11$ equations of motion, as given in \cite{Gauntlett:2002fz}, give rise to
$D=6$ equations of motion for the metric, the scalar $\lambda$ and the four-form $G$. 
These can be written in the form
\begin{align}
\label{eom6d}
0&=\nabla^2\lambda+9(\nabla\lambda)^2+4-\frac{1}{144}\ex^{-6\lambda}G^2 \, ,\nn \\
R_{ab}&=9\nabla_{ab}\lambda-9\nabla_a\lambda\nabla_b\lambda+\frac{1}{12}\ex^{-6\lambda}G^2_{ab}-4g_{ab}\,,
\end{align}
and can be obtained by extremizing the six-dimensional action
\begin{align}
\label{6dactads5}
S=\int_{M_6} \dd^6x \sqrt{g} \left\{\ex^{9\lambda}[R+90(\nabla\lambda)^2-20]-\frac{1}{48}\ex^{3\lambda}G^2\right\}\,.
\end{align}

If we impose all of the conditions for supersymmetry that were considered in reference
\cite{Gauntlett:2004zh}, then the $D=6$ equations of motion are automatically solved.
In the previous subsections we have only imposed a subset of the supersymmetry conditions
and it is interesting to examine, for certain cases, if there are additional variations to consider
in obtaining our results for the central charge and conformal dimensions of chiral primaries.

In addition to the supersymmetry conditions we have been using, we 
now impose, as additional assumptions\footnote{It is plausible that these
are not extra assumptions, but we will not pursue that further here.}, that the scalar equation of motion and
the trace of the second equation in \eqref{eom6d} are 
also satisfied. We first multiply the scalar equation of motion by $\ex^{9\lambda}$ and then integrate over $M_6$.
Then, observing 
the useful fact that $\nabla^2(\ex^{9\lambda})=9 \ex^{9\lambda}[\nabla^2\lambda+9(\nabla\lambda)^2]$ and discarding the total derivative,  
we conclude that
\begin{align}\label{usefulactlemma}
\int_{M_6} \ex^{9\lambda}\, \vol_6=\frac{1}{24}\int_{M_6} \ex^{3\lambda}G\wedge *_6 G\,.
\end{align}
In fact we can obtain this result via localization, as we show below.
We now take the trace of the second equation in \eqref{eom6d} to get an expression for the Ricci scalar, $R$, and substitute into the
action. Next, we eliminate the $\ex^{3\lambda} G^2$ terms from the resulting expression using the scalar equation of motion in \eqref{eom6d}.
We then find that the remaining scalar terms combine into a total derivative and we
obtain the following simple expression for the on-shell action
\begin{align}\label{actosfansads50}
S=-8\int_{M_6} \ex^{9\lambda}\, \vol_6=-8\int_{M_6}\Phi\,,
\end{align}
{where the second equality just uses the definition that the integral of a polyform is the integral of its top component.} 
Using the definitions \eqref{eq:M6_PhiG} and \eqref{eq:M6_Phi*G}, as well as \eqref{usefulactlemma}, it is interesting to note that
we can also write the on-shell action in terms of equivariant forms in an alternative way:
\begin{align}\label{actosfansads502}
S=-\frac{1}{3}\int_{M_6} \Phi^G\wedge \Phi^{*G}\,.
\end{align}

Now, recall that in some of the examples in the previous sections, we showed that localization gave an expression for the central charge that
still had some undetermined dependence on some variables. For example, in the case of M5-branes wrapping a Riemann surface the central
charge in \eqref{aBBBW} is off-shell, as it depends on the $b_i$, defining the Killing vector in \eqref{bbbwkv}
subject to the constraint \eqref{bepsilon}.
Similarly, for the case of M5-branes wrapping a spindle the central
charge in \eqref{a6dfinal} is off-shell, depending on the choice of Killing vector in \eqref{kvspin}.
Since the partially on-shell action in \eqref{actosfansads50} is proportional the central charge, we see that varying the central charge with
respect to these undetermined coefficients, provides a necessary condition for putting  the system on-shell. Thus, this procedure
then gives the on-shell value of the central charge as well as the conformal dimensions of the chiral primaries dual to wrapped membranes.

To conclude this section, as somewhat of an aside, we show that \eqref{usefulactlemma} can also be obtained via localization.
We start by recalling \eqref{bilinstargads5}
\begin{align}\label{bilinstargads52}
\ex^{3\lambda}* G = 3\diff (\ex^{6\lambda}\xi^\flat) - 4\ex^{6\lambda}Y\, .
\end{align}
Since $\xi^\flat$ smoothly goes to zero at a fixed point, the first term is globally exact. Thus,
for closed $M_6$ we have
\begin{align}\label{actiontrick}
\int_{M_6}\ex^{3\lambda}* G\wedge G = -4\int_{M_6}\ex^{6\lambda}Y\wedge G = 
-4\int_{M_6}\Phi^Y\wedge \Phi^G\, ,
\end{align}
where recall $\Phi^Y=\ex^{6\lambda}Y +\tfrac{1}{3}y^2$. 

At a fixed point set we have $\sin\zeta=\pm 1$, $Y'=-\sin\zeta Y=\mp Y$ and
also $2y=\ex^{3\lambda}\sin\zeta=\pm \ex^{3\lambda}$.
Localizing we then get possible contributions from four-forms, two-forms and zero-forms. Specifically:
\begin{align}
\fleft \Phi^Y\wedge \Phi^G \, \right|_{4,F} & = \tfrac{1}{3}y^2 G -\tfrac{1}{3}\ex^{9\lambda}Y\wedge Y' = -\tfrac{1}{6}y \ex^{6\lambda}Y\wedge Y -\tfrac{1}{3}(\pm 2y)\ex^{6\lambda}(\mp)Y\wedge Y\nonumber\\
& = \tfrac{1}{2}y \ex^{6\lambda}Y\wedge Y\, ,
\end{align}
where we used the bilinear \eqref{eq:M6_BilinearEqn_Useful}. We also have
\begin{align}
\fleft \Phi^Y\wedge \Phi^G \, \right|_{2,F} & = \tfrac{1}{3}y^2(-\tfrac{1}{3}\ex^{3\lambda}Y') +\tfrac{1}{9}y \ex^{6\lambda}Y = \tfrac{1}{6}y\ex^{6\lambda}Y\, ,\nn\\
\fleft \Phi^Y\wedge \Phi^G\, \right|_{0,F} & = \tfrac{1}{9}y(\tfrac{1}{3}y^2) = \tfrac{1}{27}y^3\, .
\end{align}
On the other hand
\begin{align}
\fleft \Phi \fright & =\ex^{9\lambda}\vol +\tfrac{1}{12}\ex^{9\lambda}* Y -\tfrac{1}{36}y\ex^{6\lambda}Y -\tfrac{1}{162}y^3\,\nonumber\\
& = -\tfrac{1}{12}y\ex^{6\lambda}Y\wedge Y -\tfrac{1}{36}y\ex^{6\lambda}Y -\tfrac{1}{162}y^3\nonumber\\
& =-\tfrac{1}{6} \fleft \Phi^Y\wedge \Phi^G \fright \, .
\end{align}
Assuming the fixed point 
set is non-empty, this proves that 
\begin{align}
\int_{M_6}\ex^{3\lambda}* G\wedge G = 24\int_{M_6}\Phi\, ,
\end{align}
where $\int_{M_6}\Phi=\int_{M_6}\ex^{9\lambda}\, \vol$ computes the 
central charge. On the other hand, when the fixed point set is empty, 
localization says both integrals are just zero (which is a contradiction 
for the central charge, which is manifestly positive).
This completes our localization proof of \eqref{usefulactlemma}.

\section{\texorpdfstring{$AdS_3\times M_8$}{AdS3 x M8} Solutions}\label{sec:M8}

We now consider supersymmetric $AdS_3\times M_8$ solutions of $D=11$ supergravity 
 that are dual to $\mathcal{N}=(0,2)$ SCFTs in $d=3$, 
also discussed in \cite{BenettiGenolini:2023yfe}.
These solutions have a canonical Killing vector that is dual to the R-symmetry of
the SCFT.
A general class of solutions associated\footnote{The class may also describe more general kinds of wrapped M5-branes.} 
with M5-branes wrapping a holomorphic four-cycle inside a Calabi--Yau four-fold
were classified in \cite{Gauntlett:2006qw}. 
Moreover, infinite classes of explicit solutions were also constructed in \cite{Gauntlett:2006qw},
including the particular solution found in \cite{Gauntlett:2000ng} that generalizes the Maldacena--N\'u\~nez construction. The classification results of \cite{Gauntlett:2006qw} were later extended to the most general
supersymmetric $AdS_3\times M_8$ solutions in \cite{Ashmore:2022ydf}.

In this section we will show that there are natural equivariant polyforms, 
given in
\cite{BenettiGenolini:2023yfe}, that can be constructed
from Killing spinor bilinears. Furthermore, once again they can be used to compute
the central charge of the dual SCFTs, without knowing the explicit solution. For simplicity we will focus on the 
general class of $AdS_3\times M_8$ solutions considered in \cite{Gauntlett:2006qw}, and discussed in
section 4.3 of \cite{Ashmore:2022ydf}, but we strongly suspect that our results will have a simple generalization
to the most general class of \cite{Ashmore:2022ydf}.
We will illustrate the formalism for a class of $M_8$ which are $S^4$ fibrations over a four-dimensional
base $B_4$ associated with M5-branes wrapping $B_4$. Here we will consider cases when the R-symmetry acts 
just on the $S^4$ fibre, while in \cite{BenettiGenolini:2023yfe} we consider cases where $B_4$ is toric and the R-symmetry
acts on both $S^4$ and $B_4$. 

\newcommand{\DA}{\lambda}
\newcommand{\chiA}{\epsilon}
\newcommand{\LA}{\xi}
\newcommand{\PhiA}{\Psi}
We follow the notation and conventions of \cite{Ashmore:2022ydf}, with some minimal notational changes to maintain some cohesion
with the previous sections.
The ansatz for the $D=11$ metric and four-form are given by
\begin{align}
\label{eq:AdS3Ansatz}
\diff s^2&= \ex^{2\DA}\left[ \dd s^2(AdS_3)+ \dd s^2(M_8)\right]\,,\nn\\
G&=\ex^{3\DA}F + \vol(AdS_3)\wedge \ex^{3\DA}f\,,
\end{align}
where $\DA,F$ and $f$ are a function, a four-form and a one-form on $M_8$,
respectively.
In addition $\dd s^2(AdS_3)$ is the metric on a unit radius $AdS_3$ 
and $\vol(AdS_3)$ is the corresponding volume form.
The Bianchi identity and the equation of motion for $G$ are then
\begin{align}\label{ads3beom}
\dd(\ex^{3\DA}F)&=0\, ,\qquad  \ex^{-6\DA} \dd (\ex^{6\DA}* f)+\frac{1}{2}F\wedge F=0\,,\nn\\
\dd (\ex^{3\DA}f)&=0\, ,\qquad \ \ 
\ex^{-6\DA} \dd ( \ex^{6\DA}* F)-f\wedge F=0\,,
\end{align}
where the Hodge star is with respect to the metric $\diff s^2(M_8)$.
The flux quantization condition\footnote{We are interested in large $N$ results and hence we are neglecting
the extra Pontryagin class contribution of \cite{Witten:1996md}; see also the discussion in sec 2.3 of \cite{Gauntlett:2006qw}.}
is as in \eqref{fluxnormn} and so for
any four-cycle, $C_4$, we have 
\begin{align}\label{fluxnormnads3}
\frac{1}{(2\pi\ell_p)^3}\int_{C_4} \ex^{3\DA}F \in \mathbb{Z}\, .
\end{align}

Having supersymmetric solutions that are dual to $\mathcal{N}=(0,2)$ SCFTs in $d=3$ requires that
$M_8$ admits two Majorana spinors $\chiA_i$ satisfying Killing spinor equations given in \cite{Ashmore:2022ydf}.
It is convenient to define the complex spinor $\chiA\equiv \chiA_1+\ii\chiA_2$.
For the most general class of solutions, we  can construct the following scalar bilinears
\begin{align}
\bar\chiA\chiA=1\, ,\qquad
\bar\chiA^c\chiA=1\, ,\qquad
\zeta=\bar\chiA\gamma_9\chiA\, ,\qquad
S=\bar\chiA^c\gamma_9\chiA\, ,\qquad
\end{align}
where $\zeta$ is real and $S$ is complex, and $\gamma_9 \equiv \gamma_{1 \cdots 8}$. We also have the following
one-form bilinears\footnote{In comparing with \cite{Ashmore:2022ydf} note
that $\DA=\Delta^\text{there}$, $m^\text{there}=-1/2$, $\chiA_i=\chi^\text{there}_i$ 
and  $\LA^\flat=\frac{1}{2}L^\text{there}$. We have also written $\PhiA=\Phi^\text{there}$ to avoid confusion with our notation
for polyforms. We also note that
in this section, in contrast to section \ref{sec:M6forms}, we use $\Gamma_{012\dots10}=-1$ as in \cite{Ashmore:2022ydf}.}
\begin{align}
K=\bar\chiA\gamma_{(1)}\chiA\, ,\qquad
P=\bar\chiA^c\gamma_{(1)}\chiA\, ,\qquad
\LA^\flat=-\frac{\ii}{2}\bar\chiA\gamma_9\gamma_{(1)}\chiA\, ,\qquad
\bar\chiA^c\gamma_9\gamma_{(1)}\chiA=0\, ,\qquad
\end{align}
where $K, \LA^\flat$ are real and $P$ is complex.
The vector $\LA$, dual to the one-form $\xi^\flat$, is the Killing vector that is dual to the R-symmetry and 
it generates a symmetry of the full solution: $\mathcal{L}_\LA \DA=
\mathcal{L}_\LA F=\mathcal{L}_\LA f=0$. The action of $\LA$ on $\chiA$ is given by
\begin{align}
\mathcal{L}_\LA\chiA=\frac{\ii}{2}\chiA\,,
\end{align}
and we notice that $\mathcal{L}_\LA\zeta=0$.
In addition we recall the following bilinears that are also invariant under $\LA$: 
\begin{align}
J&=-\ii\bar\chiA\gamma_{(2)}\chiA\,,\qquad \omega=-\ii\bar\chiA{\gamma_9}\gamma_{(2)}\chiA\,,\nn\\
\varphi&=-\ii\bar\chiA\gamma_{(3)}\chiA\,,\qquad \phi=\bar\chiA{\gamma_9}\gamma_{(3)}\chiA\,,\nn\\
\PhiA&=\bar\chiA\gamma_{(4)}\chiA\,,\qquad\,\, * \PhiA=\bar\chiA {\gamma_9}\gamma_{(4)}\chiA\,.
\end{align}
Further bilinears are defined in \cite{Ashmore:2022ydf}, but will not be needed
in what follows. {As in the previous section, we are able to construct the key equivariantly closed polyforms
only using a subset of the supersymmetry conditions.}

We now restrict our considerations to a sub-class of solutions by demanding
that the complex scalar and one-form bilinears which are charged under the action of $\LA$ all vanish, which means
\begin{align}
S=P=0\,,
\end{align}
and, as in \cite{Ashmore:2022ydf}, we now write
\begin{align}
\zeta=\sin\alpha\,.
\end{align} 
In this case, it is possible to write down the bilinears in terms of a local $SU(3)$ structure on $M_8$.
Introducing an orthonormal frame $\ex^i$, we define the two-form $j=\ex^{12}+\ex^{34}+\ex^{56}$ and the three-form
$\theta =
(\ex^1+\ii \ex^2)(\ex^3+\ii \ex^4)(\ex^5+\ii \ex^6)$. The bilinears we introduced above can then be written
\begin{align}\label{gstructcondsads3}
\LA^\flat&=\frac{1}{2}\cos\alpha \, \ex^7, \qquad
K=\cos\alpha \, \ex^8,\qquad
J=j+\sin\alpha\, \ex^{78},\qquad
\omega=\sin\alpha\, j+ \ex^{78},\nn\\
\varphi&=\cos\alpha \, j\wedge \ex^{8},\qquad
\phi=-\cos\alpha \,  j\wedge \ex^{7},\qquad
\PhiA=-\frac{1}{2}j\wedge j-\sin\alpha \, j\wedge \ex^{78}\,.
\end{align}
These expressions allow one to easily obtain the
expressions for the contraction of $\LA$ on the bilinears.

The bilinears satisfy the differential conditions
\begin{align}\label{diffbilinads3}
\ex^{-3\DA} \dd (\ex^{3\DA} \sin\alpha)&=f+2 K\,,\nn\\
\ex^{-3\DA} \dd (\ex^{3\DA} \LA)&= \frac{1}{2}J-\frac{1}{4}\omega\hook F+\frac{1}{4}J\hook * F\,,\nn\\
\ex^{-6\DA} \dd (\ex^{6\DA} J)&=f\wedge \omega-2\LA\hook * F\,,\nn\\
\ex^{-3\DA} \dd (\ex^{3\DA} \omega)&=-2\LA\hook F\,,\nn\\
\ex^{-6\DA} \dd( \ex^{6\DA} \phi) &=2\PhiA+\sin\alpha \, F-* F\,,\nn\\
\ex^{-3\DA} \dd( \ex^{3\DA }\PhiA)&=-K\wedge F\,.
\end{align}
Additional conditions are given in \cite{Ashmore:2022ydf}.

With this set-up we can now construct the equivariant polyforms, as in \cite{BenettiGenolini:2023yfe}.
To begin we impose the condition $\dd(\ex^{3\DA} f)=0$ in \eqref{ads3beom}
which allows us to introduce a function $a_0$, which in general is only \emph{locally}\footnote{When $H^1(M_8,\R)=0$ by definition 
there will be a global function $a_0$ satisfying \eqref{dda}.} defined, via
\begin{align}\label{dda}
\ex^{3\DA} f= \dd a_0 \,. 
\end{align}
Moreover,  using $\LA\hook  f=0$ ((2.49) of \cite{Ashmore:2022ydf})
we see that $\mathcal{L}_\xi a_0=0$ is invariant under the action of the Killing vector.
From \eqref{diffbilinads3} we can then write
\begin{align}
\dd( \ex^{3\DA}\sin\alpha-a_0)=2 \ex^{3\DA}K\,.
\end{align}
Since $\dd( \ex^{3\DA}K)=0$, as follows from the first line of \eqref{diffbilinads3}, we can introduce a local coordinate $y$ so that
 \begin{align}
y=\frac{1}{2}(\ex^{3\DA}\sin\alpha-a_0),\qquad K= \ex^{-3\DA} \dd y\,.
 \end{align}

We can now construct an equivariant form, $\Phi^F$, involving the four-form $F$:
\begin{align}\label{ads3fluxeqform2}
\Phi^F
&= \ex^{3\DA} F-\frac{1}{2}\ex^{3\DA}\omega-\frac{y}{4}\,.
\end{align}
We can also construct another equivariant form, $\Phi^{*F}$, involving the four-form $* F$:
\begin{align}\label{starFpolyads3}
\Phi^{*F}
 &= \Phi^{*F}_4 -\frac{1}{2} \left( \ex^{6\DA}J - a_0 \ex^{3\DA}\omega \right) -\frac{1}{4}y^2\,,
\end{align}
where 
 \begin{align}\label{a4phi4def}
\Phi^{*F}_4\equiv \ex^{6\DA}* F - a_0 \ex^{3\DA} F\,.
\end{align} 
We also find another 
equivariant form, with the four-form $\Psi$ as the top form, given by
\begin{align}
\Phi^\Psi
&=\ex^{6\lambda}\Psi + y \ex^{3\lambda}F - \frac{1}{2}y \ex^{3\lambda} \omega - \frac{1}{8}y^2\,,
\end{align}
which is in fact equivariantly cohomologous to $\Phi^{*F}$:
\begin{align}
\Phi^{*F}=2\Phi^\Psi-\dd_\xi( \ex^{6\lambda}\phi)\,.
\end{align}

Associated with the central charge, we can construct the equivariant polyform 
\begin{align}\label{ads3explicitphicc}
\Phi 
&= \ex^{9\DA}\, \vol_8 + \tfrac{1}{4} \ex^{9\DA} * J - \tfrac{1}{8}y \ex^{6\DA}\PhiA - \tfrac{1}{16} y^2 \ex^{3\DA}F + \tfrac{1}{32}y^2 \ex^{3\DA}\omega + \tfrac{1}{192 }y^3\,.
\end{align}
More specifically, by computing the effective three-dimensional Newton constant, one finds 
the central charge of the dual $d=2$ SCFT can be written\footnote{
The central charge is given by $c=3 L/(2G_3)$ where $L$ is the radius of the $AdS_3$ vacuum and $G_3$
is the three-dimensional Newton constant \cite{Brown:1986nw}. To compute $G_3$ and thus obtain \eqref{ads3explicitphicc}, one reduces the $D=11$ supergravity action using the ansatz \eqref{eq:AdS3Ansatz}.}
\begin{align}\label{clocformads3}
c&\equiv \frac{3}{2^5\pi^7\ell_p^9}\int_{M_8}\ex^{9\DA}\,  \vol_8
= \frac{3}{2^5\pi^7\ell_p^9}\int_{M_8}\Phi
\,,
\end{align}
{where the second equality just uses the definition that the integral of a polyform is the integral of its top component,
and hence $c$ can be evaluated using the BVAB formula.}

We now consider some properties of the fixed point set.
Since $\|K\|^2=4\|\xi^\flat\|^2=\cos^2\alpha$, at fixed points we have $\alpha=\pm \pi/2$ and the Killing spinor
is chiral/anti-chiral: $\gamma_9\epsilon=\pm\epsilon$. Since $\dd y=\ex^{3\DA}K$, at the fixed points $y$ is constant and
\begin{align}
\fleft 2y\fright=(\ex^{3\DA}\sin\alpha-a_0)=\left(\pm \ex^{3\DA}-a_0\right)\,.
\end{align}
Furthermore, from $\ex^{-3\DA} \dd (\ex^{3\DA} \omega)=-2\LA\hook F$ we have that $\ex^{3\DA} \omega$ is a closed two-form on the fixed point set
\begin{align}\label{ads3omegclosedfp}
\fleft \dd( \ex^{3\DA} \omega  \fright \mkern-4mu )=0\,.
\end{align}
It will also be helpful to note from \eqref{gstructcondsads3} we have
\begin{align}\label{omegPsifp}
\fleft \ex^{6\DA} \omega^2\fright =\fleft \ex^{6\DA} j^2\fright\,.
\end{align}

We can also compute the conformal dimension $\Delta$ of chiral primary operators in the dual~$\mathcal{N}=(0,2)$ SCFT, 
corresponding to M2-branes wrapping supersymmetric two-cycles $\Sigma_2 \subset M_8$. By following similar arguments given 
in \cite{Gauntlett:2006ai}, we claim that such M2-branes are calibrated by  $\omega$, 
\emph{i.e.} $\vol|_{\Sigma_2}=\omega|_{\Sigma_2}$, and their conformal dimension may be computed in the gravity dual via 
\begin{equation}
\label{eq:M8_ConformalDimension}
	\Delta(\Sigma_2) = \frac{1}{(2\pi)^2\ell_p^3} \int_{\Sigma_2} \ex^{3\lambda} \omega \ .
\end{equation}
By considering the restriction of the four-form flux polyform $\Phi^F$ in \eqref{ads3fluxeqform2} to $\Sigma_2$,
$(\Phi^F)|_{\Sigma_2}=-\frac{1}{2}[\ex^{3\DA}\omega+\frac{y}{2}]|_{\Sigma_2}$, we see that when
$\Sigma_2$ is $\xi$-invariant we can evaluate the integral in \eqref{eq:M8_ConformalDimension}
by localization:
\begin{align}\label{Deltaformulaads3}
\Delta(\Sigma_2) = \frac{1}{(2\pi)^2\ell_p^3}\Big(\int_{\Sigma_2}\ex^{3\lambda}\omega
+\sum_{\Sigma_0}\frac{1}{d_{\Sigma_0}}\frac{2\pi}{\epsilon_1}\frac{y}{2}\Big)\, .
\end{align}
Here $\Sigma_0\subset \Sigma_2$ are the  fixed points inside 
$\Sigma_2$ and the first term is only present (and by itself) 
when the entire $\Sigma_2$ is fixed, while the second term is present (and by itself)
when the fixed point set is $\Sigma_0\subset \Sigma_2$.

\subsection{\texorpdfstring{M5-branes wrapped on $B_4$}{S4 bundles over B4}}
\label{sec:AdS3example}

We will consider a class of solutions that were explicitly found in \cite{Benini:2013cda} (generalizing those
in \cite{Gauntlett:2000ng}). These describe the near-horizon limit of M5-branes wrapping a complex four-manifold $B_4$ inside a Calabi--Yau four-fold. 
We assume the normal bundle of $B_4$ is a bundle of the form $\mathcal{N}= \cL_1 \oplus \cL_2 \to B_4$. The total space of the bundle is Calabi--Yau if \cite{Benini:2013cda}
\begin{equation}
\label{eq:CY4Condition}
	c_1(\mathcal{L}_1) + c_1(\mathcal{L}_2) + t_1 + t_2 = 0 \in H^2(B_4,\Z) \, , 
\end{equation}
where $t_1, t_2$ are the Chern roots\footnote{A nice summary 
of the splitting principle and Chern roots may be found in \cite{bottandtu}.} of the holomorphic tangent bundle $TB_4$. Equation \eqref{eq:CY4Condition} 
is simply the condition that the first Chern class of the tangent bundle of 
the total space of $\mathcal{N}$ is zero, {\it i.e.} 
in this sense it is a (local) Calabi--Yau four-fold. 
Again, the near-horizon limit becomes the sphere bundle
\begin{equation}
	S^4 \hookrightarrow M_8 \to B_4\, ,
\end{equation}
where $S^4\subset \C_1 \oplus \C_2 \oplus \R$ and each $\C_i$ is twisted by $\cL_i$. 

For simplicity, we assume that the R-symmetry Killing vector $\xi$ acts just on the four-sphere and thus has
fixed points at the two poles.  We write it as
\begin{equation}
	\xi = \sum_{i=1}^2 b_i \partial_{\varphi_i} \, , 
\end{equation}
where we are explicitly assuming that $\partial_{\varphi_i}$ are Killing vectors of the supergravity solution.
Also, we take $\Delta\varphi_i=2\pi$ and hence each $\partial_{\varphi_i}$ rotates $\C_i$ with weight 1.
Since the sum of the weights must be $b_1+b_2=1$, as may 
be argued similarly to section \ref{sec:BBBW}, using appendices \ref{app:SpinorRegularity} and \ref{app:AdS7}, we can also write
\begin{equation}\label{b1b2resultswtsads3}
	b_1 = \frac{1}{2}(1 + \varepsilon) \, , \qquad b_2 = \frac{1}{2}(1 - \varepsilon) \, .
\end{equation} 
The fixed point set consists of two copies of $B_4^\pm$ over the two poles, 
denoted in this section by $\pm$, with normal bundles $\mathcal{N}^\pm$ (which are isomorphic to each other, up to orientations). 
We denote by $\Gamma_\alpha$ the two-cycles forming a basis of the free part of $H_2(B_4,\Z)$. 

There are several classes of four-cycles in the geometry: 
(i) The $S^4$ fibre at any  point on $B_4$, (ii)
$C_4^{(i\alpha )}$, which are the total space of $S^2_i\subset S^4$ bundles over $\Gamma_\alpha$, for $i=1,2$, and
(iii) $B_4^\pm$, which are the copies of $B_4^\pm$ at the poles of the sphere.
These four-cycles are not independent, and in particular as shown in 
equations \eqref{franklin}, \eqref{louisa} of appendix 
\ref{app:homology} satisfy the homology relations
\begin{equation}
\label{eq:HomologyRelation_AdS3_1}
	[C_4^{(1\alpha) }] = n_{2 \alpha} [S^4] \, , \qquad [C_4^{(2\alpha )}] = n_{1 \alpha} [S^4] \, , 
\end{equation}
\begin{align}
n_{i \alpha} &\equiv \int_{\Gamma_\alpha} c_1(\cL_i)\,,
\end{align}
{as well as\footnote{We will shortly provide a consistency check of 
\eqref{eq:HomologyRelation_AdS3_12} using localization.}
\begin{align}\label{eq:HomologyRelation_AdS3_12}
[B_4^+]-[B_4^-]
	&= \sum_{\alpha,\beta=1}^{b_2} I_{\alpha\beta} n_{1 \alpha} n_{2 \beta} [S^4] \equiv \langle n_1, n_2 \rangle \, [S^4] \, .
\end{align}
As in section \ref{sec:B4}, here $\Imat$ is the inverse of the intersection matrix for $B_4$, and $\langle \cdot, \cdot \rangle$ is the bilinear form defined by $\Imat$.} 

The integrals of $G$ through the four-cycles should be quantized, and we can evaluate the integrals using localization. For $S^4$ there are fixed points at the poles and we deduce
\begin{equation}
\label{eq:AdS3_NS4}
\begin{split}
	N_{S^4} 
	&= \frac{1}{(2\pi\ell_p)^3} \int_{S^4} \Phi^F 
		= - \frac{1}{(2\pi\ell_p)^3} \frac{(2\pi)^2}{4b_1 b_2}  \left( y_+ - y_- \right) \, ,
\end{split}
\end{equation}
where we used 
$\epsilon_1^+\epsilon_2^+=b_1 b_2$, $\epsilon_1^-\epsilon_2^-=-b_1 b_2$ for the products of weights on the $\R^4$ normal spaces to the 
two poles inside $S^4$.
For $C_4^{(i\alpha )}$, which have fixed point sets $\Gamma_\alpha^\pm$, we find associated fluxes 
\begin{align}
	N_{i \alpha} 
	&= \frac{1}{(2\pi\ell_p)^3} \bigg[ - \frac{\pi}{b_i} \left( c_\alpha^+ - c_\alpha^- \right) + \left(\frac{2\pi}{b_i}\right)^2\frac{1}{4} (y_+ - y_-) n_{i \alpha}  \bigg] \,.
\end{align}
Here we have used the weights 
$\epsilon_i^+=b_i $, $\epsilon_i^-=-b_i$ 
on the $\R^2$ spaces normal to the two poles 
of the $S^2_i$ fibres inside $C_4^{(i\alpha)}$ (which are the copies $\Gamma_\alpha^\pm$ of the two-submanifolds $\Gamma_\alpha$ in $B_4$), and we have also defined 
\begin{align}
\label{eq:AdS3_Defnalphai}
c_\alpha^\pm &\equiv \int_{\Gamma_\alpha^\pm} \ex^{3\lambda} \omega \,.
\end{align}
Recall from \eqref{ads3omegclosedfp} that $c_\alpha^\pm$ only depend on the homology class of ${\Gamma_\alpha^\pm}$, 
and that the normal bundles to 
$\Gamma_\alpha^\pm$ inside $C_4^{(i\alpha)}$ have opposite orientations, correlated 
with the weights $\epsilon_i^\pm$.
Notice we can solve for $c_\alpha^+-c^-_\alpha$:
\begin{equation}
\begin{split}
	c^+_\alpha - c^-_\alpha &= -  \frac{(2\pi\ell_p)^3}{\pi} \left( b_1 N_{1 \alpha} + b_2 N_{S^4} n_{1 \alpha} \right)
	=-  \frac{(2\pi\ell_p)^3}{\pi} \left(  b_2 N_{2 \alpha} + b_1 N_{S^4} n_{2 \alpha} \right) \,.
	\end{split}
\end{equation}
From \eqref{eq:HomologyRelation_AdS3_1} we have
\begin{equation}
	N_{1 \alpha} = n_{2 \alpha} N_{S^4} \, , \qquad N_{2 \alpha} = n_{1 \alpha} N_{S^4} \, ,
\end{equation}
and hence
\begin{equation}
\label{eq:AdS_Diffcalpha}
	c^+_\alpha - c^-_\alpha = -  \frac{(2\pi\ell_p)^3}{\pi} \left(  b_2 n_{1 \alpha} + b_1 n_{2 \alpha} \right) N_{S^4} \,.
\end{equation}

Continuing, we now impose flux quantization through $B_4^\pm$. To do so we recall from \eqref{diffbilinads3}, \eqref{a4phi4def}
that
we have
\begin{align}
\dd( \ex^{6\DA} \phi)&=2\ex^{6\DA}\PhiA+2 y \ex^{3\lambda} F-\Phi^{*F}_4\,.
\end{align}
Thus,
\begin{align}
	N_\pm &= \frac{1}{(2\pi\ell_p)^3} \int_{B_4^\pm} \ex^{3\lambda} F 
	=\frac{1}{(2\pi\ell_p)^3} \frac{1}{2y_\pm}\left( - \int_{B_4^\pm}2 \ex^{6\lambda}\Psi+\int_{B_4^\pm} \Phi^{*F}_4\right) \nn \\
	\label{Npmhomexp}
	&=\frac{1}{(2\pi\ell_p)^3} \frac{1}{2y_\pm}\left(
		{\langle c^\pm, c^\pm \rangle} + \int_{B_4^\pm} \Phi^{*F}_4\right)\ \, ,
\end{align}
where we used \eqref{gstructcondsads3}, \eqref{omegPsifp} and also
\begin{align}
 \int_{B_4^\pm} \ex^{6\lambda}\omega^2 = {\langle c^\pm, c^\pm \rangle}\,.
\end{align}
From the homology relation \eqref{eq:HomologyRelation_AdS3_12} we have 
\begin{align}\label{eq:HomologyRelation_AdS3_12again}
N^+-N^-
	&={ \langle n_1, n_2 \rangle \, N_{S^4} } \, . 
\end{align}

{
We now briefly pause to explain how we can obtain \eqref{eq:HomologyRelation_AdS3_12again} 
using localization without requiring the homology relation \eqref{eq:HomologyRelation_AdS3_12}, or equivalently how the homology relation is consistent with the orientation choices that we use
in the localization formulae. The idea is to trivially extend the equivariant form $\Phi^F$ in \eqref{ads3fluxeqform2} to have top-form that is the
zero six-form. We then integrate this six-form on the six-cycles $S^2_i \to B_4$, to trivially get zero, and then also evaluate it using the BVAB formula \eqref{eq:BV_v3}:
\begin{align}
	0 &= \Big[ N_+ - N_- + \frac{1}{2} \frac{1}{(2\pi\ell_p)^3} \frac{2\pi}{b_i} \langle c^+ - c^-, n_i \rangle - \frac{1}{4} \frac{1}{(2\pi\ell_p)^3} (y_+ - y_-) \frac{(2\pi)^2}{b_i^2} \langle n_i, n_i \rangle \Big] \nn\\
	&= \Big[ N_+ - N_- - \frac{1}{b_i} \langle  b_2 n_{1} + b_1 n_2 , n_i \rangle N_{S^4} + \frac{b_1 b_2}{b_i^2} \langle n_i, n_i \rangle N_{S^4} \Big] \, .
\end{align}
Here, in the first line we used 
\begin{align}\label{omegaclipmads3form}
\int_{B_4^\pm}\ex^{3\lambda}\omega\wedge  c_1(\mathcal{L}_i)= \langle c^\pm, n_i \rangle \, ,
\end{align}
and
\begin{align}\label{liplipmads3form}
\int_{B_4^\pm}c_1(\mathcal{L}_i)\wedge  c_1(\mathcal{L}_j)= \langle n_i, n_j \rangle \, ,
\end{align}
and in the second line substituted \eqref{eq:AdS3_NS4} and \eqref{eq:AdS_Diffcalpha}. Setting $i=1,2$ in \eqref{eq:BV_v3} gives \eqref{eq:HomologyRelation_AdS3_12again}.
}

Continuing, we next consider integrals of $\Phi^{*F}$. Using localization for the first two, we find 
\begin{align}\label{starfequivads3exps}
\int_{S^4}\Phi^{*F}& 
=-(2\pi)^2\frac{1}{4b_1 b_2}(y_+^2-y_-^2)\,,\nn\\
\int_{C_4^{(i\alpha )}}\Phi^{*F}&=2\int_{C_4^{(i\alpha)}}\Phi^{\Psi}
=-\frac{2\pi}{b_{{i}}}(y_+c_\alpha^+-y_-c_\alpha^{{-}})+(2\pi)^2\frac{1}{4b_{{i}}^2}(y_+^2-y_-^2)n_{i \alpha}\,,\nn\\
\int_{B_4^\pm}\Phi^{*F}&=\int_{B_4^\pm} \Phi^{*F}_4\,.
\end{align}
In the last line we are simply emphasizing that the integral of 
$\Phi^{*F}_4\equiv \ex^{6\DA}* F - a_0 \ex^{3\DA} F$
over $B_4^\pm$ will appear in various places below.
The homology relation \eqref{eq:HomologyRelation_AdS3_1} then implies
\begin{equation}
y_+ c_\alpha^+ - y_- c_\alpha^-
=\frac{2\pi }{4} \left( y_+^2 - y_-^2 \right)\left( \frac{n_{1 \alpha}}{b_1} + \frac{n_{2 \alpha}}{b_2} \right)\,.
\end{equation}
Write the left hand side as $\frac{1}{2}(y_++y_-)(c_\alpha^+-c_\alpha^-)+\frac{1}{2}(y_+-y_-)(c_\alpha^++c_\alpha^-)$
and substitute \eqref{eq:AdS_Diffcalpha}. Then substituting 
\eqref{eq:AdS3_NS4} into the right hand side, we deduce that either $y_+=y_-$ or
$c_\alpha^+=-c_\alpha^-$. The former possibility is inconsistent with $N_{S^4}\ne 0$ and so we deduce
that
\begin{equation}
\label{eq:AdS_Diffcalpha2}
	c^+_\alpha= - c^-_\alpha =-  \frac{(2\pi\ell_p)^3}{2\pi} \left(  b_2 n_{1 \alpha} + b_1 n_{2 \alpha} \right) N_{S^4}\,.
\end{equation}
To proceed, we can combine \eqref{Npmhomexp} and \eqref{eq:HomologyRelation_AdS3_12again} to get an expression
for the quantity $(y_+)^{-1}\int_{B_4^+}\Phi^{*F}_4-(y_-)^{-1}\int_{B_4^-}\Phi^{*F}_4$. We also get an expression for
$\int_{B_4^+}\Phi^{*F}_4-\int_{B_4^-}\Phi^{*F}_4$ using \eqref{eq:HomologyRelation_AdS3_12} and the first line in
\eqref{starfequivads3exps}. These can be solved for $\int_{B_4^\pm}\Phi^{*F}_4$ and after using
\eqref{eq:AdS3_NS4} and also \eqref{eq:AdS_Diffcalpha2}, we deduce
\begin{align}
	\int_{B_4^\pm} \Phi^{*F}_4 = & \pm (2\pi \ell_p)^3  y_\pm \langle n_1, n_2 \rangle N_{S^4} \nonumber\\
& 
	-  \frac{(2\pi \ell_p)^6 \ell^3}{(2\pi)^2} \langle n_{1} b_2 + n_{2} b_1,  n_{1} b_2 + n_{2} b_1 \rangle N_{S^4}^2  \, .
\end{align}
If we now substitute this as well as \eqref{eq:AdS_Diffcalpha2} into \eqref{Npmhomexp} we also deduce that
\begin{align}\label{eq:HomologyRelation_AdS3_12againfinal}
N^+=-N^-
	&= \frac{1}{2} \langle n_1, n_2 \rangle N_{S^4} \, .
\end{align}
Notice that, in contrast to \eqref{eq:filippo}, in this case it is not possible to add an additional flux number $M = (N_+ + N_-)/2$, which is set to vanish by the geometry.

We now turn to the central charge \eqref{clocformads3}. We need to compute the integral $\int_{M_8} \Phi$ using localization. 
After some computation, the BVAB formula \eqref{eq:BV_v3} gives
\begin{align}
\label{eq:AdS3_PolyformIntegralAssuming}
	\int_{M_8} \Phi &=  \frac{(2\pi)^2}{b_1b_2} \frac{1}{16} \left[ (y_+ - y_-) \langle c^+, c^+ \rangle -(2\pi\ell_p)^3 (y_+^2 N_+ -  y_-^2 N_- ) \right] \nn \\
	& \ \ \ - \frac{(2\pi)^3}{(b_1b_2)^2} \frac{1}{32} \left[ ( y_+^2 + y_-^2)  \langle c^+, n_{1} b_2 + n_{2} b_1 \rangle \right]\nn \\
	& \ \ \  +\frac{(2\pi)^4}{(b_1b_2)^3} \frac{1}{192} \left[ (y_+^3  -  y_-^3) \left( \langle n_1, n_1 \rangle b_2^2 + \langle n_1, n_2 \rangle b_1b_2 + \langle n_2, n_2 \rangle b_1^2 \right) \right]\,,
\end{align}
where we used \eqref{omegaclipmads3form}.

If we now make the additional assumption
that 
\begin{align}\label{ypmfix}
	y_+=-y_-\,,
\end{align}
things simply further. {More physically}, from the first equation in \eqref{starfequivads3exps} 
we see that this is equivalent to saying that the four-form flux associated with 
$\Phi^{*F}$, through the $S^4$ fibre, is zero: 
\begin{align}\label{fourformfluxcondassumption}
\int_{S^4}(\ex^{6\DA}* F - a_0 \ex^{3\DA} F)=0\,.
\end{align}
In contrast, the  four-form flux associated with 
$\Phi^{F}$, through the $S^4$ fibre, is $N_{S^4}$, the number of M5-branes, as in \eqref{eq:AdS3_NS4}. 
It would be interesting to know if there are solutions that do not 
satisfy \eqref{ypmfix}, \eqref{fourformfluxcondassumption} and if so what those describe {physically}. 
However, assuming this, we immediately have 
$\int_{B_4^+} \Phi^{*F}_4 =\int_{B_4^-} \Phi^{*F}_4$, and the central charge simplifies to
\begin{align}\label{cchgeads31}
c&\equiv \frac{3}{2^5\pi^7\ell_p^9}\int_{M_8}\Phi\,,\nn\\
&= - \big( \langle n_1, n_1 \rangle b_2^2 + 4 \langle n_1, n_2 \rangle b_1b_2 + \langle n_2, n_2 \rangle b_1^2 \big) N_{S^4}^3\,.
\end{align}
This is an off-shell expression for the central charge, since it still depends on $b_i$ subject to the constraint
$b_1+b_2=1$. After substituting 
\eqref{b1b2resultswtsads3} and extremizing over $\varepsilon$, we find that the extremal value
is given by 
\begin{align}\label{extremisedcads3}
c=\frac{ \langle n_1, n_1 \rangle \langle n_2, n_2 \rangle-4 \langle n_1, n_2 \rangle^2}{ \langle n_1, n_1 \rangle-4 \langle n_1, n_2 \rangle+ \langle n_2, n_2 \rangle}(-N_{S^4})^3\,,
\end{align}
and the extremal value of $\varepsilon$ is 
\begin{align}
\varepsilon=\frac{ \langle n_1, n_1 \rangle - \langle n_2, n_2 \rangle}{ \langle n_1, n_1 \rangle-4 \langle n_1, n_2 \rangle+ \langle n_2, n_2 \rangle}(-N_{S^4})^3\,.
\end{align}

We now consider M2-branes wrapped on various calibrated cycles and compute 
the dimension of the dual chiral primaries using \eqref{Deltaformulaads3}. We first consider 
the invariant submanifolds $\Gamma_\alpha^\pm$ located at the north and south poles of the $S^4$ fibre.
From \eqref{gstructcondsads3} we can infer that these will be calibrated cycles provided
that $\omega |_{\Gamma_\alpha^\pm}=\pm j |_{\Gamma_\alpha^\pm}$.
Choosing an appropriate orientation, \eqref{Deltaformulaads3} then gives the off-shell result
\begin{align}\label{DeltaBBBWads3}
\Delta(\Gamma_\alpha^+) = - \frac{1}{(2\pi)^2\ell_p^3}\int_{\Gamma_\alpha^+} 
\ex^{3\lambda}\omega 
= - \frac{c_\alpha^+}{(2\pi)^2\ell_p^3}= \left(  b_2 n_{1 \alpha} + b_1 n_{2 \alpha} \right) N_{S^4} \,,
\end{align}
where we used \eqref{eq:AdS3_Defnalphai} and \eqref{eq:AdS_Diffcalpha2}.
Similarly, choosing a suitable orientation, we have $\Delta(\Gamma_\alpha^-)=\Delta(\Gamma_\alpha^+)$.

Another possibility, is to wrap 
M2-branes on the invariant, homologically trivial, submanifolds $S^2_i$ in the $S^4$ fibre.
Now since $\xi$ is tangent to $S^2_i$, from
\eqref{gstructcondsads3}
the $S^2_i$ will be calibrated by $\omega$
provided that the volume form on $S^2_i$ is, up to sign, given by $\ex^7\wedge \ex^8$. 
Choosing a suitable orientation to get a positive result,
we can then compute $\Delta(S^2_i)$ using localization. We have fixed points at the north and south poles
of $S^2_i$, and find
\begin{align}\label{Deltaformula2last}
\Delta(S^2_i) &= \frac{1}{(2\pi)^2\ell_p^3}\Big(\frac{2\pi}{\epsilon_{i}^+}\frac{y_+}{2}+\frac{2\pi}{\epsilon_{i}^-}\frac{y_-}{2}\Big)
=\frac{2b_1 b_2}{b_i} (-N_{S^4})\, ,
\end{align}
where we used
$\epsilon_i^+=b_i$, $\epsilon_i^-=-b_i$ and also
\eqref{eq:AdS3_NS4}. Note that this result dose not use the assumption
\eqref{ypmfix}.

\subsubsection{\texorpdfstring{$B_4=KE_4^-$}{B4 KE4-}}
As an example, we consider $B_4= KE_4^-$, a K\"ahler--Einstein four-manifold with negative curvature. 
We choose a normalization of the $KE_4^-$ metric so that the K\"ahler form satisfies
$J_{KE} = - \mathcal{R} = - 2\pi c_1(\mathcal{L})$, where $\cL$ is the anti-canonical bundle. It is also useful to recall that the first Pontryagin number and the Euler number of the $KE_4^-$ are given, respectively,
in terms of the Chern roots $t_1$, $t_2$, by
\begin{align}
P_1 &= \int_{KE} (t_1^2 + t_2^2) \, , \qquad \chi = \int_{KE} t_1 t_2 \,.
\end{align}
We also have
\begin{align}
\label{eq:AdS3_KE_DiffTop}
P_1 + 2 \chi &= \int_{KE} (t_1 + t_2)^2 = \int_{KE} c_1(T KE)^2 = \int_{KE} c_1(\cL)^2 
	= \frac{1}{4\pi^2}\int_{KE} J_{KE}^2 \nn\\
	&= \frac{1}{2\pi^2} \Vol(KE) \, .
\end{align}

Now we can solve the Calabi--Yau condition \eqref{eq:CY4Condition} for the Chern roots
with a twist by writing
\begin{equation}
\label{eq:AdS3_KE_Twist}
 c_1(\mathcal{L}_1) = - \frac{1+z}{2}(t_1 + t_2) \, , \qquad c_1(\mathcal{L}_2) = - \frac{1-z}{2}(t_1 + t_2) \, .
\end{equation}
Here $t_1, t_2$ are topological invariants of $B_4=KE_4^-$, 
so that the variable $z$ we have introduced then describes 
a one-parameter family of associated local Calabi--Yau four-folds, 
via the twisting of the normal bundle $\mathcal{L}_1\oplus\mathcal{L}_2$ 
over $B_4=KE_4^-$.

We compute the formulae
\begin{align}
	 \langle n_1, n_1 \rangle  &= \frac{(1+z)^2}{4} \int_{KE} c_1(\cL) \wedge c_1(\cL) = \frac{(1+z)^2}{4} (P_1 + 2\chi) \, ,\nn\\
	 \langle n_1, n_2 \rangle &= \frac{1-z^2}{4} (P_1 + 2\chi) \, , \qquad 	 \langle n_2, n_2 \rangle  = \frac{(1-z)^2}{4} (P_1 + 2\chi) \, .
\end{align}
Substituting into the central charge \eqref{cchgeads31} and using \eqref{b1b2resultswtsads3}, we obtain
\begin{align}\label{cchgeads33}
c
&=  \frac{(P_1 + 2 \chi )}{8}  \big[ 
3-z^2-4\varepsilon z+\varepsilon^2(3z^2-1)  \big](-N_{S^4})^3
 \,,
\end{align}
which agrees with the large $N$ limit of the \emph{off-shell} expression (6.6) of \cite{Benini:2013cda}, obtained from field theory, provided we take $N^{\mathrm{there}}=-N_{S^4}>0$. After extremizing over $\varepsilon$ we get the on-shell result
\begin{align}\label{cke4-ads3}
c
&=  (P_1 + 2 \chi ) \frac{3(1-z^2)^2}{8(1-3z^2)}(-N_{S^4})^3
 \,.
\end{align}
Explicit supergravity solutions with $z=0$ were constructed in
section 4.2 of \cite{Gauntlett:2000ng}, and extended to $|z|<1/3$ in \cite{Benini:2013cda}.

It would be interesting to check the predictions for the conformal dimensions of chiral primaries associated with wrapped M2-branes given in
\eqref{DeltaBBBWads3} and \eqref{Deltaformula2last} with the explicit supergravity solutions.

\subsubsection{\texorpdfstring{$B_4 = \Sigma_1 \times \Sigma_2$}{B4 Sigma1 x Sigma2}}

We can also consider the class of solutions with $B_4\cong \Sigma_1\times \Sigma_2$, where $\Sigma_\alpha$ are Riemann surfaces
with genus $g_\alpha$, as considered in
\cite{Benini:2013cda}, extending \cite{Gauntlett:2001jj}. 
The solutions in \cite{Gauntlett:2001jj,Benini:2013cda} were constructed in $D=7$ gauged supergravity and then uplifted to $D=11$.

For this case we have
\begin{equation}
	I_{\alpha\beta} = \begin{pmatrix}
	 0 & 1 \\ 1 & 0
	\end{pmatrix} \, , \qquad \int_{\Sigma_\alpha} t_\alpha = 2(1-g_\alpha) \, .
\end{equation}
Moreover, we denote the degrees of the normal bundles $\cL_1$ and $\cL_2$ on $\Sigma_\alpha$ as $-p_\alpha$ and $-q_\alpha$, respectively. Thus, the integration of \eqref{eq:CY4Condition} on each of the $\Sigma_\alpha$ gives 
\begin{equation}
\label{eq:CY4Condition_ProductBase}
	p_\alpha +q_\alpha  = 2(1-g_\alpha) \, .
\end{equation}
Using the same parametrization as \cite{Benini:2013cda}, this 
can be solved by writing
\begin{equation}
\label{eq:agu}
	p_\alpha = \begin{cases} (1-g_\alpha) (1-\kappa_\alpha z_\alpha) & g_\alpha \neq 1 \\ - \frac{1}{2} z_\alpha & g_\alpha = 1 \end{cases} \, , \quad q_\alpha = \begin{cases} (1-g_\alpha) (1 + \kappa_\alpha z_\alpha) & g_\alpha \neq 1 \\ \frac{1}{2} z_\alpha & g_\alpha = 0 \end{cases} \, ,
\end{equation}
where $\kappa_\alpha = 0, \pm 1$ is the curvature of the Riemann surface $\Sigma_\alpha$
and $z_\alpha \in \mathbb{Q}$ (consistent with $p_\alpha,q_\alpha\in\mathbb{Z}$). Substituting into \eqref{cchgeads31} then gives the off-shell central charge
\begin{align}
\label{diffcasesads3c}
	c = \begin{cases} 
		(g_1 - 1)(g_2 - 1) \big[ 3 - \varepsilon^2 + 2\varepsilon (\kappa_1 z_1 + \kappa_2 z_2) \\ \qquad \qquad \qquad  \qquad -  \kappa_1 \kappa_2 z_1 z_2(1-3\varepsilon^2) \big] (-N_{S^4})^3\,, & \quad g_{1,2}\neq 1\, , \\
		- \frac{(g_2-1)}{2} \left[ 2 \varepsilon z_1 - \kappa_2 z_1 z_2 ( 1- 3 \varepsilon^2) \right] (-N_{S^4})^3\,, & \quad g_1 = 1, g_2 \neq 1\, , \\ 
		- \frac{1}{4} z_1 z_2 (1-3\varepsilon^2) (-N_{S^4})^3\,, & \quad g_{1,2}=1\, ,
	\end{cases} 
\end{align}
which matches the large $N$ limit of the off-shell formula in \cite{Benini:2013cda}, obtained from field theory.
The on-shell expression can be obtained by extremizing over $\varepsilon$ (or directly from \eqref{extremisedcads3}).
Note that explicit supergravity solutions were constructed in \cite{Benini:2013cda}, extending 
\cite{Gauntlett:2001jj}, when at least one of the $g_\alpha>1$. In particular, no solutions
for a four-torus, when $g_1=g_2=1$, are known to exist.

Note a special case is when we set $z_1=z_2=0$. In this case we have $p_\alpha=q_\alpha=(1-g_\alpha)$
and, with $g_\alpha>1$, for which explicit supergravity solutions are known \cite{Gauntlett:2000ng}, 
this can be viewed as a special case of the $KE_4^-$ case considered above. Indeed, setting $z_1=z_2=0$ and $\kappa_\alpha=-1$ 
in \eqref{diffcasesads3c} we get precise agreement with \eqref{cke4-ads3} after setting $z=0$, $\chi=8(1-g_1)(1-g_2)$ and $P_1=0$.

It would be interesting to check the predictions for the conformal dimensions of chiral primaries associated with wrapped M2-branes given in
\eqref{DeltaBBBWads3} and \eqref{Deltaformula2last} with the explicit supergravity solutions.

\subsection{Action calculation}
\label{sec:actioncalcads3}
Similar to section in \ref{sec:actioncalc}, we now investigate the  action.
For the general class of $AdS_3\times M_8$ solutions of $D=11$ supergravity considered in
\cite{Ashmore:2022ydf}, one can show that the $D=11$ equations of motion, as given in \cite{Gauntlett:2002fz}, give rise to
the following $D=8$ equations of motion. From the Einstein equations we get
\begin{align}
\label{eom8d}
0&=\nabla^2\lambda+9(\nabla\lambda)^2+2-\frac{1}{144}\ex^{-6\lambda}G^2 -\frac{1}{3}f^2\, ,\nn \\
R_{ab}&=9\nabla_{ab}\lambda-9\nabla_a\lambda\nabla_b\lambda+\frac{1}{12}\ex^{-6\lambda}G^2_{ab}
-\frac{1}{2}f_af_b+g_{ab}\left(-2+\frac{1}{2}f^2\right)\,,
\end{align}
along with
\begin{align}\label{ads3beom2}
\dd(\ex^{3\DA}F)&=0,\qquad  \ex^{-6\DA} \dd (\ex^{6\DA}* f)+\frac{1}{2}F\wedge F=0\,,\nn\\
\dd (\ex^{3\DA}f)&=0,\qquad 
\ex^{-6\DA} \dd ( \ex^{6\DA}* F)-f\wedge F=0\,,
\end{align}
arising from the Bianchi identity and the equation of motion for $G$ (and already seen in \eqref{ads3beom}).
These can be obtained\footnote{One should substitute $f=\ex^{-3\lambda}\dd a_0$ and $F=\ex^{-3\lambda}\dd C_3$ and then vary over 
$a_0$, $C_3$, as well as the metric and the scalar field.} by extremizing the eight-dimensional action
\begin{align}
\label{8dactads3}
S=\int_{M_8} \dd^8x \sqrt{g}\,  \ex^{9\lambda}\left(R+90(\nabla\lambda)^2-6-\frac{1}{48}F^2
+\frac{1}{2}f^2\right)+\int_{M_8}\frac{1}{2}\ex^{6\lambda}\, a_0 F\wedge F\,.
\end{align}

We now evaluate this action, partially on-shell, by utilizing the supersymmetry conditions we have been using so far, 
as well as the scalar equation of motion and the trace of the second equation in \eqref{eom8d} together with
the equation of motion for $a_0$ given in \eqref{ads3beom2}: $ \dd (\ex^{6\DA}* f)=-\frac{1}{2}\ex^{6\DA}F\wedge F$.
We can multiply the latter by $a_0$ and write 
\begin{align}\label{fstarfeq}
\frac{1}{2}a_0\ex^{6\DA}F\wedge F=\ex^{9\lambda} f\wedge *f-\diff(a_0 \ex^{6\lambda}*f)\,.
\end{align}
Substituting this into the action, and then following the same steps as in section \ref{sec:actioncalc}, we find that the partially
on-shell action can be written in the form
\begin{align}\label{actosfansads30}
S=-4\int_{M_8} \ex^{9\lambda}\, \vol_8=-4\int_{M_8}\Phi\,,
\end{align}
with $\Phi$ given in \eqref{ads3explicitphicc}.
Thus, once again, the partially on-shell action is precisely proportional to the central charge, and we again see that varying the central charge with
respect to any undetermined coefficients after carrying out localization, is a necessary condition for putting the system on-shell.

Finally, note that if we multiply the scalar equation of motion in \eqref{eom8d} by $\ex^{9\lambda}$, solve for $\ex^{9\lambda}$ and then
substitute into \eqref{actosfansads30} we can write the on-shell action as
\begin{align}\label{actosfansads503}
S&=-4\int_{M_8}\left(\frac{1}{12}\ex^{9\lambda} F\wedge *F-\frac{1}{6}\ex^{9\lambda} f\wedge *f\right)\,.
\end{align}
Then, after using \eqref{fstarfeq} and recalling the equivariant polyforms in \eqref{ads3fluxeqform2}, \eqref{starFpolyads3},
 we find that we can also write the on-shell action in the form
\begin{align}
S=\frac{1}{3}\int_{M_8}\Phi^F\wedge \Phi^{*F}\,.
\end{align}
This is the analogue of the expression that we presented in \eqref{actosfansads502}.

\section{Discussion}\label{sec:disc}
Equivariant localization has been extensively utilized both in geometry and supersymmetric quantum field theory in a wide variety of contexts, with important foundational work including \cite{Duistermaat:1982vw,BV:1982,Atiyah:1984px,Witten:1982im,Nekrasov:2002qd,Pestun:2016zxk}.
In this paper, and in \cite{BenettiGenolini:2023kxp,BenettiGenolini:2023yfe}, we have introduced a new
general calculus, based on spinor bilinears and the BVAB fixed point theorem, that allows one to compute physical quantities in AdS/CFT without having the full explicit supergravity solution, just requiring that the solution exists. 

We considered general classes of $D=11$ supergravity solutions of the form $AdS_{2k+1}\times M_{2(5-k)}$, for $k=1,2,3$, that are dual to SCFTs in $d=2k$ spacetime dimensions with an R-symmetry. The R-symmetry is dual to a Killing vector on $M_{2(5-k)}$, and can be constructed as a Killing spinor bilinear. Using various Killing spinor bilinears that are invariant under the action of the R-symmetry, we constructed equivariantly
closed polyforms and, moreover, showed how the BVAB fixed point formula can be used to compute the central charge of the dual SCFT as well as
the scaling dimension of various chiral primaries dual to wrapped membranes.
For the $AdS_{3}\times M_8$ case, we focused on a general sub-class of geometries but we expect that it is straightforward to extend to the most
general class of solutions that were analysed in \cite{Ashmore:2022ydf}.

For the $AdS_5\times M_6$ examples, there are precisely the right amount of equivariant forms allowing one to explicitly evaluate 
the fixed point contributions arising in the BVAB formula in terms of the topological data and the quantized flux. This is particularly remarkable
since the details of how the BVAB formula is used to achieve this differs in the various classes of examples that we considered. It would be interesting
to have a better understanding of this. By contrast, for the specific classes of $AdS_3\times M_8$ solutions that we considered we imposed one
additional constraint on the fixed point data ($y_+=-y_-$ in \eqref{ypmfix} or equivalently the condition on the four-form flux
\eqref{fourformfluxcondassumption}) in order to evaluate the central charge. 
If one relaxes this constraint one obtains results that depend on an undetermined quantity, and it will
be interesting to know if there are any such solutions. 
For both the $AdS_5\times M_6$ and the $AdS_3\times M_8$ cases we have 
illustrated our formalism by recovering some results for known supergravity solutions, as well calculated new results for more general classes of solutions (providing they exist). There are also other classes of solutions that can be analysed in a straightforward way; for example, it would be interesting to consider $M_8$ which are $S^2$ bundles over
$B_6$ and make contact with the explicit solutions in \cite{Gauntlett:2006qw}. 

A remarkable feature of our formalism is that we derive expressions for BPS quantities, such as the central charge, which are generically off-shell.
This allows one to directly compare with off-shell field theory results, where analogous expressions are obtained utilizing field theory extremization techniques.
In the examples we considered, the off-shell central charges are functions of some undetermined weights of the action of the R-symmetry. 
We also demonstrated that imposing a subset of the supersymmetry conditions as well as a subset of the supergravity equations of motion, one obtains
a partially on-shell action that is proportional to our expression for the central charge. Thus, extremizing our expressions for the off-shell central charge over the undetermined weights is associated with necessary conditions for putting the whole system on-shell. It would be very interesting to determine the precise class of geometries that
we are extremizing over when we partially go on-shell, analogous to what has been achieved in a different setting for Sasaki--Einstein geometry in \cite{Martelli:2005tp} and
GK geometry in \cite{Couzens:2018wnk}.

We anticipate some immediate generalizations of this work, some also discussed in \cite{BenettiGenolini:2023kxp}. 
Firstly, we expect there will be straightforward generalizations of all of the calculations
in this paper to $AdS\times M$ solutions of type IIA and type IIB supergravity when $M$ has even dimension. Second, for $AdS\times M$ solutions of 
type II and $D=11$ supergravity where $M$ has odd dimension we expect that the construction of equivariant polyforms
will be similarly straightforward. However, for this case the details of utilizing localization formulae will be different. There are known techniques to deal with localization in odd dimensions (for instance \cite{Goertsches:2015}), and we expect that it will be possible to use them to compute physical observables in the dual SCFT. This odd-dimensional case may overlap with work on Sasaki--Einstein and GK geometry, and we expect to make contact with the recent work of 
\cite{Martelli:2023oqk}, as well the extensive recent work on GK geometry starting with \cite{Couzens:2018wnk}. 
We plan to report on these topics soon.

\section*{Acknowledgements}

\noindent 
This work was supported in part by STFC grants  ST/T000791/1 and 
ST/T000864/1.
JPG is supported as a Visiting Fellow at the Perimeter Institute. 
PBG is supported by the Royal Society Grant RSWF/R3/183010.

\appendix

\section{Regularity of spinors}
\label{app:SpinorRegularity}

In this appendix we examine how spinors with a definite charge under a Killing vector behave near a fixed point, 
and how this is correlated with the chirality of the spinor. The charge $q$ of the spinor with respect to a Killing vector 
is defined, as usual, via the Lie derivative:
\begin{align}
\cL_{\xi}\epsilon 
\equiv \xi^a\nabla_a\epsilon + \frac{1}{8}(\dd\xi^{\flat})_{ab}\gamma^{ab}\epsilon= \ii q \, \epsilon\,.
\end{align}
We consider a spinor which satisfies a differential condition of the general form\footnote{Our arguments also apply if we have
$\nabla_a\epsilon=(\MM\cdot\gamma)_a\epsilon+(\mathcal{N}\cdot\gamma)_a\epsilon^c$.}
\begin{align}
\nabla_a\epsilon=(\MM\cdot\gamma)_a\epsilon\,,
\end{align}
where $\MM$ is a sum of forms. 
At a fixed point, as we make explicit below, we have $\xi^a(\MM\cdot\gamma)_a\epsilon=0$ so that
$\xi^a\nabla_a\epsilon=0$ and hence 
\begin{align}\label{eq:SpinorCharge}
\fleft \ii q \, \epsilon\fright= \frac{1}{8}(\dd\xi^{\flat})_{ab}\gamma^{ab}\epsilon \,.
\end{align}

\subsection{\texorpdfstring{$\mathbb{R}^2$}{R2}}

We first consider two-component Dirac spinors on $\mathbb{R}^2$. 
A basis for the Clifford algebra is given by $\gamma^1 = \sigma^1, \gamma^2 = \sigma^2$ and we define 
$\gamma_* = -\ii\gamma_{1}\gamma_{2} = \sigma^3$. Chiral spinors are defined via $\gamma_*\epsilon_\pm=\pm\epsilon_\pm$.

We use Cartesian coordinates $(x,y)$ and orthonormal frame $e^a=(\dd x, \dd y)$, both of which are of course globally well-defined on $\R^2$. 
We also have standard polar coordinates $(r,\varphi)$, with $\Delta\varphi=2\pi$.
We consider the Killing vector $\xi=x\partial_y-y\partial_x=\partial_\varphi$ which generates an $SO(2)$ isometry and
has a fixed point at the origin. At the fixed point, since $\xi^x=\xi^y=0$,
we have $ \xi^a(\MM\cdot\gamma)_a\epsilon=0$.
We also have $\dd\xi^{\flat}=2\dd x\wedge \dd y$ and hence at the fixed point from \eqref{eq:SpinorCharge} we deduce 
\begin{align}
\fleft \ii q \, \epsilon\fright= \frac{\ii}{2}\gamma_*{\epsilon} \, .\end{align}

We thus conclude that $|q|=\frac{1}{2}$ and furthermore that the sign of $q$ is correlated with the chirality of the spinor.
This is sensible, since the charge labels the irreducible representation of $U(1)=SO(2)$ in which the spinor transforms; but here the $SO(2)$ is also the ``Lorentz'' group, so the charge labels the irreducible representations of the ``Lorentz'' group, whose elements are the chiral spinors.

\subsection{\texorpdfstring{$\mathbb{R}^4$}{R4}}

We may consider spinors on $\R^4$ similarly. We use Cartesian coordinates $(x_1,y_1,x_2,y_2)$ and a frame
$e^a=(\dd x_1,\dd y_1, \dd x_2 ,dy_2)$. Writing $\R^4=\R^2\oplus\R^2$, 
we can also introduce polar coordinates $(r_1,\varphi_1,r_2,\varphi_2)$ with $\Delta\varphi_i=2\pi$ in order to avoid conical singularities at the planes 
$r_i = 0$.

We consider the Killing vector $\xi$ to be a linear combination of rotations in the two planes:
\begin{align}
\xi = b_1\partial_{\varphi_1} + b_2\partial_{\varphi_2} = \sum_{i=1}^2 b_i\left( - y_i \partial_{x_i} + x_i\partial_{y_i}\right) \, .
\end{align}
We compute, in the Cartesian frame,
\begin{align}\label{eq:dxb}
\diff\xi^{\flat} = 2\begin{pmatrix}
0 & b_1 & 0 & 0 \\
-b_1 & 0 & 0 & 0 \\
0 & 0 & 0 & b_2 \\
0 & 0 & -b_2 & 0
\end{pmatrix} \, .
\end{align}
The norm squared of the Killing vector is $\|\xi\|^2=  b_1^2\, r_1^2+b_2^2\, r_2^2$. Thus, when both
$b_i\ne 0$ the norm vanishes at the origin of $\R^4$, and the fixed point set is a nut. If one of the $b_i=0$, then the fixed point set of 
$\xi$ is a two-plane bolt.

A basis for the Clifford algebra is
$\gamma^1 = \sigma^1 \otimes \identity_2$,
 $\gamma^2 = \sigma^2 \otimes \identity_2$,
 $\gamma^3 = \sigma^3 \otimes \sigma^1$, 
 $\gamma^4 = \sigma^3 \otimes \sigma^2$.
We define the chirality operator $\gamma_* \equiv \gamma^{1234} = - \sigma^3 \otimes \sigma^3$ so that
chiral spinors are given, in the Cartesian frame, by
\begin{equation}
	 \epsilon_+ = \begin{pmatrix}
		0 \\ \epsilon_2 \\ \epsilon_3 \\ 0
	\end{pmatrix} \, , \qquad \epsilon_- = \begin{pmatrix}
		\epsilon_1 \\ 0 \\ 0 \\ \epsilon_4 
	\end{pmatrix}\,.
\end{equation}
We now compute the Lie derivative along $\xi$. Notice that both for a nut and a bolt we have $ \xi^a(\MM\cdot\gamma)_a\epsilon=0$.
Thus at a fixed point, from \eqref{eq:SpinorCharge} we deduce 
\begin{equation}\label{r4actspin}
	\fleft \ii q \epsilon \fright
	= \frac{\ii}{2} \begin{pmatrix}
	(b_1 + b_2) \epsilon_1 \\ (b_1-b_2) \epsilon_2 \\ - (b_1 - b_2) \epsilon_3 \\ - (b_1+b_2) \epsilon_4 
	\end{pmatrix} \, .
\end{equation}

We thus conclude $|q|=\frac{1}{2}|b_1\pm b_2|$ and there is a further correlation with the chirality of the spinor. For example,
suppose the spinor has charge $q=+1/2$, as we assume in the bulk of the paper, with
\begin{align}
\cL_\xi \epsilon = \frac{\ii}{2} \epsilon\,.
\end{align}
If we have a nut, {\it i.e.} both $b_i\ne0$, then we see that the spinor must either have $b_1 + b_2 = \pm 1$ and have negative chirality, or it must have $b_1 - b_2 = \pm 1$ and have positive chirality. For a bolt, with one of the $b_i=0$, say $b_1=0$, we must have $b_2=1$ but in this case the spinor can still
have either chirality.

\section{\texorpdfstring{$AdS_7\times M_4$}{AdS7xM4} solutions}\label{app:AdS7}

In this appendix we consider the case of $AdS_7 \times M_4$ solutions of $D=11$ supergravity. There is a unique solution of this type, which is the Freund--Rubin solution $AdS_7\times S^4$ (up to a quotient of $S^4$), 
so it is straightforward to obtain the central charge by direct construction. Nevertheless, it is interesting to see
how the central charge can be obtained using the method of equivariant localization, without full knowledge of the solution.

\subsection{Equivariantly closed forms}
The metric is a warped product
\begin{equation}
\label{eq:AdS7_AnsatzMetric}
    \diff s^2 = \ex^{2\lambda} [ \diff s^2 (AdS_7) + \diff s^2 (M_4)] \, , 
\end{equation}
where $AdS_7$ has unit radius, and compatibility with the $AdS$ symmetries implies that
$\lambda$ is a function on $M_4$ and the four-form is given by
\begin{equation}
\label{eq:AdS7_AnsatzForm}
    G = \cG \, \vol(M_4) \, ,
\end{equation}
for some constant $\cG$. 

We can decompose the $D=11$ Clifford algebra 
${\rm Cliff}(10,1) \cong {\rm Cliff}(6,1) \otimes {\rm Cliff}(4,0)$ by writing
\begin{equation}
\label{eq:M4_CliffSplitting}
	\Gamma_a = \rho_a \otimes \gamma_{5} \, , \qquad \Gamma_m = \identity \otimes \gamma_m \, ,
\end{equation}
where 
$\rho_a$
generate ${\rm Cliff}(6,1)$ with $\rho_{0123456}=+1 $, the Hermitian $\gamma_m$
generate ${\rm Cliff}(4,0)$
and $\gamma_5 \equiv \gamma_{1234}$ with $\gamma_5^2 = \identity$. 
We decompose the eleven-dimensional spinor $\epsilon_{11}$ by writing
\begin{equation}
\label{eq:M4_Spinor}
	\epsilon_{11} = \ex^{\lambda/2}\psi \otimes \epsilon\, ,
\end{equation}
where $\epsilon$ is a spinor on $M_{4}$ and $\psi$ is a Killing spinor on $AdS_{7}$ satisfying
$D_\alpha\psi= - \frac{1}{2} \rho_\alpha \psi$. 
This ansatz implies that $\epsilon$ satisfies the following equations on $M_4$:
\begin{align}
\label{eq:M4_InternalSpaceKSE}
	0 &= \Big[ \nabla_m + \frac{1}{2} ( 1 - \tfrac{1}{2}{  \ex^{-3\lambda} \cG}) \gamma_m \gamma_5  \Big] \epsilon \, , \nn\\
    	0 &= \Big[ \gamma^m \partial_m \lambda - ( 1 - \tfrac{1}{6}{ \ex^{-3\lambda} \cG} ) \gamma_5 \Big] \epsilon \, .
\end{align}

We can now construct various bilinears using the spinor $\epsilon$ and consider the algebraic and differential equations that follow from \eqref{eq:M4_InternalSpaceKSE}. It is not difficult to deduce that $\lambda$ must be constant, so we set $\ex^{\lambda}\equiv L$. This then fixes the flux to be
\begin{equation}
	 \cG = 6 \ex^{3\lambda} \, .
\end{equation}
The first equation in \eqref{eq:M4_InternalSpaceKSE} then becomes the Killing spinor
equation $\nabla_m\epsilon  = \gamma_m \gamma_5\epsilon$ and hence we deduce \cite{Bar:1993gpi} that
$M_4$ must be $S^4$ (or a quotient thereof) and $\diff s^2(M_4)=\frac{1}{4} \diff s^2(S^4)$ where $\diff s^2(S^4)$ has unit
radius and volume $8\pi^2/3$. Thus, the only supersymmetric 
$AdS_7$  solution of $D=11$ supergravity is locally the Freund--Rubin solution $AdS_7\times S^4$. It is
then straightforward to carry out flux quantization and compute the central charge from the explicit solution.

Here, instead, we want to compute the central charge using Killing spinor bilinears and localization. Many of the expressions
have direct generalizations to the examples considered in the main text.
To proceed we 
first note that \eqref{eq:M4_InternalSpaceKSE} implies $\overline{\epsilon}\epsilon$ is a constant and we normalize
so that $\overline{\epsilon}\epsilon=1$. We define the following differential forms on $M_4$:
\begin{align}
	1 &= \overline{\epsilon}\epsilon \, , & \sin\zeta &\equiv \overline{\epsilon} \gamma_5 \epsilon \, , & K &\equiv \overline{\epsilon} \gamma_{(1)} \epsilon \, , \nn\\
	\xi^\flat &\equiv -\tfrac{\ii}{4} \overline{\epsilon} \gamma_{(1)} \gamma_{5} \epsilon \, , & Y &\equiv - \ii \overline{\epsilon} \gamma_{(2)} \epsilon \, , & Y' &\equiv \ii \overline{\epsilon} \gamma_{(2)} \gamma_5 \epsilon \, ,
\end{align}
where $\gamma_{(r)} \equiv \frac{1}{r!} \gamma_{\mu_1\cdots \mu_r} \diff x^{\mu_1} \wedge \cdots \wedge \diff x^{\mu_r}$. 
From \eqref{eq:M4_InternalSpaceKSE} it follows that the vector $\xi$, dual to the one-form $\xi^\flat$, is a Killing vector
and we have chosen a normalization so that  $\epsilon$ has charge $\frac{1}{2}$ under the action of $\xi$:
\begin{align}\label{normksads7}
\cL_\xi \epsilon = \frac{\ii}{2}\epsilon\,.
\end{align}
The square norm of the Killing vector is given by
\begin{equation}
\label{eq:M4_Normxi}
	16 \lVert \xi \rVert^2 = \cos^2\zeta = \lVert K \rVert^2 \, .
\end{equation}
Thus, the fixed points of $\xi$ correspond to $\sin\zeta = \pm 1$ and we 
observe that
at such a fixed point the spinor is chiral/anti-chiral, satisfying $\gamma_5\epsilon=\pm\epsilon$.

We can also show that
\begin{align}
    \xi \hook G &= 6\, \ex^{3\lambda}  \xi \hook \vol =  -\tfrac{1}{4} \diff (\ex^{3\lambda}  Y')  \, ,\nn\\
    \xi \hook Y' &= -\tfrac{1}{4} K \, , \qquad \diff \sin\zeta = - 2 K \, ,
\end{align}
which implies that the polyform, extending the four-form $G$,
\begin{equation}
\label{eq:M4_PhiG}
    \Phi^G = G - \tfrac{1}{4}\, \ex^{3\lambda}  Y' -\tfrac{1}{32}\, \ex^{3\lambda}  \sin\zeta\,,
\end{equation}
is equivariantly closed: $\diff_\xi\Phi^G = 0$. This allows us to use the equivariant localization formula to compute the flux quantization of the four-form.

The holographic central charge of the dual field theory is given by \cite{Henningson:1998gx, Bastianelli:2000hi}
\begin{equation}
\label{eq:CentralCharge_6d}
	a =  \frac{3}{112 \pi^5\ell_p^9} \int_{M_{4}} \ex^{9\lambda}\,  \vol(M_{4}) \, .
\end{equation}
For this example, the integrand is proportional to the four-form $G$, since $\lambda$ is constant. Thus, we can immediately
adapt \eqref{eq:M4_PhiG} to write down the equivariant completion of the integrand of the central charge:
\begin{equation}
\label{eq:M4_Phi}
	\Phi = \ex^{9\lambda}\,  \vol(M_4) - \tfrac{1}{24}\, \ex^{9\lambda} * Y -  \tfrac{1}{192}\, \ex^{9\lambda} \sin\zeta  \, .
\end{equation}
Here, we have written $Y' =  *\, Y$ to be consistent with the results in other dimensions.

\subsection{Localization}

With these polyforms, we are able to evaluate the central charge of the field theory dual to any $M_4$ using localization formulae after fixing the topology of $M_4$. For this case we know that
$M_4=S^4$, with a round metric, is the only solution, so we proceed with just the assumption that \emph{topologically} $M_4=S^4$
and that $\xi$ is a Killing vector field of the form
\begin{equation}
	\xi = b_1 \partial_{\varphi_1} + b_2 \partial_{\varphi_2} \, ,
\end{equation}
where $\varphi_i$ are the polar angles on $\C_i$ if we view $S^4 \subset \C_1 \oplus \C_2 \oplus \R$. The weights $b_i$ are arbitrary, and the fixed points of $\xi$ are at the two poles, which we label $N$ and $S$. 
We observe that $\epsilon_1^N\epsilon_2^N=b_1b_2$ and 
$\epsilon_1^S\epsilon_2^S=-b_1b_2$, with the change of relative 
orientation at the poles as in the 
 example considered in the $S^4$ example in section \ref{sec:s4example}.

Quantization of the flux through $S^4$ can be imposed by applying the localization formula to \eqref{eq:M4_PhiG}:
\begin{equation}
	N \equiv \frac{1}{(2\pi\ell_p)^3}\int_{S^4} G = - \frac{1}{(2\pi)^3}\frac{(2\pi)^2}{b_1b_2} \frac{1}{32} \left( \frac{L}{\ell_p} \right)^3 \left( \sin\zeta_N - \sin\zeta_S \right) \, .
\end{equation}
For $N\neq 0$, we must take $\sin\zeta_N = -\sin\zeta_S$, since $|\sin\zeta|=1$ at a fixed point.
The central charge \eqref{eq:CentralCharge_6d} can be similarly computed using \eqref{eq:M4_Phi} and we obtain
\begin{align}\label{accs4ex}
	a &=  \frac{3}{112 \pi^5 \ell_P^9} \int_{S^4} \Phi = -  \frac{1}{2^{9}\cdot 7 \pi^5} \frac{(2\pi)^2}{b_1b_2} \left( \frac{L}{\ell_p} \right)^9 \sin\zeta_N\nn \\ 
	&= \frac{256}{7} (b_1b_2)^2 N^3 \, .
\end{align}

Now, importantly, we need to consider regularity of the spinor at the two poles. As we explain in appendix \ref{app:SpinorRegularity}, there
is a relationship between the chirality of the spinor and the weights at the poles and the charge of the Killing spinor. 
Suppose\footnote{One can argue similarly when $\sin\zeta_N ={+}1$.}
$\sin\zeta_N ={-}1$; then the spinor at the north pole has {negative} chirality and from appendix \ref{app:SpinorRegularity},
given the  normalization in \eqref{normksads7}, we deduce that 
$(\epsilon_1+\epsilon_2)_N=b_1+b_2=\pm 1$. By changing the sign of both
$b_i$ if necessary, without changing the orientation on $S^4$, 
without loss of generality we can assume 
\begin{align}\label{biconstads7}
b_1+b_2=1\,.
\end{align}
At the south pole $\sin\zeta_S ={+}1$, and the spinor has positive chirality.

With the constraint \eqref{biconstads7}, the result for the central charge in \eqref{accs4ex}
matches the large $N$ limit of the \emph{off-shell} trial central charge of the six-dimensional $(2,0)$ theory, with mixing due to the $U(1)^2$ maximal torus in the $SO(5)$ R-symmetry. Carrying out the extremization over the weights $b_i$ subject to \eqref{biconstads7}, we deduce that
the extremal on-shell weights are $b_1 = b_2 = \frac{1}{2}$ and the central charge is $a=16/7 N^3$.  
Note\footnote{If we had chosen $\sin\zeta_N =+1$ above, we find $b_1-b_2=\pm 1$ and after extremization $b_1 b_2<0$ and so $N<0$, but still with the extremized central charge $a>0$.} also that
at the extremal values $b_1 b_2>0$ and so $N>0$.
We have thus recovered the results
that can alternatively be obtained from direct evaluation in the Freund--Rubin solution. 
We emphasize that our new derivation did not require the explicit metric on $S^4$ and, moreover, since we only utilised a subset of
the information contained in the Killing spinors, we obtained an off-shell expression for the central charge.

\section{Homology relations from Poincar\'e duals}
\label{app:homology}

In the main text we have used a number of homology relations 
for manifolds which are the total spaces of even-dimensional sphere 
bundles. These may be proven  using the 
explicit differential form construction of Poincar\'e duals in \cite{bottandtu}, 
together with a simple additional ingredient. We mainly focus on the total spaces of $S^2$ bundles over a four-manifold $B_4$, as in section \ref{sec:B4}, to illustrate the 
general method. We then briefly indicate 
the extension to $S^4$ bundles that appear in sections \ref{sec:BBBW} and \ref{sec:AdS3example}. 

Consider a six-manifold $M_6$ which is the total space of an $S^2$ bundle 
over a four-manifold $B_4$, with projection map $\pi:M_6\rightarrow B_4$. More specifically, we are interested 
in  unit sphere bundles inside the $\R^3$ bundle $\mathcal{L}\oplus\R\rightarrow B_4$, 
where $\mathcal{L}$ is a complex line bundle. Choosing 
coordinates $(z,x)$ on the $\R^3=\C\oplus \R$ fibre, the unit 
sphere is $|z|^2+x^2=1$, with the north and south pole 
sections $B_4^N$, $B_4^S$ at $\{z=0, x=\mp 1\}$, respectively. 
We choose orientations 
such that the normal bundle to the north pole section $B_4^N$ inside 
$M_6$ is $\mathcal{L}$, so that the normal bundle to the south pole section 
$B_4^S$ is then $\mathcal{L}^{-1}$.\footnote{Comparing to section 
\ref{sec:B4} we then have $B_4^+=B_4^S$, $B_4^-=B_4^N$ -- see \eqref{Lpm}.} 

\begin{figure}
  \centering
    \includegraphics[width=0.5\textwidth]{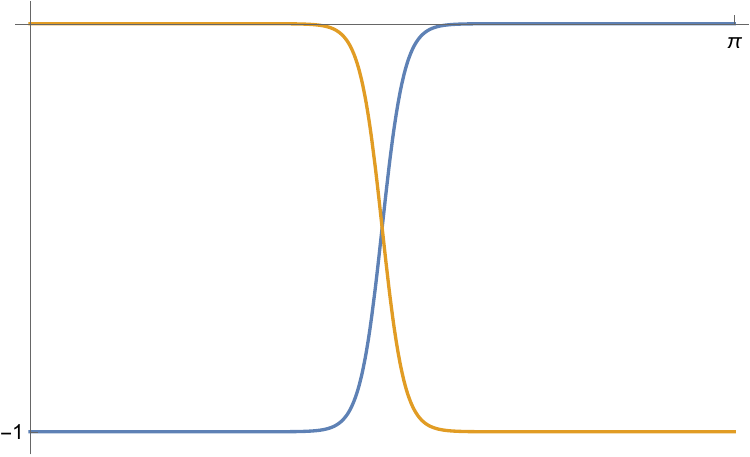}
 \caption{The function $\rho_N$ (in blue) and $\rho_S$ (in orange), satisfying 
$\rho_N+\rho_S=-1$.}
\label{fig:rho}
\end{figure}

We may then write down the following explicit representative 
of the Poincar\'e dual to $B_4^N$, using the construction in \cite{bottandtu}:
\begin{align}\label{PsiN}
\Psi_N \equiv \mathrm{PD}\left[B_4^N\right]\equiv  \frac{1}{2\pi}\diff [\rho_N(\diff \psi - \pi^*A)]\, ,
\end{align}
which is closed, but not exact.
Here $\psi$ is an azimuthal coordinate for the two-sphere fibre, with period $2\pi$, and  $A$ is a local connection form for $\mathcal{L}$, with curvature $F=\diff A$ 
and first Chern class $c_1(\mathcal{L})=[F/2\pi]\in H^2(B_4)$. The combination 
$(\diff \psi- \pi^*A)/2\pi$ is called the \emph{global angular form} 
in \cite{bottandtu}, and is a global one-form on $M_6\setminus (B_4^N\cup B_4^S)$, {\it i.e.} away from the north and south pole sections. 
The function $\rho_N$ is taken to depend only on the 
polar angle ``$\vartheta$'' on the two-sphere, which near to the 
north pole $\vartheta=0$ is also a radial coordinate away from $B_4^N$. 
$\rho_N$ has the property that it is $-1$ near to $\vartheta=0$, and is $0$ 
near to the south pole $\vartheta=\pi$ -- see Figure \ref{fig:rho}. 
 Similarly $\pi-\vartheta$ is a radial 
coordinate away from $B_4^S$, which has Poincar\'e dual
\begin{align}\label{PsiS}
\Psi_S \equiv \mathrm{PD}\left[B_4^S\right] \equiv-\frac{1}{2\pi}\diff [\rho_S(\diff \psi - \pi^*A)]\, .
\end{align}
Here the minus sign is due to the fact that $\diff(\pi-\vartheta)=-\diff\vartheta$. It will be convenient to take $\rho_S$ to satisfy
\begin{align}\label{rhosum}
\rho_N + \rho_S = -1 \, ,
\end{align}
although, as with the choice of $\rho_N$, the cohomology class
 of $\Psi_S$ in $H^2(M_6)$ is independent of the precise form of 
$\rho_S$. Given \eqref{rhosum}, one easily checks that
\begin{align}\label{Pdual}
\Psi_N - \Psi_S = \frac{1}{2\pi}\pi^*F 
\quad \Rightarrow \quad [\Psi_N]-[\Psi_S] = \pi^* c_1(\mathcal{L})\in H^2(M_6)\, .
\end{align}

As in section \ref{sec:B4}, we may introduce a basis $\Gamma_\alpha$ for the 
free part of $H_2(B_4,\Z)$, and furthermore introduce 
Poincar\'e duals $\Psi_\alpha\equiv \mathrm{PD}\left[\Gamma_\alpha\right]\in H^2(B_4)$. Given a closed two-form 
$\omega$ on $B_4$, we may expand in a basis in two different ways:
\begin{align}
c_\alpha\equiv \int_{\Gamma_\alpha}\omega\, , \qquad \omega\equiv 
\sum_{\alpha=1}^{b_2}\hat{c}_\alpha \Psi_\alpha\, ,
\end{align}
where $b_2\equiv \dim H_2(B_4,\R)$. Given the defining property of the 
Poincar\'e dual $\Psi_\alpha$ we immediately have
\begin{align}
c_\alpha = \int_{B_4}\omega\wedge \Psi_\alpha = \sum_{\beta=1}^{b_2} \Qmat_{\alpha\beta} \, \hat{c}_\beta\, ,
\end{align}
where we have introduced the \emph{intersection form}
\begin{align}
\Qmat_{\alpha\beta}\equiv \int_{B_4}\Psi_\alpha\wedge \Psi_\beta = [\Gamma_\alpha] \cap [\Gamma_\beta]\, ,
\end{align}
which is an integer-valued unimodular symmetric matrix. For example, we may then compute
\begin{align}
\int_{B_4}\omega\wedge\omega = \sum_{\alpha,\beta=1}^{b_2}\Qmat_{\alpha\beta}\, \hat{c}_\alpha 
\hat{c}_\beta  = \sum_{\alpha,\beta=1}^{b_2}Q^{-1}_{\alpha\beta}\, c_\alpha c_\beta = \sum_{\alpha,\beta=1}^{b_2}\Imat_{\alpha\beta}\, c_\alpha c_\beta \, ,
\end{align}
where we have introduced\footnote{Notice we can introduce a dual basis 
$\hat{\Psi}_\alpha$ satisfying $\int_{B_4}\hat{\Psi}_\alpha\wedge \Psi_\beta=\delta_{\alpha\beta}$, so that $\omega = \sum_{\alpha=1}^{b_2}c_\alpha \hat{\Psi}_\alpha$ and $\hat{\Psi}_\alpha = \sum_{\beta=1}^{b_2}Q^{-1}_{\alpha\beta}\Psi_\beta$.}
\begin{align}
\Imat\equiv \Qmat^{-1}\, .
\end{align}

With this notation in hand, we may then expand
\begin{align}
c_1(\mathcal{L})=\sum_{\alpha=1}^{b_2} \hat{n}_\alpha \Psi_\alpha\, ,
\end{align}
where  $\hat{n}_\alpha\in\Z$. Given 
a closed four-form $G$ on $M_6$, using \eqref{Pdual} we then compute
\begin{align}\label{homrelproof}
\int_{B_4^N}G -\int_{B_4^S}G & = \int_{M_6}(\Psi_N-\Psi_S)\wedge G  \nonumber\\
& 
= \sum_{\alpha=1}^{b_2}\hat{n}_\alpha \int_{M_6}\pi^*\Psi_\alpha\wedge G
= \sum_{\alpha=1}^{b_2} \hat{n}_\alpha \int_{C_4^{(\alpha)}} G\nonumber\\
& = \sum_{\alpha,\beta =1}^{b_2}\Imat_{\alpha\beta}n_\alpha \int_{C_4^{(\beta)}} G\, .
\end{align}
Here $C_4^{(\alpha)}$ represent four-cycles in $M_6$, and 
are the total spaces of the $S^2$ bundle restricted to the $\Gamma_\alpha\subset B_4$, 
again as in section \ref{sec:B4}. The equality \eqref{homrelproof} then 
precisely proves the homology relation \eqref{homrel}, used 
in the main text.

Next consider the two-cycles in $M_6$ defined by taking copies 
of $\Gamma_\alpha\subset B_4$ inside the north and south pole 
sections $B_4^N$, $B_4^S$. We denote these by $\Gamma_\alpha^N$, 
$\Gamma_\alpha^S$, respectively, with homology classes 
$[\Gamma_\alpha^N],[\Gamma_\alpha^S]\in H_2(M_6,\Z)$. These
have Poincar\'e duals
\begin{align}
\mathrm{PD}\left[\Gamma_\alpha^N\right]= \pi^*\Psi_\alpha\wedge \Psi_N\, , \quad 
\mathrm{PD}\left[\Gamma_\alpha^S\right]= \pi^*\Psi_\alpha\wedge \Psi_S \, \in \, H^4(M_6)\, ,
\end{align}
and from \eqref{Pdual} we compute
\begin{align}\label{PdualGamma}
\mathrm{PD}\left[\Gamma_\alpha^N\right]- \mathrm{PD}\left[\Gamma_\alpha^S\right] = \pi^*[
\Psi_\alpha\wedge c_1(\mathcal{L})]\, .
\end{align}
On the other hand,  by definition
\begin{align}
\label{eq:DefinitionnAlpha}
\int_{B_4}\Psi_\alpha\wedge c_1(\mathcal{L}) = n_\alpha\, ,
\end{align}
and the Poincar\'e dual of \eqref{PdualGamma} gives
\begin{align}\label{m6casenpmnmhom}
[\Gamma_\alpha^N]-[\Gamma_\alpha^S] = n_\alpha [S^2_{\mathrm{fibre}}] \, \in \, H_2(M_6,\Z)\, ,
\end{align}
which is the homology relation \eqref{twocycles} used in the main text.
To see \eqref{m6casenpmnmhom}, we simply integrate an arbitrary closed two-form on both sides and use the fact
that $\Psi_\alpha\wedge c_1(\mathcal{L})$ is proportional to the volume form of $B_4$.

We may similarly analyse other sphere bundles, and in the main text 
$S^4$ bundles appear in several classes of examples. Consider 
a total space $M$ with projection map $\pi:M\rightarrow B$, where 
$M$ is the unit sphere bundle inside the $\R^5$ bundle 
$\mathcal{L}_1\oplus\mathcal{L}_2\oplus\R$, where $\mathcal{L}_i$ 
are two complex line bundles. Choosing coordinates $(z_1,z_2,x)$ 
on the $\R^5=\C_1\oplus\C_2\oplus\R$ fibre, the unit sphere is 
$|z_1|^2+|z_2|^2+|x|^2=1$, with the north and south pole sections 
$B^N$, $B^S$ at $\{z_1=z_2=0,x=\mp 1\}$, respectively. 

In section 
\ref{sec:BBBW} we consider the case where $B=\Sigma_g$ is a Riemann 
surface, and defined four-cycles $C_4^{(i)}$ as the total spaces of 
$S^2_i$ bundles over $\Sigma_g$. Specifically, $C_4^{(1)}$ is 
defined as the locus $\{z_2=0\}$, whose fibre is then a unit sphere inside 
$\C_1\oplus\R\subset \R^5$. Its normal bundle inside $M$ is $\mathcal{L}_2$, 
with Poincar\'e dual $\mathrm{PD}\big[C_4^{(1)}\big]$ which may 
be constructed analogously to \eqref{PsiN}. But also then
$\mathrm{PD}\big[C_4^{(1)}\big]=\pi^*c_1(\mathcal{L}_2)$, 
which taking the Poincar\'e dual immediately gives
\begin{align}
\big[C_4^{(1)}\big] = p_2 [S^4_{\mathrm{fibre}}]\, ,
\end{align}
where $p_i\equiv \int_{\Sigma_g}c_1(\mathcal{L}_i)$. This 
proves \eqref{hom4} in the main text, with an analogous computation 
holding for the cycle $C_4^{(2)}$. 

Finally, we may construct Poincar\'e duals to the north and south pole 
sections analogously to \eqref{PsiN}. For example, we may write
\begin{align}
\Psi_N \equiv \mathrm{PD}\left[B^N\right]\equiv  \diff (\rho_N \eta)\, , \quad \mbox{where}\quad \diff \eta =-\pi^* e(\mathcal{N}B^N)\, .
\end{align}
Here $\eta$ is a global angular three-form for the normal bundle 
$\mathcal{N}B^N$ of $B^N\hookrightarrow M$, and $e(\mathcal{N}B^N)$ is the Euler class. 
Using a similar construction at the south pole section, 
and again taking \eqref{rhosum}, one similarly shows
\begin{align}\label{Pdual_S4}
[\Psi_N] - [\Psi_S] =\pi^* e(\mathcal{N}B^N)=\pi^*e(\mathcal{L}_1\oplus\mathcal{L}_2) = \pi^*\left(
c_1(\mathcal{L}_1)\wedge c_1(\mathcal{L}_2) \right) \in H^4(M)\, .
\end{align}
 Note that this is not in contradiction with the comment below \eqref{N12}, as in that case the base $\Sigma_g$ is two-dimensional and the right hand side of \eqref{Pdual_S4} is zero.

In section \ref{sec:AdS3example} we consider the case where $B=B_4$ is a four-manifold. The relation \eqref{Pdual_S4} paired with a closed four-form $G$ on $M_8$ gives a relation analogous to \eqref{homrelproof}
\begin{align}\label{franklin}
	\int_{B_4^N} G - \int_{B_4^S}G &= \int_{M_8}(\Psi_N - \Psi_S) \wedge G \nn \\ 
	&= \sum_{\alpha,\beta=1}^{b_2} I_{\alpha\beta} n_{1\alpha} n_{2\beta} \int_{S^4_{\rm fibre}} G \, , 
\end{align}
where $n_{1\alpha}$, $n_{2\alpha}$ are the coefficients of the expansion of $c_1(\cL_{1,2})$ on $\Gamma_\alpha$, as in \eqref{eq:DefinitionnAlpha}. Furthermore, we consider the four-cycles $C_4^{(i\alpha)}$ obtained as the total spaces of the $S^2_i$ bundles over $\Gamma_\alpha$. We then have PD$\, [C^{(1\alpha)}_4] = \pi^* \Psi_\alpha \wedge c_1(\cL_2)$ (and the analogous relation for $C^{(2\alpha)}_4$), and upon taking the Poincaré dual
\begin{equation}\label{louisa}
	[C^{(1\alpha)}_4] = n_{2\alpha} [S^4_{\rm fibre}] \, .
\end{equation}

\bibliographystyle{utphys} 
 \bibliography{helical}{}

\end{document}